\pdfoutput=1 % To be processed as a pdf latex.
\documentclass[11pt,a4paper,oneside,onecolumn]{report}
\usepackage{amsmath,amssymb,amsfonts,amsthm,amscd,graphicx,color,keyval,slashed,braket,bm,geometry,hyperref,relsize}
\usepackage[nottoc]{tocbibind} % Include the bibliography in the table of contents.
\DeclareMathAlphabet\mathbfcal{OMS}{cmsy}{b}{n}
\newcommand{\SpeAbs}[1]{{\Huge\abstract{{\normalsize{#1}}}}} % To increase the font size of the title, while keeping the text unchanged.
\newcommand{\NewTableofContents}{\hypersetup{pageanchor=false}\tableofcontents\hypersetup{pageanchor=true}} % To avoid conflict with page numbering.
\newcommand{\SpeAcknowledgments}[1]{{\chapter*{Acknowledgments}{#1}}} % The acknowledgments part to put in the beginning (in english and in french).
\newcommand{\ita}[1]{\textit{#1}} % To put titles in italic, mainly in the bibliography.
\newcommand{\bibtitle}[1]{\item[]\hspace{-\labelwidth}\hspace{-\labelsep}\textbf{\Large #1}} % To put titles inside the bibliography.
\newcommand\trmi[1]{{\mbox{\scriptsize #1}}} % Subscript name.
\newcommand{\beq}{\begin{equation}} % Begin an equation.
\newcommand{\eeq}{\end{equation}} % End an equation.
\newcommand{\beqa}{\begin{eqnarray}} % Begin an array of equation.
\newcommand{\eeqa}{\end{eqnarray}} % End an array of equation.
\newcommand{\aE}[1]{\alpha_\trmi{E#1}} % Aleph functions.
\def\sumint{\hbox{$\sum$}\!\!\!\!\!\!\!\int} % Sum-Integral symbol.
\newcommand{\lmsb}{\bar{\Lambda}} % MSbar scheme, renormalization scale.
\newcommand{\lQCD}{\Lambda_{\rm\overline{MS}}} % QCD scale in the MSbar scheme.
\newcommand{\vect}[1]{\mathbf{#1}} % Bold vectors.
\newcommand\intec[3]{\oint_{\lower+.25em\hbox{$\mbox{\tiny $#1$}$}}\kern-#2 \mbox{\Large $\frac{\mathop{{\rm d}\omega}}{2\pi i}$}\kern-#3\mbox{}} % Contour integral along the path #1.
\newcommand\disc{{\rm Disc}\ } % Disc() symbol.
\def\inteddwbis{\int_{\kern-0.25em \lower.5em \hbox{\mbox{\scriptsize $0$}}}^{\kern-0.25em \lower-.5em\hbox{\mbox{\scriptsize $k$}}}\kern-0.50em\mbox{\normalsize $\mathop{{\rm d}\omega}$}\ } % Another 1d momentum integration over w (going with another 1d momentum integration over k).
\newcommand\ImPart{\rm {Im}\ } % Imaginary part.
\newcommand\RePart{\rm {Re}\ } % Real part.
\newcommand\MSbar{{\overline{\mbox{\trmi{MS}}}}} % MSbar symbol.
\newcommand\inteddwters[2]{\int_{\kern-0.25em \lower.5em\hbox {\mbox{\scriptsize $#1$}}}^{\kern-0.25em \lower-.5em \hbox{\mbox{\scriptsize $#2$}}}\kern-0.75em\mbox{\normalsize $\mathop{{\rm d}\omega_\trmi{E}}$}} % Again another 1d momentum integration over w (going with another 1d momentum 
\def\intedk{\int_{\kern-0.25em \lower.5em \hbox{\mbox{\scriptsize $0$}}}^{\kern-0.25em \lower-.5em\hbox {\mbox{\scriptsize $\infty$}}}\kern-0.75em\mbox{\normalsize $\mathop{{\rm d}k}$}\ } % 1d momentum integration over k.
\def\inteddw{\int_{\kern-0.25em \lower.5em \hbox{\mbox{\scriptsize $0$}}}^{\kern-0.25em \lower-.5em \hbox{\mbox{\scriptsize $\infty$}}}\kern-0.75em\mbox{\normalsize $\mathop{{\rm d}\omega}$}\ } % 1d momentum integration over w (going with another 1d momentum integration over k).
\def\inteddk{\int_{\kern-0.25em \lower.5em \hbox{\mbox{\scriptsize $\omega$}}}^{\kern-0.25em \lower-.5em \hbox{\mbox{\scriptsize $\infty$}}}\kern-0.75em\mbox{\normalsize $\mathop{{\rm d}k}$}\ } % 1d momentum integration over k (going with another 1d momentum integration over w).
\def\intdwdk{\int_{\kern-0.25em \lower.5em\hbox{\mbox{\scriptsize $0$}}}^{\kern-0.25em \lower-.5em\hbox{\mbox{\scriptsize $\infty$}}}\kern-0.75em \mbox{\normalsize $\mathop{{\rm d}\omega}$}\kern-0.15em\int_{\kern-0.25em \lower.5em\hbox{\mbox{\scriptsize $\omega$}}}^{\kern-0.25em \lower-.5em \hbox{\mbox{\scriptsize $\infty$}}}\kern-0.75em\mbox{\normalsize $\mathop{{\rm d}k}$}\,} % combination of 1d momentum integration over k and 1d momentum integration over w, to save some space.
\newcommand\arctanh{\text{arctanh}} % Inverse hyperbolic tangent.
\newcommand{\intedki}[1]{\int\kern-0.0em\mbox{\normalsize $\mathop{{\rm d^{3-2\epsilon}}k_{#1}}$}\ } % 1d momentum integration over k_i.
\begin{document}
%%%%%%%%%%%%%%%%%%%%%%%%%%%%%%%%%%%%%%%%%%%%%%%%%%
%                    HEADING.
%%%%%%%%%%%%%%%%%%%%%%%%%%%%%%%%%%%%%%%%%%%%%%%%%%
\setcounter{page}{1}
\newpage
\thispagestyle{empty}
\vspace*{1truecm}
\begin{center}
{\Large{Ph.D. thesis}}\\*
\vspace*{0.5truecm}
{\Large{Faculty of Physics}}\\*
\vspace*{0.2truecm}
{\Large{Bielefeld University}}
\end{center}
\vspace*{4truecm}
\begin{center}
{\bfseries\huge{Probing the finite density equation of state of}}\\*
\vspace*{0.25truecm}
{\bfseries\huge{QCD via resummed perturbation theory}}\\*
\vspace*{0.8truecm}
{\LARGE{Sylvain Mogliacci}}
\end{center}
\vspace*{5.6truecm}
\begin{center}
{\itshape\Large{In attainment of the academic degree}}\\*
\vspace*{0.25truecm}
{\itshape\Large{Doctor rerum naturalium}}\\*
\end{center}
\vspace*{1.6truecm}
\begin{center}
{\Large{Bielefeld}}\\*
\vspace*{0.25truecm}
{\Large{June 2014}}\\*
\end{center}
%%%%%%%%%%%%%%%%%%%%%%%%%%%%%%%%%%%%%%%%%%%%%%%%%%
%     ABSTRACT, TABLE of CONTENT, and PREFACE.
%%%%%%%%%%%%%%%%%%%%%%%%%%%%%%%%%%%%%%%%%%%%%%%%%%
\setcounter{page}{1} % To exclude the front page in the page numbering.
\SpeAbs{In this Ph.D. thesis, the primary goal is to present a recent investigation of the finite density thermodynamics of hot and dense quark-gluon plasma. As we are interested in a temperature regime, in which naive perturbation theory is known to lose its predictive power, we clearly need to use a refined approach. To this end, we adopt a resummed perturbation theory point of view and employ two different frameworks. We first use hard-thermal-loop perturbation theory (HLTpt) at leading order to obtain the pressure for nonvanishing quark chemical potentials, and next, inspired by dimensional reduction, resum the known four-loop weak coupling expansion for the quantity.

We present and analyze our findings for various cumulants of conserved charges. This provides us with information, through correlations and fluctuations, on the degrees of freedom effectively present in the quark-gluon plasma right above the deconfinement transition. Moreover, we compare our results with state-of-the-art lattice Monte Carlo simulations as well as with a recent three-loop mass truncated HTLpt calculation. We obtain very good agreement between the two different perturbative schemes, as well as between them and lattice data, down to surprisingly low temperatures right above the phase transition. We also quantitatively test the convergence of an approximation, which is used in higher order loop calculations in HTLpt. This method based on expansions in mass parameters, is unavoidable beyond leading order, thus motivating our investigation. We find the ensuing convergence to be very fast, validating its use in higher order computations.}
\SpeAcknowledgments{First and foremost, I would like to express the most profound gratitude to my supervisor Aleksi Vuorinen who supported me at every stage of my doctoral study, gave me a space for development but at the same time offering much needed guidance. He provided me with an acute efficiency to solve technical problems, but he was also very eager to discuss conceptual aspects of the problems.

I am also truly indebted to my closest collaborators. Jens Oluf Andersen showed interest and gave assistance in various forms throughout my doctoral work. He kindly hosted me firstly in Copenhagen at the Niels Bohr Institute, supporting my work on the Ph.D. project. Later, he showed the same encouragement towards my professional development by hosting me at the physics department of NTNU in Trondheim, for a collaboration on another project. A special thanks is due for his careful reading of this manuscript on short notice. In addition, I would like to warmly thank Nan Su for great moral support as well as numerous inspiring discussions, from early stages of the present work. I would also like to express my sincere acknowledgments to Michael Strickland who encouraged me and believed in my ability to cope with the tasks.

I express my appreciation to Tom\'{a}\v{s} Brauner for his guidance in my early and uncertain steps of the Ph.D. study, which helped me immensely. Furthermore, I wish to show my gratitude to both Igor Kondrashuk and York Schr\"{o}der for their collaboration and support in a very interesting but separate project, which we plan to pursue in the future as well.

I would also like to mention my gratitude for many inspiring discussions with visitors, present and former members of the department of physics such as Dietrich B\"{o}deker, Paolo Castorina, Nirupam Dutta, Ioan Ghi\c{s}oiu, Toru Kojo, Martin Kr\v{s}\v{s}\'{a}k, Edwin Laermann, Mikko Laine, P\'{e}ter Petreczky, Fabrizio Pucci, Christian Schmidt and Sayantan Sharma to name a few. These discussions certainly inspired and shaped many of the ideas I have, which I will certainly try to develop further.

Many thanks go to the secretaries of the department of physics, Gudrun Eickmeyer and Susi Reder, for all the help in making my life easier in a new country. In particular, I am indebted to Gudrun who provided me with technical and moral support, beyond any expectation, with joyfulness day after day. I would also like to thank the Sofja Kovalevskaja program for providing all the necessary facilities for study and research. And I wish to thank the Bielefeld Graduate School in Theoretical Sciences, for providing me with a six weeks mobility grant which enabled me to visit the Institute of Physics at the University of Helsinki, to which I am grateful for the warm welcome.

At last but not least, I would like to say that I am endlessly indebted to my family, especially my parents and sister Marina for their unlimited support along the years. My father, Serge, for so much enhancing my curiosity and my mother, Christiane, for so many moral support whenever it was the most needed. Of course, I could not end the acknowledgments without expressing my deepest gratitude to my fianc\'{e}e -- soon to be wife -- Rada Jan\v{c}i\'{c}, who truly inspired my work from the very first day we met in Bielefeld. She endured my ``doctoral moods'', on a daily basis, in a manner which forces the respect and provided me with support whenever it was needed.} % Page of acknowledgments.
\clearpage
\begin{center}
\vspace*{\fill}
{\LARGE{``Il n'est de combat qui soit perdu d'avance...''}}
\vspace*{\fill}
\end{center}
\NewTableofContents % Table of contents.
%\SpePreface{} % Page for preface.
%\addtocontents{toc}{\vskip-30pt} % To reduce the space in between the title ``Table of contents'' and the first item (makes it fit on one page), when adding the preface.
%\addcontentsline{toc}{chapter}{Preface} % Then, add the preface to the table of contents.
%%%%%%%%%%%%%%%%%%%%%%%%%%%%%%%%%%%%%%%%%%%%%%%%%%
%          BEGINNING of the MAIN TEXT.
%%%%%%%%%%%%%%%%%%%%%%%%%%%%%%%%%%%%%%%%%%%%%%%%%%
\chapter{Introduction}\label{chapter:Introduction}

Improving our understanding of strongly interacting matter under extreme conditions is known to be very important for a variety of reasons. To name a few, heavy ion collisions, early universe thermodynamics and the physics of compact stars clearly call for a better understanding of gauge theories at finite temperature and/or density. See e.g.~\cite{QCDReview} for a recent review dedicated to the challenges and perspectives of strongly coupled gauge theories. Consequently, the determination of the phase diagram of QCD has received a lot of attention during the past couple of decades. One quantity of central interest is the pressure, and in particularly its weak coupling expansion, which has been under extensive work in the past two decades. Indeed, it is known to have very bad convergence features, when naively tackled~\cite{ConvergenceWeakCouplingExpansion}. It is therefore very important to try to improve the situation, in particular because such perturbative calculations provide first principle crosschecks of lattice QCD results and also approach the problem in a radically different way.

The importance of these efforts stems both from a desire to obtain precise determinations of the observables in question and from gaining qualitative understanding of the properties of the plasma. While the former of these goals is typically better addressed by means of lattice Monte Carlo simulations\footnote{As far as temperatures reachable by modern experiments are concerned.}, the latter clearly needs to be approached using analytical methods as well as field theoretical models. This need is highlighted by experimental developments in recent years that have stressed the need to understand strongly interacting systems close to the deconfinment transition region and somewhat above it. The quark-gluon plasma created in heavy ion collisions\footnote{See e.g.~\cite{SatzCumulants,MullerQCDMatter} for recent reviews.} at temperatures somewhat above the pseudo-critical temperature of the deconfinement transition $T_\trmi{c}=154\pm 9$MeV~\cite{AbsoluteScale,BazavovTc} has been seen to have somewhat unexpected properties, in contradiction with naive perturbative expectations. Therefore, it is very important to study, whether this plasma is better described through the machinery of weakly coupled gauge theory or perhaps something radically different.

In addition to increasing qualitative understanding about the quark-gluon plasma, perturbative methods are important to study the phase diagram of the theory at nonzero density\footnote{See e.g.~\cite{RHIC,LHC} for current and~\cite{FAIR1,FAIR2,NICA} for future experiments.}, where lattice Monte Carlo simulations are straightforwardly inapplicable. This is due to the so-called sign problem~\cite{SignPbForcrand,SignPbGupta}, which stems from the complexity of the lattice action, making importance sampling techniques impossible to implement. Various approaches have been taken to resolve this problem\footnote{See e.g.~\cite{AartsComplexLangevin} for one of the main directions.}, but the most fruitful one so far consists of simply Taylor expanding the pressure in powers of the chemical potentials. This reduces the problem to the determination of cumulants of the partition function, which can furthermore be a very good probe of the changes in the degrees of freedom of the system. Most importantly, these quantities are manageable on the lattice, as they are evaluated at zero density. Such a technique is, however, only applicable at densities moderate  compared to the typical temperatures in a deconfined plasma. Note that recent lattice studies of these quantities can be found in~\cite{WBLatticeData1,BielefeldLatticeDataHighT,BielefeldLatticeData1,BielefeldLatticeData2,BorsanyiDeltaP}.

The limitation of lattice Monte Carlo techniques to small densities motivates us to approach the problem of fluctuations and correlations of conserved charges, i.e. the finite density part of the equation of state, via a resummed perturbation theory point of view, for which there is no sign problem. Notice that perturbative results can in principle be straightforwardly extended to very large values of the chemical potentials compared to the temperature. Besides, given that performing lattice simulations far above the pseudo-critical\footnote{We recall that at zero density, the transition is of crossover nature.} deconfinement transition temperature $T_\trmi{c}$ is highly nontrivial, it is important to have complementary methods to bridge the gap between the low and asymptotically high temperature regions.

Let us finally list some of the analytic calculations performed during the past few years as attempts to accomplish the above goals. Those use techniques such as unresummed perturbation theory~\cite{AleksiFirstPaperPressure,AleksiFirstPaperQNS,IppAllMuPressure}, various hard-thermal-loop motivated approaches~\cite{BlaizotQNS1,BlaizotQNS2,MustafaQNS1,MustafaQNS2,MustafaQNS3}, hard-thermal-loop perturbation theory~\cite{BaierRedlich,JensMikeDenseEoS,sylvain1,sylvain2,sylvain3,sylvain4,HTLptFiniteMuTwoLoop,HTLptFiniteMuTwoLoop2,HTLptFiniteMUThreeLoop1,HTLptFiniteMUThreeLoop2}, and the large-$N_\trmi{f}$ limit of QCD~\cite{Ipp1,Ipp2}.

Before proceeding to the outline of the thesis, we further point out that all of our calculations have been carried out in the limit of vanishing bare quark masses. In the case of the dimensional reduction framework, introduced shortly, we have explicitly checked that the results are not affected by the light quark masses in a noticeable way. As to our HTLpt calculations, we would like to refer to~\cite{Seipt1} for a study of the one-loop quark self-energy and gluon polarization functions including their mass dependence, which arrived at similar conclusions. Moreover, we would like to emphasize the fact that we are going to focus on the strong interaction only, disregarding the electroweak force, driven by the fact that we are working at energy scales which are negligible compared to the typical mass scale of this interaction.

This dissertation is organized as follows. In Chapter~\ref{chapter:Aspects}, we first explain the basics of quantum field theory at finite temperature and chemical potentials, focusing on the path integral formulation of the partition function, renormalization and the running of the coupling. In Chapter~\ref{chapter:Thermodynamics_Hot_Dense}, we then introduce some key points of thermodynamics, in particular those relevant for finite density. Next, in Chapter~\ref{chapter:Resummed_Perturbative_QCD}, we introduce both of the resummation frameworks that will be used later, before looking into the details of our exact one-loop HTLpt calculation in Chapter~\ref{chapter:Exact_LO_HTLpt}. We then finish by analyzing our findings in Chapter~\ref{chapter:Results}. In addition, the reader can find several complementary sections in the appendices; these include one on our notation in appendix~\ref{appendix:Notation}, another on the matching coefficients of Electrostatic QCD relevant to our dimensional reduction framework in appendix~\ref{appendix:EQCD_Matching_Coefficients}, and a third for the evaluation of the HTL sum-integrals, needed in the mass truncated approximation of the exact one-loop HTLpt pressure, in appendix~\ref{appendix:HTL_Master_Sum-Integrals}.

\chapter{Aspects of thermal field theory}\label{chapter:Aspects}

The present chapter focuses on the basics of quantum field theory at finite temperature and density. It intends to introduce the general context of our study in an informal -- yet rigorous -- way. We choose to emphasize a few aspects only, mainly those important for this study. Readers interested in these aspects, or simply wishing to revive their memory on the topic, are invited to go through this chapter carefully. Others can simply proceed to the next one, which aims at discussing thermodynamics in the light of our studies. For completeness, we would like to refer the reader to the standard textbooks~\cite{KapustaBook,LeBellacBook}, respectively more appropriate for the imaginary- and real-time formalisms. We also refer to excellent review articles such as~\cite{LandsmanReview,JensMikeReview,AntonReview}, the two latter giving a more up-to-date perspective. We further encourage to complement these readings with tutorials such as~\cite{MikkoTutorial}, where a number of intermediate results are provided.

In the following, we first introduce a fundamental object known as the partition function, which plays a crucial role in the thermodynamic description of the system. We then connect this partition function to the path integral formulation of quantum field theory in the imaginary-time formalism. Next, after reviewing the machinery of perturbation theory for evaluating the path integral, we comment on some aspects of such an expansion. Finally, after elaborating on renormalization in the vacuum, we briefly comment on the situation at finite temperature and density, regarding the choice of parameters and the running of the coupling, when evaluating a thermodynamic quantity.

\section{Path integral representation of the partition function}\label{section:Path_Integral}

We start from an explicitly time-independent Hamiltonian operator $\hat{H}$ in Hilbert space, which defines our system. In order to probe the finite density regime of the theory, we also consider conserved charge operators $\hat{Q}_{f}$, which are assumed to commute among themselves as well as with the Hamiltonian. In QCD, these conserved charges will mostly be chosen to be the up, down and strange quark ones. Indeed, the applicability of our work is restricted to the deconfined phase, where these degrees of freedom are likely to be the most natural ones. However, one can equivalently consider the baryon, electric charge and strangeness conserved numbers, which we will do when discussing the degrees of freedom just above the phase transition in Section~\ref{section:Three_flavors}. We are then left to describe the equilibrium state of our system, in the rest frame of a dense heat bath, employing the effective Hamiltonian density
\beq\label{ShiftedHamiltonian}
\mathcal{\hat{H}}\longrightarrow\mathcal{\hat{H}}_\trmi{$\mu_\trmi{f}$}\equiv\mathcal{\hat{H}}-\sum_f\mu_f\mathcal{\hat{Q}}_{f}\,,
\eeq
where the sum runs over $N_\trmi{f}$ fermion flavors. The $\mu_f$ chemical potentials are proportional to the Lagrange multipliers of the averaged charges, i.e. the particle numbers of the system. These quantities will be properly defined in Section~\ref{section:Correlations_Fluctuations}, as they are very important for our study. Note also that the temperature multiplies the Hamiltonian. This bounded operator is a functional of all the fields present in our theory, as well as their conjugate momenta, and is assumed to be at most quadratic in the latter\footnote{There are numerous physical issues as well as possible solutions~\cite{SimonHigerOrderDerivative} associated with theories having higher order derivatives in the fields, despite their mathematical consistency. We can mention, e.g., violation of unitarity due to the absence of a lower bound in the energy spectrum. We shall avoid such theories as we are interested in QCD.}.

From now on, we consider our theory in $d+1$ dimensions, where $d$ is arbitrary for the sake of regularizing possible ultraviolet divergences. We refer the reader to Section~\ref{section:Renormalization_Running} for more details on the subject. We can then write the grand canonical density operator $\hat{\rho}$, normalized to be of unit trace, together with the partition function $\mathcal{Z}$ as
{\allowdisplaybreaks
\beqa
\hat{\rho}\left(T,\left\{\mu_f\right\};V\right)&\equiv&\mathcal{Z}^{-1}\exp\left[-\beta\!\int\!\mathop{{{\rm d}^d}\vect{x}}\ \mathcal{\hat{H}}_\trmi{$\mu_\trmi{f}$}\right]\,, \label{DensityOperator}\\
\mathcal{Z}\left(T,\left\{\mu_f\right\};V\right)&\equiv&\mbox{Tr}_{{}_{\mathcal{P}}}\ \exp\left[-\beta\!\int\!\mathop{{{\rm d}^d}\vect{x}}\ \mathcal{\hat{H}}_\trmi{$\mu_\trmi{f}$}\right]=\sum_{\phi\in\mathcal{P}}\,\Bra{\phi}\exp\left[-\beta\!\int\!\mathop{{{\rm d}^d}\vect{x}}\ \mathcal{\hat{H}}_\trmi{$\mu_\trmi{f}$}\right]\Ket{\phi}\,, \label{PartitionFunction}
\eeqa}
\hspace{-0.12cm}the latter playing a central role in our thermodynamic study. Note that the trace is taken over all the possible and distinct physical states of our system to which our notation $\mathcal{P}$ refers. The set of eigenfunctions forms a complete and orthonormal basis in the Hilbert space. A direct consequence of the constraint on the trace is that the gauge must be fixed when dealing with a gauge theory, in order to avoid any possible over-counting. We will mention later, how to make explicit this highly nontrivial constraint, in the operator formalism. Let us however point out already now that it can be elegantly implemented in the functional integral representation by using the Faddeev-Popov trick as we will see later in this section.

From~(\ref{DensityOperator}), it follows that $\hat{\rho}$ defines a normalized thermal average, such that for any operator $\hat{\vartheta}$, we have
\beq\label{ThermalAverages}
\left\langle\hat{\vartheta}\right\rangle\equiv\mbox{Tr}_{{}_{\mathcal{P}}}\ \Big(\hat{\vartheta}\cdot\hat{\rho}\Big)=\mathcal{Z}^{-1}\ \mbox{Tr}_{{}_{\mathcal{P}}}\ \left(\hat{\vartheta}\cdot\exp\left[-\beta\!\int\!\mathop{{{\rm d}^d}\vect{x}}\ \mathcal{\hat{H}}_\trmi{$\mu_\trmi{f}$}\right]\right)\,.
\eeq
Note that normalizing the trace by any finite constant (here by the partition function itself) does not change the physics\footnote{One can encounter some confusion in the literature about this point, when claiming that an infinite constant multiplying the partition function is irrelevant. To be rigorous, the partition function itself must be well defined, hence free of any divergences. Only when using its path integral representation, one can encounter some infinities during intermediate steps of its evaluation. However, this is merely an artifact of a naive definition for the corresponding volume elements, which in principle should be determined so that every functional integral is well defined on its own.}. To see this, one can use the path integral representation that we are about to introduce, and see that any finite multiplicative number can be expressed via an additional functional integration over a fictitious quadratic field. The Feynman rules associated to such a field, at any order in perturbation theory, lead to power divergent loop momentum integrals, at least in the vacuum. Those are taken care of, during the renormalization procedure, and usually set to zero within dimensional regularization~\cite{RKUnzPathInteAndVolElement}. At finite temperature, however, the situation gets more complicated. Indeed, the Gibbs-Duhem relation tells us that the pressure, related to the partition function, is a differentiable function of the temperature and the chemical potentials,
\beqa
{\rm d}P&=&\mathcal{S}\,{\rm d}T+\sum_{f}\,\mathcal{N}_f\,{\rm d}\mu_f\,,
\eeqa
where $\mathcal{S}$ and $\mathcal{N}_f$, being defined in Section~\ref{section:Bulk_Thermodynamics_Fundamental_Relations}, are respectively the entropy and particle number densities of the system. Therefore, the overall normalization constant must be medium independent so that it does not contribute to the thermodynamics.

We now turn to defining the so-called functional integral representation of the partition function for quantum field theories, which generalizes the Wiener integration relevant to the path integral formulation of quantum mechanics\footnote{See~\cite{FeynmanPathIntegral} for the original reference by R.~P.~Feynman, and~\cite{GelFandYaglomWienerMeasure} for a recent and rigorous review on the topic.}. There are a few subtleties in the definition of functional integrations that need a careful treatment to ensure that one obtains a well defined formulation of the partition function. In particular, two of the major difficulties are related to the definition of a proper volume element (that is, a proper generalization of the Lebesgue measure), as well as the choice of a relevant domain of integration. The former point is rather mathematical in essence, and has not much effect on the physics in the present case of flat space-time (see~\cite{RKUnzPathInteAndVolElement} for more details on the effect of a curved space-time on the volume elements). The latter point on the other hand is deeply connected to the gauge symmetry, and is present in gauge theories as a restriction on the group volume. We will now start by presenting the former, and use a neutral scalar field theory to formally connect the partition function with functional integrals. Then we will turn to gauge theories in order to elaborate on the latter point, and in particular apply the developed machinery to QCD.

In order to be able to introduce the first problem in a simple fashion, let us restrict ourselves to Lebesgue integrals, and consider the following Gaussian integral in $D$ dimensions
\beq
\int_{\mathbb{R}^D}\mathop{{{\rm d}^D}\vect{x}}\ \ e^{-\pi\frac{\left|\vect{x}\right|^2}{a}}\,=\,a^{D/2}\,.
\eeq
From this simple example, it is easy to see the difficulty in taking the limit $D\rightarrow\infty$. Indeed, for a positive definite $a$, the limit $a^{D/2}\rightarrow a^{\infty}$ is clearly untraceable (either vanishing or infinite) for $a\neq 1$. A simple solution to this scaling problem is to introduce the following volume element
\beq
\mathcal{\widetilde{D}}_a\vect{x}\,\equiv\,a^{-D/2}\mathop{{\rm d}x_1\,...\,{\rm d}x_D}\,.
\eeq
Consequently, the Gaussian integral becomes continuous in the $a$-parameter in the limit $D\rightarrow\infty$
\beq
\int_{\mathbb{R}^D}\mathcal{\widetilde{D}}_a\vect{x}\ \ e^{-\pi\frac{\left|\vect{x}\right|^2}{a}}\,=\,1\,.
\eeq
However, the volume of the domain of integration still has a somewhat restrictive nature, as can be seen through an analogy with the following integral
\beq
\int_{\vect{x}\in\left[\vect{a},\vect{b}\right]\subset\mathbb{R}^\infty}\mathcal{\widetilde{D}}_C\vect{x}\,=\,C^\infty\,\prod_{n=1}^{\infty}(b_n-a_n)\,,
\eeq
which only makes sense for $(b_n-a_n)=1$, or at best for a convergent infinite product, in addition to being still untraceable for $C\neq 1$. A way out of these difficulties is to introduce a damping factor $\mathop{{\rm d}x_n}\longrightarrow p(x_n)\,\mathop{{\rm d}x_n}$, such that the probability distribution
\beq
\mathcal{D}_{(p)}\vect{x}^{(D)}\,\equiv\,\prod_{n=1}^{D}p(x_n)\,\mathop{{\rm d}x_n}\,,
\eeq
can be extended to $\mathbb{R}^\infty$. Notice that in the present case, the Gaussian damping factor is a good solution, and leads to a perfectly well defined integral, giving a normalized volume
\beq
\int_{\mathbb{R}^\infty}\mathcal{D}_{(e^{-\pi\,\left|\vect{x}\right|^2})}\vect{x}^{(\infty)}\,=\,1\,.
\eeq
Note also that in $D=\infty$ dimensions, the damping factor cannot be pulled apart from the volume element. Hence, in the corresponding path integral for a quantum field theory, there is no interplay between infinite multiplying constants, relative to the momentum and field functional integrations.

Next, we point out that the situation is much richer in functional spaces, but as we shall be rather brief in our presentation, we refer the reader to the references~\cite{DeVittMoretteProceedings,DeVittMoretteBook,DeVittMorettePaper}. The present discussion is based on these references, from which one can find all the necessary details on the modern and rigorous theory of functional integrations. It should also be noted that various interesting damping factors and hence volume elements can be built. However, since we are interested in weak coupling expansions, we shall restrict ourselves to the Gaussian ones. Thus, it can be shown that the quadratic part of the Hamiltonian unequally defines the volume elements to be used, via Fourier transforms. In the following, all volume elements will be defined according to the above references, as it is a rigorous generalization of the limit definition for the Wiener measure in quantum mechanics.

We now turn to a neutral scalar field theory\footnote{Our discussion remains generally applicable to a multi-component field theory, despite the fact that we suppress the possible indices of summation.}, and recall the functional integral representation of the transition amplitude in the vacuum, for going from a state $\Ket{\phi_\trmi{i}}$ at the time $t_\trmi{i}$ to a state $\Ket{\phi_\trmi{f}}$ at the time $t_\trmi{f}$. The result, first suggested by Feynman~\cite{FeynmanPathIntegralMatrixElement}, reads
\beq\label{TransitionElement}
\!\!\!\!\!\!\!\Bra{\phi_\trmi{f}}e^{-i\,(t_\trmi{f}-t_\trmi{i})\,\mathcal{\hat{H}}}\Ket{\phi_\trmi{i}}\propto\!\!\int_{\phi(\vect{x},t_\trmi{i})=\phi_\trmi{i}}^{\phi(\vect{x},t_\trmi{f})=\phi_\trmi{f}}\!\!\mathcal{D}\phi(x)\!\!\int\mathcal{D}\pi(x)\,\exp\left[i\!\int_{t_\trmi{i}}^{t_\trmi{f}}\!\!\mathop{{{\rm d}}t}\!\!\int\!\mathop{{{\rm d}^d}\vect{x}}\left(\pi(\vect{x},t)\,\frac{\partial \phi(\vect{x},t)}{\partial t}-\mathcal{H}(\phi,\pi)\right)\right]\!,\!\!\!\!
\eeq
where we adopt the above definition for the volume elements, motivating the symbol of proportionality. We notice that the domain of integration over the momentum fields $\pi(x)$ is unrestricted, but this is not a problem as the volume element basically takes care of regularizing the integral.

Motivated by the above relation, we also recall the Lie-Trotter-Kato product formula, which states that for any bounded operators $A$ and $B$, the following limit exists\footnote{There are different prerequisites for working in Minkowskian or Euclidean space-time. For the present purpose of a thermodynamical study, this formula holds.}
\beq
e^{A+B}=\lim_{N\rightarrow\infty}\,\left(e^{A/N}\,e^{B/N}\right)^N\,.
\eeq
Now, the connection to our partition function is quite straightforward. Simply apply the above to the left hand side of~(\ref{TransitionElement}), perform a change of variables corresponding to Wick-rotating the Hamiltonian operator $t\rightarrow -i\tau$, and set the field $\phi_\trmi{i}$ at $\tau=0$ to be equal to $\phi_\trmi{f}\equiv\phi(\vect{x},\beta)$ at $\tau=\beta$. Summing over all possible states, which corresponds to take the trace, we have
\beq
\mathcal{Z}_{\phi}=\lim_{N\rightarrow\infty}\,\mbox{Tr}\ \left(\exp\left[-\beta/N\!\int\!\mathop{{{\rm d}^d}\vect{x}}\ \mathcal{\hat{H}}\right]\right)^N\,,
\eeq
where we dropped the notation $\mathcal{P}$ for the trace, as it is in the whole Hilbert space. Finally, with repeated uses of the standard completeness relation and (\ref{TransitionElement}), we get
\beq\label{PathScalarFieldPhiAndPi}
\mathcal{Z}_{\phi}\propto\int_{\phi_\trmi{periodic}}\!\!\!\!\!\!\!\!\!\!\mathcal{D}\phi(x)\int\mathcal{D}\pi(x)\,\exp\left[\int_{\mathcal{C}_{\beta}}\!\mathop{{{\rm d}}\tau}\int\!\mathop{{{\rm d}^d}\vect{x}}\left(i\,\pi(\vect{x},\tau)\,\frac{\partial \phi(\vect{x},\tau)}{\partial \tau}-\mathcal{H}(\phi,\pi)\right)\right]\,,
\eeq
where $\mathcal{C}_{\beta}$ is a Euclidean time path in the complex plane, yet to be defined, which goes from $0$ to $\beta$. Note that the domain of integration over the fields $\phi(x)$, except for the periodicity constraint $\phi(\vect{x},0)=\phi(\vect{x},\beta)$, should not be restricted\footnote{As we will mention later, this is not the case with gauge theories.}.

Assuming a quadratic Hamiltonian in the momentum fields, it is easy to complete the square and perform the integration over them. We then get the canonical path integral formulation for the partition function of our original neutral scalar field theory
\beq\label{CanonicalPathScalarField}
\mathcal{Z}_{\phi}=\int_{\phi_\trmi{periodic}}\!\!\!\!\!\!\!\!\!\!\mathcal{D}\phi(x)\,\exp\left[-\int_{\mathcal{C}_{\beta}}\!\mathop{{{\rm d}}\tau}\int\!\mathop{{{\rm d}^d}\vect{x}}\bigg(\mathcal{L_\trmi{eff}}(\phi,i\partial\phi/\partial\tau)\bigg)\right]\,,
\eeq
where $\mathcal{L_\trmi{eff}}$ is nothing but the Lagrangian density of our theory in Euclidean space-time, to which we will eventually add some ghost terms. Those arise if the Hamiltonian piece, which is quadratic in $\pi$, is $\phi$-dependent (presence of derivative interactions), which is not the case in the present scalar field theory example. Notice also that in principle the quadratic integration over the momentum fields brings a multiplying constant, which is finite here. This can and should be included back into the effective Lagrangian via fictitious additional ghost fields. The latter, as already mentioned in this chapter, lead to power divergent loops which are set to zero under dimensional regularization in the vacuum. This is the reason why such a constant is usually dropped, as it is not relevant upon renormalization. This also applies to the finite constant of proportionality which connects the sign $\propto$ in~(\ref{PathScalarFieldPhiAndPi}) to the equal sign of~(\ref{CanonicalPathScalarField}). However, at finite temperature, such constants can be medium-dependent and should not be, a priori, ignored regarding the thermodynamics. Yet, it was shown\footnote{Using a different definition for the volume elements, and hence carrying some intermediate infinities. This, in fine, does not affect the conclusion of our discussion.} in~\cite{BernardFunctionalPiIntegration} that this medium dependence is only ``virtual'', as the remaining functional integration over the fields $\phi(x)$ takes care of its cancellation. This can be understood as a medium dependence of each volume element separately, which is not surprising considering that the quadratic part of our Hamiltonian~(\ref{PathScalarFieldPhiAndPi}) defines them, while their product is not medium dependent. For simplicity, we will avoid lingering over this detail, and just skip such an intermediate fictitious contribution to the effective Lagrangian density. From now on, unless explicitly stated otherwise, $\mathcal{L_\trmi{eff}}$ will be understood in this way.

Having dealt with the proper definition of the volume elements, we now turn to the second point, which is adopting a suitable definition for the domain of integration. To that end, we consider a non-Abelian Yang-Mills gauge theory, coupled to $N_\trmi{f}$ fermions with chemical potentials as described by the Hamiltonian density in~(\ref{ShiftedHamiltonian}). We first note that due to the gauge symmetry, this time the Hilbert space counts a certain number of unphysical states. As they are equivalent to each other via gauge transformations, a gauge must be fixed properly when evaluating the partition function\footnote{Or any matrix element, even in the vacuum.}. In our partition function~(\ref{PartitionFunction}) within the operator formalism, one can insert a projection operator onto the space of physical states~\cite{GrossPisarskiYaffeQCDInstantons}. However, we shall use here the path integral representation to perform this task explicitly. The goal is then to select a unique representative from each gauge orbit, which ideally intersect the gauge condition only once.

To demonstrate the above procedure, let us first write the canonical partition function for our gauge theory in a naive way, that is without fixing the gauge\footnote{That is to say without any restriction on the domain of integration, which would correspond to span only distinct physical states in the Hilbert space.}
\beq\label{PreCanonicalPathGaugeTheory}
\!\!\!\!\!\!\mathcal{Z}_{A,\psi}\stackrel{A_\mu^G}{\mathlarger{\mathlarger{\approx}}}\int_{A_\mu{}_\trmi{periodic}}\!\!\!\!\!\!\!\!\!\!\!\!\mathcal{D}A_\mu(x)\int_{\bar{\psi},\psi_\trmi{anti-periodic}}\!\!\!\!\!\!\!\!\!\!\!\!\!\!\!\!\!\!\!\!\mathcal{D}\bar{\psi}(x)\,\mathcal{D}\psi(x)\,\exp\left[-\int_{\mathcal{C}_{\beta}}\!\mathop{{{\rm d}}\tau}\int\!\mathop{{{\rm d}^d}\vect{x}}\bigg(\mathcal{L^{\prime\prime}_\trmi{eff}}\,(A_\mu,\bar{\psi},\psi)-\bar{\psi}\,\gamma_0\,\bm{\mu}\,\psi\bigg)\right]\!,\!\!\!
\eeq
where $\bm{\mu}$ is the matrix of all the chemical potentials, assumed to have a diagonal structure in flavor space. Note that in the present case, the approximation symbol is nothing more than a notation for ``equal up to the restriction relative to the gauge over-counting''. The symbols $A_\mu(x)$, $\bar{\psi}(x)$ and $\psi(x)$ denote the gauge and fermionic fields, while the boundary conditions state respectively $A_\mu(\vect{x},0)=A_\mu(\vect{x},\beta)$, $\bar{\psi}(\vect{x},0)=-\bar{\psi}(\vect{x},\beta)$ and $\psi(\vect{x},0)=-\psi(\vect{x},\beta)$. Note that the antiperiodicity conditions for the fermionic fields is due to the Grassmann nature of these variables\footnote{Indeed, the generators of an infinite dimensional Grassmann algebra anticommute, and hence obey the identity $\theta^2(x)=0$. This is motivated by the Pauli principle itself, thanks to the half-integer spin nature of the fermionic fields.}. Those variables admit different, yet well-defined, rules of integration. However, we are not going to elaborate on this matter, as the procedure to define the corresponding volume elements is not changed. For completeness, we merely refer the reader to the excellent textbook~\cite{BertlmannBook}, recalling also the differences in the definitions of the volume elements.

Selecting a unique representative from each gauge orbit is a highly nontrivial task. In fact, we shall assume here -- without a strong argument -- that our gauge fixing condition will be intersected by every gauge orbit only once. Also, in line with perturbation theory, we will deal with infinitesimal fluctuations of the gauge potential only. Consequently, it is possible to implement the Faddeev-Popov trick~\cite{FaddeevPopovTrick}, as for small fluctuations the corresponding determinant remains positive~\cite{AssumptionForGhostsZwanziger}. Having said that, we shall ignore the so-called Gribov ambiguity~\cite{GribovAmbiguity} for the present purpose of performing resummations of the weak coupling expansion\footnote{See~\cite{AssumptionForGhostsZwanziger} for a recent review on the topic, bearing in mind that such a procedure introduces a dimensionful parameter, for which a rigorous fixing method is still not fully understood. Note also that another alternative for gauge theories without Faddeev-Popov ghosts can be found in~\cite{DeVittMoretteProceedings}.}.

Let us now come back to our initial, still ill-defined path integral formulation~(\ref{PreCanonicalPathGaugeTheory}) of a non-Abelian Yang-Mills gauge theory coupled to $N_\trmi{f}$ quarks, and consider a covariant gauge fixing condition,
\beq
f\left[A_\mu^G\right]\equiv \partial^\mu A_\mu^G(x)-c\,(x)=0\,,
\eeq
where $c\,(x)$ is an unspecified function, while the superscript $G$ denotes a gauge field to which we have applied the gauge transformation
\beq
A_\mu\rightarrow A_\mu^G\equiv G^{-1} A_\mu G+G^{-1} \partial_\mu G\,.
\eeq
We then use the identity
\beq
\int_{G{}_\trmi{periodic}}\!\!\!\!\!\!\!\!\mathcal{D}G(x)\,\,\,\Delta_{{}_\trmi{FP}}\left[A_\mu^G\right]\,\,\delta\left(f\left[A_\mu^G\right]\right)=1\,
\eeq
for which we have defined the Faddeev-Popov determinant
\beq
\Delta_{{}_\trmi{FP}}\left[A_\mu^G\right]=\det\,\left(\frac{\delta f\left[A_\mu^G\right]}{\delta G}\right)\,,
\eeq
and where the periodicity condition comes from the fact that gauge transformations need to be periodic in the temporal direction. Having done so, we rewrite our path integral~(\ref{PreCanonicalPathGaugeTheory}) as
\beqa\label{PreBisCanonicalPathGaugeTheory}
\mathcal{Z}_{A,\psi}\stackrel{A_\mu^G}{\mathlarger{\mathlarger{\approx}}}\int_{A_\mu{}_\trmi{periodic}}\!\!\!\!\!\!\!\!\!\!\!\!\!\!\!\!&&\!\!\!\!\!\!\!\!\mathcal{D}A_\mu(x)\int_{\bar{\psi},\psi_\trmi{anti-periodic}}\!\!\!\!\!\!\!\!\!\!\!\!\!\!\!\!\!\!\!\mathcal{D}\bar{\psi}(x)\,\mathcal{D}\psi(x)\,\int_{G{}_\trmi{periodic}}\!\!\!\!\!\!\!\!\mathcal{D}G(x)\nonumber \\
&&\!\!\!\!\!\!\Delta_{{}_\trmi{FP}}\left[A_\mu^G\right]\,\delta\left(f\left[A_\mu^G\right]\right)\,\exp\left[-\int_{\mathcal{C}_{\beta}}\!\mathop{{{\rm d}}\tau}\int\!\mathop{{{\rm d}^d}\vect{x}}\bigg(\mathcal{L^{\prime\prime}_\trmi{eff}}\,(A_\mu,\bar{\psi},\psi)-\bar{\psi}\,\gamma_0\,\bm{\mu}\,\psi\bigg)\right]\!.
\eeqa

The next step amounts to using the gauge invariance of the above effective action, together with the ones of the group volume element $\mathcal{D}G(x)$ and volume elements $\mathcal{D}A_\mu(x)$, $\mathcal{D}\bar{\psi}(x)$, and $\mathcal{D}\psi(x)$\footnote{While the invariance of $\mathcal{D}G(x)$ is obvious, the three last are merely direct consequences of the volume elements' definitions using the quadratic part of the action, hence of the invariance of the action itself.}. By performing a gauge transformation $A_\mu^G\rightarrow A_\mu$ ($G\rightarrow G^{-1}$), we factor out the normalized group volume
\beq
\int_{G{}_\trmi{periodic}}\!\!\!\!\!\!\!\!\mathcal{D}G(x)\,,
\eeq
which can be reabsorbed into the effective Lagrangian, following the procedure that we already mentioned. We can now use the equality symbol, having suppressed the over-counting, and write
\beqa\label{PreTersCanonicalPathGaugeTheory}
\mathcal{Z}_{A,\psi}=\int_{A_\mu{}_\trmi{periodic}}\!\!\!\!\!\!\!\!\!\!\!\!\!\!\!\!&&\!\!\!\!\!\!\!\!\mathcal{D}A_\mu(x)\int_{\bar{\psi},\psi_\trmi{anti-periodic}}\!\!\!\!\!\!\!\!\!\!\!\!\!\!\!\!\!\!\!\mathcal{D}\bar{\psi}(x)\,\mathcal{D}\psi(x)\nonumber \\
&&\!\!\!\!\!\!\Delta_{{}_\trmi{FP}}\left[A_\mu\right]\,\delta\left(f\left[A_\mu\right]\right)\,\exp\left[-\int_{\mathcal{C}_{\beta}}\!\mathop{{{\rm d}}\tau}\int\!\mathop{{{\rm d}^d}\vect{x}}\bigg(\mathcal{L^{\prime}_\trmi{eff}}\,(A_\mu,\bar{\psi},\psi)-\bar{\psi}\,\gamma_0\,\bm{\mu}\,\psi\bigg)\right]\,.
\eeqa
Finally, we simply have to re-express the Faddeev-Popov condition, $\Delta_{{}_\trmi{FP}}\left[A_\mu\right]\,\delta\left(f\left[A_\mu\right]\right)$, in a more convenient way for calculations. For the sake of shortening our presentation, we again refer the reader to~\cite{BertlmannBook} for more details on this procedure. It should be noted, however, that the Faddeev-Popov determinant can be put in a functional form over Grassmann ghost variables $\bar{\eta}$ and $\eta$, that have periodic boundary conditions due to the periodic gauge invariance of the partition function itself.

Following the standard Faddeev-Popov procedure, we finally end up with the functional integral
\beqa\label{CanonicalPathGaugeTheory}
\mathcal{Z}_{A,\psi}=\int_{A_\mu{}_\trmi{periodic}}\!\!\!\!\!\!\!\!\!\!\!\!\!\!\!\!&&\!\!\!\!\!\!\!\!\mathcal{D}A_\mu(x)\int_{\bar{\psi},\psi_\trmi{anti-periodic}}\!\!\!\!\!\!\!\!\!\!\!\!\!\!\!\!\!\!\!\mathcal{D}\bar{\psi}(x)\,\mathcal{D}\psi(x)\int_{\bar{\eta},\eta_\trmi{periodic}}\!\!\!\!\!\!\!\!\!\!\!\!\!\!\!\!\!\!\!\mathcal{D}\bar{\eta}(x)\,\mathcal{D}\eta(x)\,\,\,\,\,\,\,\,\,\,\,\,\,\,\,\,\,\,\,\,\,\,\,\,\,\,\,\,\,\,\,\,\,\,\,\,\,\,\,\nonumber \\
&&\,\,\,\,\,\,\,\,\,\,\,\,\,\exp\left[-\int_{\mathcal{C}_{\beta}}\!\mathop{{{\rm d}}\tau}\int\!\mathop{{{\rm d}^d}\vect{x}}\bigg(\mathcal{L_\trmi{eff}}\,(A_\mu,\bar{\psi},\psi,\bar{\eta},\eta)\bigg)\right]\,,
\eeqa
where we did not write explicitly the possible dependence of the effective Lagrangian on additional fictitious ghost fields that have no effect upon renormalization. The effective Lagrangian for QCD appearing here, comprising the gauge-fixing and Faddeev-Popov terms, reads~(for a more careful derivation see e.g.~\cite{AleksiThesis})
\beq\label{EffectiveLagrangianQCD}
\left.\mathcal{L_\trmi{eff}}(A_\mu,\bar{\psi},\psi,\bar{\eta},\eta)\right|_\trmi{QCD}=\mathcal{L_\trmi{QCD}}(A_\mu,\bar{\psi},\psi)+\frac{(\partial_\mu A_\mu^a)^2}{2\xi}-\bar{\psi}\,\gamma_0\,\bm{\mu}\,\psi+\bar{\eta}^a\left(\partial^2\delta^{ab}+g\,f^{abc}A_\mu^c\partial_\mu\right)\eta^b\,,
\eeq
where $\mathcal{L_\trmi{QCD}}$ denotes the original bare Lagrangian of QCD in Euclidean space-time, $f^{abc}$ stand for the antisymmetric structure coefficients of SU($N_\trmi{c}$), and $g$ is the gauge coupling. Moreover, $\xi$ is a gauge-fixing parameter upon which no physical quantity should depend on. Its appearance follows from introducing a $c(x)$-dependent prefactor to the partition function before integration. The delta functional coming along with the determinant makes then the $c(x)$ disappear from the path integral.

Last but not least, what remains is to specify our choice for the integration path $\mathcal{C}_{\beta}$ in the time variable $\tau$. In principle, different choices of the contour lead to different formulations of the theory as well as different sets of Feynman rules, all being physically equivalent~\cite{MatsumotoChoiceContour}. In practice, however, there are a limited number of convenient choices from a computational point of view. Two of the most widely used choices are the Matsubara imaginary-time~\cite{MatsubaraFormalism}, and the Keldysh real-time~\cite{KeldyshFormalism} formalisms. These have Green's functions primarily defined for imaginary and real time values\footnote{More precisely, for values along their path $\mathcal{C}_{\beta}$.}, respectively. While the real-time formalism is more suited for dynamical processes\footnote{As a consequence of the equivalence between the different formalisms, it is, in principle, possible to analytically continue any Green's functions as functions of the time variable, from an imaginary value back to the real axis. However, in practice, this is often cumbersome~\cite{CunibertiAnalyticalContinuationGreenFunction}. This is the reason why it is preferable to choose the real-time formalism from the beginning, when interested in real-time dynamics.}, the imaginary one has the advantage of more closely following methods from vacuum quantum field theory. The practical difference is that the temporal component of the four-momentum becomes discrete in the imaginary time case. This is simply a consequence of the compactification of the temporal direction, and basically leads to sum-integrals instead of integrals when evaluating Feynman diagrams. In this work, we choose the Matsubara formalism, i.e. the Euclidean time variable and $\mathcal{C}_{\beta}$ going from $0$ to $\beta$ via a straight line. In addition, note that the analytical properties of the path integral require that the real part of $\mathcal{C}_{\beta}$ is monotonically increasing~\cite{LandsmanReview,AntonReview}, which is trivially satisfied in the present case.

From now on, the form of the partition function~(\ref{CanonicalPathGaugeTheory}) with the effective Lagrangian~(\ref{EffectiveLagrangianQCD}) together with the above contour will be our reference point for the thermodynamics of QCD. More precisely, the resummation framework inspired by dimensional reduction will use this as a starting point, while hard-thermal-loop perturbation theory will further include an improvement term in its Lagrangian density. See Sections~\ref{section:Resummed_DR} and~\ref{section:HTLpt}, respectively, for more details on the two setups.

Having defined our partition function, we can now turn to its perturbative evaluation, which the next section is devoted to.

\section{Perturbative evaluation of the partition function}\label{section:Perturbation_Theory}

In order to review the basics of perturbative expansions, let us return to a neutral scalar field theory, which we assume to contain a quartic interaction term. We then start from the path integral representation of the partition function given by~(\ref{CanonicalPathScalarField}), together with the following Euclidean action
\beq
S_\phi=S_{\phi_0}+S_{\phi_\trmi{\tiny I}}\equiv\int_{\mathcal{C}_{\beta}}\!\mathop{{{\rm d}}\tau}\int\!\mathop{{{\rm d}^d}\vect{x}}\left(\frac{1}{2}\partial^\mu\phi\,\partial_\mu\phi+\frac{1}{2}m^2\phi^2\right)+\lambda\,\int_{\mathcal{C}_{\beta}}\!\mathop{{{\rm d}}\tau}\int\!\mathop{{{\rm d}^d}\vect{x}}\left(\frac{1}{4}\,\phi^4\right)\,,
\eeq
where the subscripts $\phi_0$ and $\phi_\trmi{\tiny I}$ refer to the free (always quadratic) and interaction parts, respectively.

Before proceeding, let us generalize the thermal average of~(\ref{ThermalAverages}) for neutral scalar fields, using functional integrals rather than a trace in the Hilbert space
\beq\label{ExpectationValues}
\left\langle\vartheta\right\rangle_{X_\trmi{S}}\equiv\frac{\int_{\phi_\trmi{periodic}}\!\!\!\!\!\!\!\!\!\!\mathcal{D}\phi(x)\,\Big\{\vartheta(\phi)\,\,\exp\left[-X_\trmi{S}(\phi)\Big]\right\}}{\int_{\phi_\trmi{periodic}}\!\!\!\!\!\!\!\!\!\!\mathcal{D}\phi(x)\,\exp\left[-X_\trmi{S}(\phi)\right]}\,.
\eeq
We see that~(\ref{ThermalAverages}) is equivalent to the above, provided that $X_\trmi{S}(\phi)=S_\phi$. From a functional point of view, integrals such as the above are very difficult to tackle analytically. In fact, only a few\footnote{Excluding equivariant cohomological localization of the path integral, most applicable to topological quantum field theories~\cite{SzaboLocalization}. Apart from analytical techniques, Monte Carlo importance sampling methods for a discretized version of the theory, aka Lattice Field Theory, can be used in the non-perturbative regime of quantum field theory. See~\cite{LatticeBook} for a recent account on lattice gauge theories.}, mostly Gaussian types of integrations (i.e. for $X_\trmi{S}(\phi)$ quadratic in the fields), are known in a closed form.

Perturbation theory consists of formally expanding the exponential containing the interaction part of the action in powers of the coupling, and then integrating each term separately\footnote{We refer to the end of this section for a discussion on the consequences of such an interchange of operations.}. Notice that besides being a convenient way for approximating complicated path integrals, in the case of QCD this is clearly physically motivated by asymptotic freedom~\cite{AsymptoFreedom1GrossWilczek,AsymptoFreedom2Politzer}. It allows us to approximate the partition function by a sum of Gaussian functional integrals, for which analytical solutions can be found. Using the definition of expectation values~(\ref{ExpectationValues}), it is then straightforward to rewrite the logarithm of the partition function\footnote{Its logarithm is indeed a quantity of central interest for thermodynamics, as we will see in Section~\ref{section:Bulk_Thermodynamics_Fundamental_Relations}.} as
\beqa\label{LogPartitionFunctionTaylorExpansion}
\log\,\mathcal{Z}_{\phi}=\log\,\mathcal{Z}_{\phi_0}&-&\lambda\left\{\left\langle\widetilde{S}_{\phi_\trmi{\tiny I}}\right\rangle_{S_{\phi_0}}\right\}\!\!\!\!\!\,\,\,\,\,\,+\,\,\frac{\lambda^2}{2}\left\{\left\langle\widetilde{S}_{\phi_\trmi{\tiny I}}^2\right\rangle_{S_{\phi_0}\!\!\!\!\!\!\!\!}-\left\langle\widetilde{S}_{\phi_\trmi{\tiny I}}\right\rangle_{S_{\phi_0}}^2\right\}\,\nonumber \\
&-&\frac{\lambda^3}{6}\left\{\left\langle\widetilde{S}_{\phi_\trmi{\tiny I}}^3\right\rangle_{S_{\phi_0}\!\!\!\!\!\!\!\!}-3\left\langle\widetilde{S}_{\phi_\trmi{\tiny I}}\right\rangle_{S_{\phi_0}\!\!\!\!}\left\langle\widetilde{S}_{\phi_\trmi{\tiny I}}^2\right\rangle_{S_{\phi_0}\!\!\!\!\!\!\!\!}+2\left\langle\widetilde{S}_{\phi_\trmi{\tiny I}}\right\rangle_{S_{\phi_0}}^3\right\}+{\cal O}\left(\lambda^4\right)\,,
\eeqa
where $\mathcal{Z}_{\phi_0}$ refers to the known free partition function, while $\widetilde{S}_{\phi_\trmi{\tiny I}}\equiv S_{\phi_\trmi{\tiny I}}/\lambda$.  This expansion, assuming Borel summability, is naively expected to be a good approximation of the full partition function  for $\lambda << 1$, when sufficiently many terms of it are considered. In the next chapters, we will see that this is, however, not always the case. It is also good to notice already now that unlike in our dimensional reduction framework, which uses the coupling as an expansion parameter, hard-thermal-loop perturbation theory introduces a different formal expansion parameter, which leads to a reorganization of the perturbative series.

Following the above path of approximation, one ends up having to evaluate expressions such as
\beqa
\hspace*{-4.5truecm}\left\langle\widetilde{S}_{\phi_\trmi{\tiny I}}^n\right\rangle_{S_{\phi_0}\!\!\!}&=&\frac{1}{4^n}\int_{\mathcal{C}_{\beta}}\!\mathop{{{\rm d}}\tau_1}\int\!\mathop{{{\rm d}^d}\vect{x}_1}...\int_{\mathcal{C}_{\beta}}\!\mathop{{{\rm d}}\tau_n}\int\!\mathop{{{\rm d}^d}\vect{x}_n}\,\,\Bigl\langle\phi^4(x_1)...\phi^4(x_n)\Bigr\rangle_{S_{\phi_0}\!\!\!}\, \nonumber \\
&\propto&\int_{\mathcal{C}_{\beta}}\!\mathop{{{\rm d}}\tau_1}\int\!\mathop{{{\rm d}^d}\vect{x}_1}...\int_{\mathcal{C}_{\beta}}\!\mathop{{{\rm d}}\tau_n}\int\!\mathop{{{\rm d}^d}\vect{x}_n}\,\,\left(\int_{\phi_\trmi{periodic}}\!\!\!\!\!\!\!\!\!\!\mathcal{D}\phi(x)\,\Big\{\phi^4(x_1)...\phi^4(x_n)\,\,\exp\left[-S_{\phi_0}\Big]\right\}\right)\,,
\eeqa
where $n$ is a positive integer. The symbol of proportionality comes from the fact that we have omitted the trivial denominator of~(\ref{ExpectationValues}), as well as the $1/4^n$ prefactor. The computation of such an expression can be greatly simplified thanks to the so-called Wick theorem~\cite{WickTheorem} and its generalization to finite temperature by C.~Bloch and C.~De Dominicis~\cite{BlochDeDominicisTh}. This theorem allows for a very convenient algebraic reduction: The above expectation value of $n$ products of four fields, evaluated at the same space-time point, is written as a sum over all combinations of $2n$ products of expectation values of two fields. This operation is called Wick contraction and reads
\beqa
\hspace*{-2truecm}\Bigl\langle\phi^4(x_1)...\phi^4(x_n)\Bigr\rangle_{S_{\phi_0}\!\!\!}&=&\Bigl\langle\phi(x_1)\phi(x_1)\phi(x_1)\phi(x_1)...\phi(x_n)\phi(x_n)\phi(x_n)\phi(x_n)\Bigr\rangle_{S_{\phi_0}\!\!\!}\, \\
&=&\Bigl\langle\phi(x_1)\,\phi(x_2)\Bigr\rangle_{S_{\phi_0}\!\!\!}...\,\,\Bigl\langle\phi(x_{n-1})\,\phi(x_n)\Bigr\rangle_{S_{\phi_0}\!\!\!}+\ \left\{\begin{subarray}{c} (4n-1)!!\,\,-\,\,1 \\ \mbox{permutations}\end{subarray}\right\}\,, \label{contractionWicktheorem}
\eeqa
in which we can recognize two different types of contractions. The first ones connect fields at two different space-time points and are called connected contractions. The second ones on the other hand depend only on one space-time variable and are called disconnected contractions. In the present case, when contracting $n$ products of four fields, it can be shown that all terms containing disconnected contractions are exactly canceled in the full expansion~(\ref{LogPartitionFunctionTaylorExpansion}).

In general, we can define an object called an $n$-point Green's function by
\beq\label{NPointGreenFunction}
\mathcal{G}^\trmi{\tiny$(n)$}(x_1, ...,x_n)\equiv \left\langle\phi(x_1)...\phi(x_n)\right\rangle_{S_{\phi}}\,.
\eeq
Following the above procedure of taking Wick contractions, its perturbative evaluation reduces to products of the much simpler free two-point functions\footnote{We set aside the evaluation of momentum sum-integrals that appear upon Fourier transforming the remaining space-time integrals, which convolute products of two-point Green's functions. This will be treated in our later chapters.}
\beq
\int_{\phi_\trmi{periodic}}\!\!\!\!\!\!\!\!\!\!\mathcal{D}\phi(x)\,\Big\{\phi(x_i)\phi(x_j)\,\,\exp\left[-S_{\phi_0}\Big]\right\}\,.
\eeq
These are the basic building blocks of all Feynman diagrams, called the free propagators.

After having seen the main steps in perturbatively evaluating the partition function, we will make a few more comments, before to proceed to the next section dedicated to the renormalization procedure.

On a positive note, we see that unlike with Monte Carlo methods, the presence of nonvanishing chemical potentials is not a problem in perturbation theory. Indeed, nonzero chemical potentials lead to a complex fermionic determinant, which renders Monte Carlo sampling integrals highly oscillatory, i.e.~ to the sign problem~\cite{SignPbForcrand,SignPbGupta}. On the other hand, with perturbative expansions the functional integrals are performed analytically, the volume element being defined upon the quadratic part of the action without any problems.

Regarding a more technical aspect, let us study the convergence of the perturbative expansion; we do this for the logarithm of the partition function, but the discussion can be extended to the expectation value of any physical quantity. Interchanging the order between performing a series expansion and a functional integral is nontrivial, and could lead to a divergent series. This phenomenon arises quite often in quantum field theories, and even leads to the factorial growth of the Feynman diagrams. However, as troubling as a divergent series might sound, it is not necessarily a problem in practice\footnote{We shall recall, e.g., that divergent asymptotic series are as useful as rigorous tools in applied mathematics~\cite{OvidiuAsymptoticsandBorelSumma}, as well as in physics; see~\cite{RenormalonsBeneke} for an excellent review on the renormalon phenomenon.}, as long as the series in question has suitable asymptotic properties. Indeed, such a series, when truncated, can be a very good approximation to the full quantity already at the few first orders~\cite{OvidiuAsymptoticsandBorelSumma,AsymptoticSchulman}. However, we are neither going to elaborate on the possible Borel summability of the perturbative expansions, nor on related improvements such as resummations inspired from Borel transforms~\cite{BorelResummation1,BorelResummation2}. Indeed, the main  goal of this thesis is to improve the apparent convergence of the weak coupling expansion of the pressure of QCD\footnote{In the sense of trying to optimize the successive approximations of the full quantity, as the order of truncation increases.} as well as its derivatives, using arguments motivated by \textit{physics} rather than mathematics. For further readings on the topic of asymptotic series relevant to QCD, see for example~\cite{QCD2AsymptoticZhitnitsky}. We also recommend the textbook~\cite{ZinnJustinLargeOrderPT} for a collection of important works on the large order behavior of perturbation theory in quantum mechanics and quantum field theories.

Next, we move on to discuss the renormalization procedure at nonzero temperature and density.

\section{Renormalization and the running of the coupling}\label{section:Renormalization_Running}

We shall now focus our discussion to QCD. In addition, we adopt a strict perturbative point of view, and refer the reader to~\cite{BergesNonPerturbativeRG} for an introduction to nonperturbative renormalization. Notice also that in this section we only deal with ultraviolet divergences even when not explicitly stated; the physics of infrared divergences is altogether different and will be returned to in later chapters.

In short, the renormalization procedure of vacuum quantum field theory enables one to eliminate ultraviolet divergent contributions when computing momentum loop integrals. It is however not just a trick, but relies in a deep way on the renormalization group invariance~\cite{WilsonRGInvariance} of the theory, as we are going to see shortly. Before entering the details of the procedure, we however first simply note that field theories can be classified according to their renormalizability in four dimensions. This is done by relating the large momentum behavior of Feynman diagrams to the interaction terms in the action of the theory~\cite{ItzyksonAndZuber}. The superficial degree of divergence of a given vertex $\omega_\trmi{v}$, which counts the total number of fields ($b_\trmi{S}$ for bosons and $f_\trmi{S}$ for fermions) and field derivatives $\delta_\trmi{S}$ entering it, is defined as
\beq
\omega_\trmi{v}\equiv b_\trmi{S}+\frac{3}{2}\,f_\trmi{S}+\delta_\trmi{S}\,,
\eeq
and enables the following classification\footnote{The present classification is valid for monomial types of interactions only.}:
{\begin{itemize}\itemsep0em
\item[\tiny$\bullet$] A theory, in which all interaction terms satisfy $\omega_\trmi{v}<4$, is said to be super-renormalizable,
\item[\tiny$\bullet$] A theory, in which all interaction terms satisfy $\omega_\trmi{v}\leq 4$, but at least one has $\omega_\trmi{v}=4$, is said to be renormalizable,
\item[\tiny$\bullet$] A theory, in which at least an interaction term has $\omega_\trmi{v}>4$, is said to be non-renormalizable,
\end{itemize}}

Following the above classification, it is obvious that QCD is a renormalizable field theory. It is also worth mentioning that the three-dimensional effective field theory of high temperature QCD known as Electrostatic QCD (EQCD), which is needed for the evaluation of the pressure of the full theory to order $g^6 \log g$, is in turn super-renormalizable\footnote{As defined in~(\ref{EQCDLagrangian}), enough to evaluate the pressure to four-loop order. Beyond this, non-renormalizable field operators enter the Lagrangian; see Section~\ref{section:Resummed_DR} for more details.}. Regarding hard-thermal-loop perturbation theory, we refer the reader to Section~\ref{section:Renormalizing_Result} for more details and discussion.

In order to set the stage for the problem of renormalizing our theory, the first step is to regulate all the potentially divergent expressions encountered in practical calculations. This is done by generalizing the momentum integrals to be functions of a regularizing parameter, provided that the original expressions are recovered upon taking an appropriate limit. In practice, of course, this limit is only taken after having renormalized the theory, that is, after having removed all potentially divergent contributions by means of the renormalization group invariance. There exist various regularization schemes, each of them making use of both a regulating parameter and an associated energy (or regularization) scale\footnote{While in most schemes, the regulating parameter itself corresponds to an energy scale, in dimensional regularization the two are different and the energy scale appears for dimensional reasons.}. Examples of widely used regularization schemes include a sharp three-momentum ultraviolet cut-off, where the momentum integrals are bounded by a finite scale, or the Pauli-Villars regularization~\cite{PauliVillarsReg}, which includes a fictitious massive field whose mass plays the role of the cut-off. One can also use a lattice regularization which discretizes the space-time, hereby provided a natural cut-off, or the Schwinger regularization which uses an integral representation for the propagators, with a finite bound provided by the cut-off.

In the present study, we shall, however, use the dimensional regularization scheme~\cite{tHooftVeltmanDimReg}, as it allows for analytical regularization. This is achieved by altering the momentum integration measure, as the name suggests, by trading the integer dimensionality of the space-time $d+1=4$ for a complex one $d+1=4-2\epsilon$, where $\epsilon$ is a complex regulator, which at the end of the calculation is typically sent to zero\footnote{Most often, this limit is reached from above on the real axis, as far as ultraviolet divergences are concerned.}. The analyticity of the regularization then relies upon certain constraints on the real part of the dimensional regulator $\epsilon$. A careful analysis shows that this allows for setting power-like divergences to zero\footnote{By formally treating the ultraviolet and infrared limit of the integrand in different manners, that is assuming two different conditions for the real part of the dimensional regulator $\epsilon$, the sum of the two contributions vanishes~\cite{LeibbrandtDimReg,SmirnovBook}.}, while the logarithmic ones appear as poles in the regulator $\epsilon$, and have to be removed using the renormalization procedure.

On general grounds, it is worth noting that the choice of regularization should not affect the renormalization itself. The final result must be independent of the energy scale introduced in the course of the regularization, as we will see later. Of course, it might happen that some of the symmetries are violated at intermediate stages of the calculation. For example, the lattice regularization breaks rotational invariance while the sharp three-momentum cut-off breaks Lorentz invariance. However, this is not a problem since, provided that the procedure has been carried out properly, the renormalized theory shall possess all the symmetries originally present in the unrenormalized bare action\footnote{Except when dealing with an anomaly, where a symmetry of the classical action is broken by quantum corrections~\cite{BertlmannBook}.}. In particular, dimensional regularization preserves all the symmetries during the intermediate stages of the calculation, including the gauge symmetry. Note also that this regularization scheme suggests a very simple way of subtracting the divergences, as we are going to see in the following. However, the drawback in using such a scheme is that the treatment of tensor-like objects, such as the Levi-Civita tensor which is originally defined in integer dimensions, appears to involve a great deal of care\footnote{Hence, dimensional regularization of theories involving operators with the $\gamma_5$ matrix is perhaps not the best choice.} when  generalizing to continuous complex dimensions~\cite{LeibbrandtDimReg}.

As we already mentioned, one aspect of dimensional regularization is the introduction of another parameter, independent of the regulator $\epsilon$. More precisely, changing the dimensionality of the space-time makes the action dimensionful, and requires the introduction of an energy scale, say $\Lambda_\trmi{reg}$, in order to keep it dimensionless. This can be done in two different but equivalent ways. The first way is to multiply the action by the scale raised to an appropriate power, i.e. by $\Lambda_\trmi{reg}^{2\epsilon}$. The second way is by adjusting the mass dimension of the fields so that the kinetic terms have the proper dimension. However, the latter requires a rescaling of the coupling constant by $\Lambda_\trmi{reg}^{-2\epsilon}$, and a subsequent change in the Feynman rules. For the sake of simplifying our discussion, since the renormalization procedure already involves a certain rescaling of the coupling, we find the former approach more consistent, and will stick to it from now on. We shall then change all the three-momentum measures according to
\beq
\int\,\kern-0.5em\frac{\mathop{{\rm d}^{3}\!}\nolimits {\bf k}}{(2\pi)^{3}}\longrightarrow \Lambda_\trmi{reg}^{2\epsilon}\,\int\,\kern-0.5em\frac{\mathop{{\rm d}^{3-2\epsilon}\!}\nolimits {\bf k}}{(2\pi)^{3-2\epsilon}}\,.
\eeq
We then notice that by doing so, there is a simple way of subtracting the potentially divergent contributions within the renormalization procedure. This is known as the minimal subtraction scheme ($\trmi{MS}$). It consists of Laurent expanding a given physical result around $\epsilon=0$, after having performed all the momentum integrations, then subtracting only the poles, i.e. the negative powers of $\epsilon$. Our following discussion on the actual procedure of renormalization shall assume for readability that we work in this subtraction scheme. We will only at the end motivate our choice for a slightly modified -- yet more convenient -- scheme of subtraction. Having regularized our theory, we can now move on to the renormalization procedure itself.

The usual way of getting rid of the divergences is by introducing to the Lagrangian of the theory a series of terms, monomial in the fields, which allow for the cancellation of the ultraviolet divergent contributions order by order in perturbation theory. Renormalization group invariance then states the existence of a certain class of field theories (see the above classification), in which the number of counter-terms remains finite, yet allowing for the cancellation of all the divergent pieces to all orders in perturbation theory. This is certainly not a trivial statement, considering in particular how a theory which would need an infinite number of counter-terms could be interpreted as a fundamental one\footnote{Not to confuse with effective field theories, where non-renormalizability is not a problem. Those are in general interpreted as effective theories of ``more fundamental'' and renormalizable ones, up to some energy scale at which the need of the latter becomes unavoidable.}. Note also that even after the renormalization of all the fields has been performed, it is not guaranteed that composite operators will be finite: Those might need counter-terms by their own. The same applies to gauge fixing terms in the bare Lagrangian of gauge theories, where the so-called Slavnov-Taylor identities restrict the form of the allowed counter-terms.

The introduction of counter-terms is, however, not sufficient for carrying out the renormalization procedure, and in particular one needs to determine their coefficients for the purpose of practical computations. This is done by means of normalization conditions that need to be imposed on the divergent Green's functions $\mathcal{G}^{(n)}$ and proper vertices $\Gamma^{(n)}$ of the theory\footnote{Note that $\Gamma^{(2)}$ is proportional to the inverse Green's function $\mathcal{G}^{(2)}$, and that the $\Gamma^{(n>2)}$ are related to the one-particle irreducible $n$-point Green's functions, whose external momenta have been set to zero.}. Properly implemented, order by order in perturbation theory, these conditions not only fix the divergences encountered but also the finite parts of the counter-terms. For example, consider a massive scalar field theory with a quartic interaction and a bare coupling $g_0$ as well as a bare mass $m_0$. The renormalization conditions amount to fixing the renormalized $2$-point vertex so that $\Gamma^{(2)}_\trmi{R}(k^2=0)=-m^2$, together with its derivative satisfying $\mbox{d\,}\Gamma^{(2)}_\trmi{R}(k^2=0)/\mbox{d\,}k^2=1$ as well as the renormalized $4$-point proper vertex obeying $\Gamma^{(4)}_\trmi{R}(k_1^2=0,k_2^2=0,k_3^2=0,k_4^2=0)=-g$.

Provided that the structure of each counter-term is the same as the corresponding piece in the bare Lagrangian, the operation of introducing the counter-terms can be re-interpreted as a redefinition of the parameters and fields of the bare theory. This redefinition is then performed via rescaling, and the factors by which one rescales various parameters correspond to the coefficients of the counter-terms that have to be determined order by order in perturbation theory. Within dimensional regularization, those admit a double expansion: A Taylor series in powers of the coupling and a Laurent expansion around $\epsilon=0$. In such a situation, it should be noted that renormalization turns to a transformation which is dictated by a certain scale invariance, under the symmetry of the renormalization group. For example, with a given $n$-point proper vertex\footnote{The present discussion extends equally well to the Green's functions.}, this transformation reduces to the following scaling relation
\beq
\Gamma^{(n)}_\trmi{R}(k_1,...k_n;g,m^2)=Z^{n/2}(g,\Lambda/\Lambda_\trmi{reg})\,\,\Gamma^{(n)}(k_1,...k_n;g_0,m_0^2,\Lambda)\,,
\eeq
for which we have introduced another energy scale $\Lambda$, known as the renormalization scale. This time, the scale is not connected to the regularization procedure but denotes the energy at which the renormalization is performed and the potentially divergent contributions are removed. Note that for simplicity here and in the rest of the manuscript, we shall set $\Lambda_\trmi{reg}\equiv \Lambda$ without any loss of generality. Indeed, as the only dimensionless combinations from the above relation are $g$, $\Lambda/\Lambda_\trmi{reg}$, $k/\Lambda_\trmi{reg}$, and $k/\Lambda$, we see that the renormalization constant $Z$ can only be a function of the coupling and the ratio $\Lambda/\Lambda_\trmi{reg}$, which was shown in the above equation. However, it turns out that while expanding in powers of $\epsilon$, the regularization scale gets canceled, at every order in $g$, through such a combination
\beq
\log \left(\frac{k^2}{\Lambda_\trmi{reg}^2}\right) + \log \left(\frac{\Lambda_\trmi{reg}^2}{\Lambda^2}\right) = \log \left(\frac{k^2}{\Lambda^2}\right)\,.
\eeq
We shall then avoid the distinction between the two formally different scales from now on.

It should be noted, in addition, that the renormalized proper vertex $\Gamma^{(n)}_\trmi{R}(k_1,...k_n;g,m^2)$ must not depend on the renormalization scale, even though the parameters of the theory do. This leads us to the so-called Callan-Symanzik equation, which is nothing but an implementation of this statement, i.e. $\mbox{d\,}\Gamma^{(n)}_\trmi{R}(k_1,...k_n;g,m^2)/\mbox{d\,}\Lambda=0$. By means of chain rules we arrive at the following equation
\beq
\left(\Lambda\,\frac{\partial\,}{\partial\,\Lambda}+\Lambda\,\frac{\mbox{d\,}g}{\mbox{d\,}\Lambda}\,\frac{\partial\,}{\partial\,g}+\Lambda\,\frac{\mbox{d\,}m}{\mbox{d\,}\Lambda}\,\frac{\partial\,}{\partial\,m}\right)\,\,\Gamma^{(n)}_\trmi{R}(k_1,...k_n;g,m^2)=n\,\,\frac{\Lambda}{2}\,\,\frac{\mbox{d\,}\log Z}{\mbox{d\,}\Lambda}\,\,\Gamma^{(n)}_\trmi{R}(k_1,...k_n;g,m^2)\,.
\eeq
It is important to note that each of the terms in this equality obeys an equation which stems out of the renormalization group, and renormalized parameters such as the coupling $g$ or the mass $m$ will have to depend on the renormalization scale in exactly such a way that the above equation holds, order by order in perturbation theory. Concretely, e.g. the coefficient of the term containing a derivative with respect to the gauge coupling is nothing but the so-called beta function $\beta(g)$. These coefficients need to be determined to a given order in perturbation theory, and the ensuing equation solved to give the corresponding running of the parameter in question. For example with the $\beta(g)$ function of QCD, we obtain at leading order
\beq
\beta\trmi{QCD}(g)\equiv\Lambda\,\frac{\mbox{d\,}g}{\mbox{d\,}\Lambda}=-g^3\,\mbox{b}_\trmi{0}+{\cal O}\left(g^5\right)\,,
\eeq
where $\mbox{b}_\trmi{0}=(11\,N_\trmi{c}-2N_\trmi{f})/(48\,\pi^2)$ is the first coefficient of the perturbative series. This yields the leading order perturbative running of the coupling,
\beq
g^2_\trmi{1-loop}(\Lambda)=\frac{11\,N_\trmi{c}-2N_\trmi{f}}{24\,\pi^2}\,\,\Big/\log \left(\frac{\Lambda}{\Lambda_\trmi{QCD}}\right)\,.
\eeq

In the above one-loop solution for the coupling, we see the appearance of a new scale, $\Lambda_\trmi{QCD}$. From a technical point of view, this is merely a constant of integration. However, from a physical point of view, it is a fundamental parameter of the theory that has to be determined from experimental input, or nonperturbatively via lattice Monte Carlo simulations. We see, indeed, that it sets the scale (a couple of hundred MeV) where perturbation theory is meant to break down due to the unphysical Landau pole in the running of the coupling. Notice also that the first coefficient in the perturbative series of the QCD beta function is clearly negative for $N_\trmi{f}<33/2$. This signals that the charge, i.e. the strength of the coupling, decreases at short distances or high energies, which was first noticed in~\cite{AsymptoFreedom1GrossWilczek,AsymptoFreedom2Politzer} for QCD and named asymptotic freedom. We finally refer the reader to~\cite{RunningCouplingQCDReview} for more details about the beta function, the running of the coupling, as well as the link between the former and the fundamental scale of QCD $\Lambda_\trmi{QCD}$.

Next, one runs into the issue that when evaluated to any finite order, perturbative results carry dependence on the value of the renormalization scale, which is usually chosen to have a typical value relevant for the physical process under study. In the thermodynamical context, one typically ends up using physical arguments to fix the scale as some function of the temperature and/or the chemical potentials, asymptotically behaving linearly with the temperature. We refer the reader to Section~\ref{section:Fixing_parameters} for more discussion concerning our choice in the present investigation.

Finally, as mentioned previously, there exist a very convenient way for subtracting divergences within dimensional regularization. This is known as the modified minimal subtraction scheme ($\MSbar$)~\cite{BardeenMutaMSbar}, and allows one to get rid of finite terms such as $\log (4\pi)\!-\!\gamma_\trmi{\tiny E}$ in the perturbative results, always coming along with the poles in $\epsilon$. This choice is perfectly fine, as it is consistent with the renormalization group invariance, and does nothing but exploit the freedom in defining the running of the coupling when solving the renormalization group equations. In turn, this argument boils down to the existence of an equivalence class of theories, related by finite renormalizations, and means nothing but that changing $\Lambda$ into $\Lambda^{\prime}$ amounts to a finite redefinition and renormalization of the (already finite) parameters and fields~\cite{ItzyksonAndZuber}. As a consequence, the physical result shall not be affected by such a choice, when summing over all perturbative orders. However, when truncating at some finite order, a dependence on the subtraction scheme might appear via a dependence of the solution to the renormalization group equations. As we will deal only with one- and two-loop runnings of the coupling in massless QCD, we shall not be worried about this, knowing that the perturbative solutions to the QCD beta function are independent of the subtraction scheme through two-loop order. Finally, we note that the $\MSbar$ scheme can easily be implemented using the $\trmi{MS}$ one thanks to the following identity
\beq\label{MSMSbarRelation}
\lmsb\equiv\Lambda\,e^{(\log 4\pi-\gamma_\trmi{\tiny E})/2}\,,
\eeq
which relates the corresponding renormalization scales. We shall then, from the very beginning, define our three-momentum integrals using
\beq
\int\,\kern-0.5em\frac{\mathop{{\rm d}^{3}\!}\nolimits {\bf k}}{(2\pi)^{3}}\longrightarrow \left(\frac{\lmsb^2\,e^{\gamma_\trmi{\tiny E}}}{4\pi}\right)^{\epsilon}\,\int\,\kern-0.5em\frac{\mathop{{\rm d}^{3-2\epsilon}\!}\nolimits {\bf k}}{(2\pi)^{3-2\epsilon}}\,.
\eeq

We are now finally ready to introduce in the next chapter our original work concentrating on the bulk equilibrium properties of the quark-gluon plasma.

\chapter{Hot and dense thermodynamics}\label{chapter:Thermodynamics_Hot_Dense}

This chapter introduces the bulk equilibrium properties of a hot and dense system, starting from the partition function of the theory. In particular, it introduces a number of physical quantities obtained from partial derivatives of the pressure with respect to the chemical potentials. These quantities give access to various cumulants of conserved charges, and are of interest for the phase diagram of QCD.

This chapter is meant to provide a comprehensive picture, but focuses on aspects relevant for our study. For more details on the thermodynamics of quantum fields, or in general thermodynamics, we refer the reader to the textbooks~\cite{Bogolyubov2Stat} and~\cite{LandauLifshitzStat}. We also refer to the reviews~\cite{SatzCumulants,KochCumulants} regarding fluctuations and correlations of conserved charges in the context of heavy ion collisions.

In the following, we first briefly derive fundamental quantities and thermodynamic relations relevant for bulk properties. Then, we discuss the issue of thermodynamic consistency within our approach, anticipating the introduction of two different frameworks that we will use, before to give more details about them in the next two chapters. Finally, we introduce the concept of correlations and fluctuations of globally conserved quantum numbers, thereby motivating our study.

\section{Bulk thermodynamics and fundamental relations}\label{section:Bulk_Thermodynamics_Fundamental_Relations}

Let us start from the partition function of QCD $\mathcal{Z}_\trmi{QCD}$, as defined in~(\ref{CanonicalPathGaugeTheory}). This object, as we are going to demonstrate, turns out to almost fully describe the thermodynamic equilibrium of a hot and dense system in the grand canonical ensemble. In other words, the system can freely exchange an arbitrary amount of heat and particles with its surroundings\footnote{In this context, ``surroundings'' has to be understood neither in the local (microscopic) nor in the global (macroscopic) sense, but rather in a mesoscopic one.}. We shall work out the corresponding thermodynamic definitions for various basic physical quantities, as well as the relations between them. Note that we are going to first consider a system with a finite volume $V$, and later take the limit relevant for large volumes\footnote{We shall nevertheless ignore shear effects from the very beginning, assuming the volume to be large enough.}. In addition, we would like to point out that despite the fact that all of our formulas refer to the rest frame of the heat bath, an explicit covariant formulation can always be obtained with the help of the four-velocity vector of the rest frame.

As we are only interested in bulk thermodynamic effects, that is we ignore possible surface effects, various thermodynamic quantities can be obtained from the partition function itself via the relations
\beqa
\mathcal{P}_\trmi{QCD}&\equiv&T\,\,\frac{\partial\log\mathcal{Z}_\trmi{QCD}}{\partial V}\,,\\
\mathcal{S}&\equiv&\frac{1}{V}\,\,\frac{\partial\,T\log\mathcal{Z}_\trmi{QCD}}{\partial T}\,,\\
\mathcal{N}_f&\equiv&\frac{T}{V}\,\,\frac{\partial\log\mathcal{Z}_\trmi{QCD}}{\partial \mu_f}\,.
\eeqa
where $\mathcal{P}_\trmi{QCD}$, $\mathcal{S}$ and $\mathcal{N}_f$ stand for the pressure, entropy and particle number densities of the system, respectively. Then, taking the infinite volume limit $V\rightarrow\infty$, the pressure of the system reduces to
\beq\label{PressurePartitionFunctionInfiniteV}
\mathcal{P}_\trmi{QCD}=\frac{T}{V}\,\,\log\mathcal{Z}_\trmi{QCD}\,,
\eeq
which leads to the following simple relations for the entropy and particle number densities
\beqa
\mathcal{S}&=&\frac{\partial\,\mathcal{P}_\trmi{QCD}}{\partial T}\,,\\
\mathcal{N}_f&=&\frac{\partial\,\mathcal{P}_\trmi{QCD}}{\partial \mu_f}\,.
\eeqa

It is customary to call the relation between to pressure and its first derivatives, as defined above, the equation of state of the system. In order to establish this relation, we first make use of the definition of the grand canonical density operator~(\ref{DensityOperator}), together with some fundamental properties of the entropy (such as the additivity), given the well known statistical relation
\beq
\mathcal{S}=-\left\langle\log\hat{\rho}\right\rangle_{S_\trmi{QCD}}\,.
\eeq
Here, we recall that $\left\langle...\right\rangle_{S_\trmi{QCD}}$ is the thermal average~(\ref{ExpectationValues}) but for the QCD action $S_\trmi{QCD}$, obtained using the Lagrangian density~(\ref{EffectiveLagrangianQCD}). We then get the so-called equation of state
\beq\label{EquationOfState}
\mathcal{E}+\mathcal{P}_\trmi{QCD}=T\,\mathcal{S}+\sum_f\,\mu_f\,\mathcal{N}_f\,,
\eeq
which gives the energy density of the system $\mathcal{E}$, otherwise defined as
\beq\label{StatDefEnergyDensity}
\mathcal{E}\equiv\frac{1}{V}\,\,\left\langle\hat{H}_\trmi{QCD}\right\rangle_{S_\trmi{QCD}}\,.
\eeq

We now move on to discuss thermodynamic consistency in the next section, considering in particular the frameworks we are going to use in our computations.

\section{Thermodynamic consistency}\label{section:Thermodynamic_Consistency}

In this section, we choose to simplify the discussion by setting the chemical potentials to zero\footnote{It should be noted, however, that the generalization of all of the following results to finite density is straightforward.}. In this limit, the equation of state reduces to
\beq\label{EquationOfStateZeroMu}
\mathcal{E}+\mathcal{P}_\trmi{QCD}=T\,\frac{\mbox{d}\,\mathcal{P}_\trmi{QCD}}{\mbox{d}\,T}\,,
\eeq
and many other thermodynamic relations simplify similarly. Our discussion follows to some extent the reference~\cite{ThermoConsistencyOriginalGorensteinNanYang}, which contain a comprehensive treatment of the topic, as well as to~\cite{PeshierSolThermoConsistency,RebhanVariantSolThermoConsistency}, which offer other ways to resolve the problem we are about to explain.

In short, the problem of thermodynamic consistency means nothing but that the statistical definition of the energy density~(\ref{StatDefEnergyDensity}) does not match the expression obtained from the pressure and its first derivative when using the equation of state~(\ref{EquationOfStateZeroMu}). In order to make this statement more precise, let us first consider a system, in which the temperature appears only explicitly, i.e. which has no temperature dependent effective parameters whatsoever and whose Hamiltonian operator is temperature independent as well. Then, by rewriting the corresponding~(\ref{StatDefEnergyDensity}) for this system, using the trace rather than a thermal average, we get
\beq\label{StatDefEnergyDensityWithTrace}
\mathcal{E}(T)=\frac{\mathcal{Z}^{-1}(T)}{V}\,\,\mbox{Tr} \left(\hat{H}\,\,e^{-\hat{H}/T}\right)\,,
\eeq
which is indeed the statistical definition of the energy density. Note that above and in the following, we assume the possible gauge freedoms to have been fixed, and drop any index referring to this procedure for the sake of readability\footnote{See Section~\ref{section:Path_Integral} for details on the gauge fixing procedure.}. Now, using the formal definition of the partition function~(\ref{PartitionFunction}), we have
\beq\label{StatDefPrssureWithTrace}
\mathcal{P}(T)=\frac{T}{V}\,\,\log\,\,\mbox{Tr} \left(e^{-\hat{H}/T}\right)\,.
\eeq
It is then trivial to obtain the equation of state of our system in the form of~(\ref{EquationOfStateZeroMu}) by differentiating the above and using~(\ref{StatDefEnergyDensityWithTrace}).

Next, we look at a system that has temperature dependent parameters, in which case we quickly run into problems. For the sake of argument, let us assume that the only $T$-dependent parameter is an effective mass parameter $M\equiv M(T)$. Thus, simply by means of the chain rule, the differentiation of the corresponding~(\ref{StatDefPrssureWithTrace}) now leads to
\beqa
T\,\frac{\mbox{d}\,\mathcal{P}(T,M)}{\mbox{d}\,T}&=&\mathcal{P}+\frac{\mathcal{Z}^{-1}(T,M)}{V}\,\,\mbox{Tr} \bigg(\hat{H}(M)\,\,e^{-\hat{H}(M)/T}\bigg)\, \nonumber \\
&-&\frac{T/V}{\mathcal{Z}}\,\,\frac{\mbox{d}\,M(T)}{\mbox{d}\,T}\,\,\mbox{Tr} \left(\frac{\partial\hat{H}(M)}{\partial M}\,\,e^{-\hat{H}(M)/T}\right) \,, \\
&=&\mathcal{P}+\mathcal{E}-\frac{T/V}{\mathcal{Z}}\,\,\frac{\mbox{d}\,M(T)}{\mbox{d}\,T}\,\,\mbox{Tr} \left(\frac{\partial\hat{H}(M)}{\partial M}\,\,e^{-\hat{H}(M)/T}\right) \,,
\eeqa
and we see that the last term on the right hand side of the last equation invalidates the canonical equation of state ~(\ref{EquationOfStateZeroMu}), leading to a mismatch in the expression for the energy density. This observation is of direct relevance to our forthcoming frameworks, given that both our effective parameters $m_\trmi{E}(T)$, $g_3(T)$, $m_\trmi{D}(T)$, $m_\trmi{q$_f$}(T)$ and the coupling $g(T)$ will be medium dependent\footnote{See Section~\ref{section:Resummed_DR} and~\ref{section:HTLpt} for more details on the dimensional reduction and HTLpt setups, respectively. We recall that in our investigation, these effective parameters can also be chemical potential dependent.}. The former dependence is in part explicit while the latter arises from the choice of the renormalization scale.

Regarding both our frameworks and considering the above, we see that all possible additional terms will be proportional to some derivatives with respect to the temperature of either the effective parameters or the coupling. For example, such a derivative reads $\mbox{d}^n g(T)/\,\mbox{d}T^n$ with $n\geq 1$. As far as the latter dependence is concerned, knowing the functional form of the running of the coupling\footnote{We recall that we typically consider the one-loop perturbative running.}, it is obvious that these derivatives decrease with increasing $n$, and
\beq
T^n\,\frac{\mbox{d}^n g(T)}{\mbox{d}T^n}\ll T\,\frac{\mbox{d} g(T)}{\mbox{d}T}\sim g^2(T) \,.
\eeq
This is the reason why we can safely ignore the higher derivative contributions in the present investigation. As to terms proportional to derivatives of the effective parameters, e.g. $\mbox{d}^n m_\trmi{E}(T)/\mbox{d}T^n$ or $\mbox{d}^n m_\trmi{D}(T)/\mbox{d}T^n$ for the dimensional reduction and HTLpt frameworks, respectively, the situation is similar in the sense that one can always show the violation of thermodynamic consistency to be of higher order in $g$ than the order to which the perturbative calculation is performed.

\section{Correlations and fluctuations of conserved charges}\label{section:Correlations_Fluctuations}

Considering that the partition function of QCD can be written in the form\footnote{Again, we assume the gauge fixing procedure to have been carried out.}
\beqa\label{SplittingPartitionFunction}
\mathcal{Z}_\trmi{QCD}&=&\mbox{Tr}\ \exp\left[-\beta\!\int\!\mathop{{{\rm d}^d}\vect{x}}\ \bigg(\left.\mathcal{\hat{H}}_\trmi{QCD}\right|_{\mu_f=0}-\sum_f\mu_f\mathcal{\hat{Q}}_{f}\bigg)\right]\,,
\eeqa
it might be interesting to investigate the physical quantities that the quantum operators $\mathcal{\hat{Q}}_{f}$ give when thermally averaged. For example, the mean
\beq
\left\langle \mathcal{\hat{Q}}_f \right\rangle_{S_\trmi{QCD}}\,,
\eeq
and the (co)variance
\beq
\left\langle \left(\mathcal{\hat{Q}}_f-\left\langle \mathcal{\hat{Q}}_f \right\rangle_{S_\trmi{QCD}}\right) \cdot \left(\mathcal{\hat{Q}}_g-\left\langle \mathcal{\hat{Q}}_g \right\rangle_{S_\trmi{QCD}}\right) \right\rangle_{S_\trmi{QCD}}\,,
\eeq
measure nothing but the fluctuation and correlation of conserved number densities. The latter is called correlation simply because it involves more than one operator, but reduces to a fluctuation (here a variance) if we chose $\mathcal{\hat{Q}}_g=\mathcal{\hat{Q}}_f$.

For the present purpose of defining quantities of the above type, we only consider the up, down and strange quark conserved numbers, as the splitting of the Hamiltonian in~(\ref{SplittingPartitionFunction}) is then trivial. However, notice that we could also express the partition function in terms of the baryon, electric charge and strangeness numbers\footnote{Indeed, both sets of operators come with their chemical potentials which are related via linear relations to the up, down and strange quark chemical potentials. Describing fluctuations of cumulants with quark numbers is then equivalent to doing so with baryon number.}. It is then clear that when considering the covariance of two charges, says $\mathcal{\hat{Q}}_\trmi{u}$ and $\mathcal{\hat{Q}}_\trmi{d}$, we probe the correlation between the two corresponding flavors which has been argued to contain information about possible bound-state survival in the plasma above $T_c$~\cite{BOUNDSTATESURVIVAL1,BOUNDSTATESURVIVAL2}. On the other hand, fluctuations of charges tell us how the system reacts to small increases of density. Also, the existence and location of a possible critical point on the phase diagram of QCD can be investigated using the behavior of certain cumulants. For example, the variance of the baryon number $\chi_\trmi{B2}\left(T,\mu_\trmi{B}\right)$ as a function of the baryon chemical potential $\mu_\trmi{B}$ is expected to display a sharp peak at the critical point $\left(T_\trmi{c},\mu_\trmi{B,c}\right)$ due to its sensitivity with respect to changes of density in the system. What is usually done on the lattice is to Taylor expand such a quantity, leading to higher order cumulants at vanishing chemical potentials, which can then be computed by means of Monte Carlo simulations. We refer the reader to~\cite{HandLatticeDatasu6u2} for a thorough discussion on such an investigation in the case of two flavors.

It is of course obvious that the mean and (co)variance defined above can be obtained via successive differentiations of the logarithm of the partition function,
\beqa
\left\langle \mathcal{\hat{Q}}_f \right\rangle_{S_\trmi{QCD}}&=&T\,\frac{\partial}{\partial \mu_f} \log \mathcal{Z}_\trmi{QCD} \, , \label{Mean} \\
\left\langle \left(\mathcal{\hat{Q}}_f-\left\langle \mathcal{\hat{Q}}_f \right\rangle_{S_\trmi{QCD}}\right) \cdot \left(\mathcal{\hat{Q}}_g-\left\langle \mathcal{\hat{Q}}_g \right\rangle_{S_\trmi{QCD}}\right) \right\rangle_{S_\trmi{QCD}}&=&T^2\,\frac{\partial^2}{\partial \mu_f \partial \mu_g} \log \mathcal{Z}_\trmi{QCD} \,,
\eeqa
as does a plethora of higher order fluctuations and correlations. Besides, from~(\ref{PressurePartitionFunctionInfiniteV}) we see that the practical computation of such quantities simply boils down to taking partial derivatives of the pressure with respect to various chemical potentials, each of them corresponding to a conserved charge. These quantities are typically referred to as susceptibilities, and we define them, here for quark numbers only, via the generic formula
\beq
\chi_\trmi{u\!\;\!\! i\;d\!\;\!\! j\;s\!\;\!\! k\; ...}\left(T,\left\{\mu_f\right\}\right) \equiv \frac{\partial^{i+j+k+...}\,\, \mathcal{P}_\trmi{QCD}\left(T,\left\{\mu_f\right\}\right)}
{\partial\mu_\trmi{u}^i\, \partial\mu_\trmi{d}^j \, \partial\mu_\trmi{s}^k\, ...} \ ,
\eeq
where we have not considered the effect of possible medium dependent effective parameters\footnote{See the previous section for a detailed discussion on the effects of the medium dependence for such parameters.}.

It is good to note that in Chapter~\ref{chapter:Results}, we will only investigate diagonal susceptibilities at vanishing chemical potentials. Considering that we set from the beginning the quark masses to zero, these susceptibilities moreover always refer to those of the light quark flavors. Indeed, within our perturbative investigations, cumulants related to the strange flavor have the same values as the light quark ones. Finally, the following relations between the different cumulants often turn out useful
\beqa
\chi_\trmi{B2}&=& \Big(\chi_\trmi{u2}+\chi_\trmi{d2}+\chi_\trmi{s2}+2\,\chi_\trmi{ud}+2\,\chi_\trmi{ds}+2\,\chi_\trmi{us}\Big)/9 \,, \\
\chi_\trmi{B4}&=& \Big(\chi_\trmi{u4}+\chi_\trmi{d4}+\chi_\trmi{s4}+4\,\chi_\trmi{u3\,d}+4\,\chi_\trmi{u3\,s}+4\,\chi_\trmi{d3\,u}+4\,\chi_\trmi{d3\,s}+4\,\chi_\trmi{s3\,u}+4\,\chi_\trmi{s3\,d} \nonumber \\
&&\,\,\,+6\,\chi_\trmi{u2\,d2}+6\,\chi_\trmi{d2\,s2}+6\,\chi_\trmi{u2\,s2}+12\,\chi_\trmi{u2\,ds}+12\,\chi_\trmi{d2\,us}+12\,\chi_\trmi{s2\,ud}\Big)/81 \,.
\eeqa
They are particularly relevant for Figures~\ref{fig:ChiB4Nf3} and~\ref{fig:Ratiosu4Overu2Andb4Overb2Nf3}, as those involve baryon numbers.

Before entering more details regarding the two perturbative resummation frameworks that will be used in our study, we would like to stress some references on statistical QCD and heavy ion collisions to the reader. In particular, we have found ~\cite{SatzCumulants,KochCumulants} very useful regarding many aspects of the above discussions.

\chapter{Resummed perturbative thermal QCD}\label{chapter:Resummed_Perturbative_QCD}

In this chapter, our goal is to motivate the need for resummed perturbative QCD, as well as to describe the two setups that will be used in our thermodynamic analysis. The resummed framework inspired by dimensional reduction is presented first together with the corresponding result for the pressure. As it is a rather specialized topic, we will give all the needed references along the main text.

We start by briefly reviewing some of the difficulties, known as infrared problems, in evaluating thermodynamic quantities within the unresummed weak coupling expansion. We also present the ways out and the possible improvements, before introducing in detail the frameworks of resummation that we will use in deriving our results in Chapter~\ref{chapter:Results}. These are motivated by the phenomenon of dimensional reduction of high-temperature QCD~\cite{DRPhenomenon2}, as well as by the hard thermal loop limit of high temperature QCD~\cite{FrenkelTaylorHTL,BraatenPisarskiHTL}.

\section{Naive weak coupling expansion in thermal QCD}\label{section:Weak_Coupling}

We shall first focus on the infrared divergences in thermal field theories, which arise when using purely unresummed perturbation theory\footnote{Which we will also call ``naive'' perturbation theory.}. The so-called infrared catastrophe was first discovered by A.~D.~Lind\'{e}~\cite{LindeIRcatastrophe}. Years after, E.~Braaten suggested a way out, which combines both effective field theory methods and Monte Carlo simulations~\cite{BraatenSolIRcatastrophe}.

Recalling that, diagrammatically, perturbative QCD amounts to expanding functional integrals in an even power series of the coupling, a general problem appears when dealing with massless bosonic fields\footnote{Only bosonic fields are affected by infrared divergences, as the fermionic ones have nonvanishing Matsubara frequencies. This does, however, not exclude the generation of thermal masses and the existence of soft (plasminos) or even ultra-soft~\cite{DaisukeUltrasoftFermionicModes} modes.}. Indeed, the absence of a mass term in the corresponding propagators seems problematic as far as the low momentum regions of the loop integrations are concerned. This problem can be understood via the zero Matsubara mode of a massless bosonic field, the temporal component of the momentum turning into a discrete variable that has to be summed over\footnote{That is to say for the $n=0$ term of the Matsubara sum, which for bosonic fields has a zero frequency.}. Hence, for such a massless propagator it is obvious that the integrand can be singular in a region where the three-momentum tends to zero. Note that in general, this type of divergence cannot be avoided, unlike the ultraviolet ones. Despite the fact that it is a quite generic feature of bosonic massless field theories, it has more severe consequences in gauge theories, which we will now investigate.

Although this problem exists for any perturbative quantity given a high enough order in $g$, let us take a look at a $(1\!+\!l)$-loop vacuum diagram contributing to the pressure. For example, consider a ``ladder'' graph with $3l$ propagators and $2l$ three-gluon vertices, and introduce an infrared cut-off $M$ for the propagators. Neglecting possible ultraviolet divergences, its zero mode structure behaves as~\cite{KapustaBook}
\beq
B^{(1+l)}_{n=0}\propto g^{2l}\,T^4\,\intedki{1}...\,\intedki{1+l}  \frac{k^{2l}}{\Big(k^2+\left(M/T\right)^2\Big)^{3l}}\,,
\eeq
which in three spatial dimensions is obviously finite in the limit $M\rightarrow 0$, provided that $1\leq l\leq 2$. However, still in $3\!+\!1$ dimensions, the diagram becomes logarithmically divergent already at four loops
\beq
B^{(4)}_{n=0}/T^4\sim\,g^{6}\log (T/M)\,,
\eeq
and even worse, for $l>3$, it is power divergent
\beq\label{PowerLikeIRDiv}
B^{(1+l)}_{n=0}/T^4\sim\,g^{6} \left(g^2\,T/M\right)^{l-3}\,.
\eeq

Of course, the previous reasoning applies only for bare propagators, and the electric screening phenomenon\footnote{The screening of the electric charges refers to the temporal components of the gauge fields whose two point Green's functions exhibit a nonvanishing thermal scale of order $g\,T$.} makes the corresponding propagators effectively acquire a thermal mass of order $g\,T$. This calls for resumming a certain type of (ring) diagrams to all orders in perturbation theory. Consequently, such a resummation may well be a way out if only there would be no more scales generated, or if all the generated scales would be at least of order $g\,T$. Indeed, it can be seen from~(\ref{PowerLikeIRDiv}) that for thermal masses $M\sim g^2\,T$, all diagrams beyond four loops contribute at the same order $g^6$. And unfortuntely it turns out that in non-Abelian Yang-Mills theories, this scale indeed gets generated by the spatial component of the gauge fields. This new scale, being of order $g^2\,T$~\cite{GrossPisarskiYaffeQCDInstantons}, leads to an infrared catastrophe in the sense that the unresummed perturbative computation of the pressure breaks down at the order $g^6$.

Besides making naive perturbative expansions inapplicable beyond a certain order, straightforwardly resumming the infrared divergences also tend to induce very bad convergence features for the perturbative series, and is in fact not possible for the magnetic sector. We refer the reader to~\cite{ConvergenceWeakCouplingExpansion} for more details on the apparent convergence of perturbative thermal QCD. As the present work is not meant to be accurate beyond $g^6\,\log g$, we shall, however, not deal with the infrared catastrophe itself. Instead, we rather make use of two frameworks which considerably improve the convergence properties of the QCD pressure already at this order.

In the following two sections, to which we point for references on the topics, we will introduce two ways of curing the convergence problems. First, we resum certain higher order contributions by means of a weak coupling expansion within EQCD, motivated by the phenomenon of dimensional reduction. Then, we use hard-thermal-loop perturbation theory, where the expansion point of perturbation theory is shifted to an ideal gas of massive thermal quasiparticles. The latter provides a substantial improvement in the convergence properties of the weak coupling expansion, despite the fact it is formally not meant to be manageable from four loops onwards\footnote{As such a framework makes only sense within a perturbative expansion, the way out of the infrared catastrophe suggested by E.~Braaten in~\cite{BraatenSolIRcatastrophe} is of course not applicable. In any case, based on the asymptotic freedom properties of QCD~\cite{AsymptoFreedom1GrossWilczek,AsymptoFreedom2Politzer}, corrections of order $g^6$ and beyond are expected to be negligible at very high temperatures.}. The former, on the other hand, makes use of the very same effective field theory, which sets the stage for solving the infrared catastrophe\footnote{Being only interested in the pressure to $g^6\,\log g$, we shall ignore Magnetostatic QCD (MQCD) which starts to contribute at the order $g^6$ and encodes the fundamentally non-perturbative contributions to the pressure.}, and more interestingly in the present context, allows for weak coupling calculations to be carried out to high orders in perturbation theory. Moreover, let us point out that both approaches rely on resummations of a certain class of higher order diagrams. Finally, before further introducing these frameworks, we refer the interested reader to the reviews~\cite{AntonReview,JensMikeReview,BlaizotThermo,BlaizotHTLResummation1} for more details about various other resummation schemes and higher order contributions.

\section{Resummation inspired by dimensional reduction}\label{section:Resummed_DR}

In the imaginary time formalism of thermal QCD, the compact temporal direction clearly shrinks as $1/T$ at high temperatures~\cite{DRPhenomenon2,DRPhenomenon1}. This phenomenon, known as dimensional reduction, effectively renders the system three-dimensional. Taking advantage of it, a way out of the perturbative infrared catastrophe, as proposed in~\cite{BraatenSolIRcatastrophe}, consists of using two three-dimensional effective field theories~\cite{KajantieEQCD,BraatenEQCD} in order to exactly reproduce the infrared sector of QCD\footnote{The ultraviolet sector is, on the other hand, probed using the full theory itself.}. As a consequence of dimensional reduction, and based on the Appelquist-Carrazone decoupling theorem~\cite{AppelquistCarrazoneDecoupling}, it can be shown that the relevant degrees of freedom for full QCD contributing at length scales of order $1/(gT)$ and larger reduce at very high temperature to the zero Matsubara modes of the gluon fields.

The first of these effective theories (EQCD; see Section~\ref{section:Renormalization_Running} for the acronym) can be tackled via perturbation theory. It encodes the dynamics of the length scales from order $1/(gT)$, and should be matched to the full theory using some observables (generally the long-distance behavior of Green's functions). The second one (MQCD; see the previous section for the acronym) is by essence non-perturbative as it contains the correct contributions from the non-perturbative scale $g^2 T$, and turns to encode the dynamics of the length scale of order $1/(g^2 T)$. The coefficients of this theory also need to be matched to the full theory, but in order to compute the non-perturbative contributions, one has to use three-dimensional lattice Monte Carlo simulations\footnote{Of course, lattice QCD in four dimensions can be used in the first place, in order to compute thermodynamic observables. However, the necessary computational resources needed to simulate a three-dimensional theory are much more modest than the ones needed in simulations of four-dimensional QCD.}. Doing so, the most delicate part resides in the high loop order matching procedure between the lattice and whatever continuum regularization schemes used~\cite{DiRenzoMatchingLattice3dQCD}. Note that, in turn, the modes relevant to length scales of order $1/(gT)$ and larger are responsible for all the aforementioned infrared problems in QCD. This makes those effective theories particularly relevant for the perturbative determination of various thermodynamic (static) quantities. However, when performing a nonperturbative study of the infrared sector of QCD, it should be noted that those effective theories explicitly break the Z($N_\trmi{c}$) center symmetry present in four-dimensional Yang-Mills theories. This can nevertheless be remedied by means of generalizing the degrees of freedom of the effective theory from the temporal gauge field to coarse grained Wilson loop operators~\cite{AleksiYaffeZ3EffectiveTheory,DeForcrandZEffectiveTheory,TwoColorQCDTomas}. As we are only interested in weak coupling expansions, which correspond to expanding the functionals around the trivial Z($N_\trmi{c}$) vacuum, we shall not be worried about this symmetry, which is anyway explicitly broken by the inclusion of quarks.

As previously mentioned, we shall now focus on EQCD only. The Lagrangian density of this effective theory is obtained by integrating out the hard modes from QCD, i.e. the ones having thermal masses of order $T$. This leads to a three-dimensional SU($N_\trmi{c}$) Yang-Mills theory coupled to an adjoint Higgs field $A_0$, which corresponds at leading order to the zero Matsubara mode of the four dimensional temporal gauge field (for more details on that procedure, see~\cite{KajantieEQCD,BraatenEQCD}). Skipping the possible higher order non-renormalizable field operators (cast in the $\delta{\mathcal L}_\trmi{E}$ piece below), which enter the weak coupling expansion of the pressure beyond order $g^6$, the EQCD Lagrangian reads
\beqa\label{EQCDLagrangian}
{\mathcal L}_{\trmi{EQCD}}&\equiv&\frac{1}{2}\,{\rm Tr}\left[G_{ij}^2\right]+{\rm Tr}\!\left[(D_i\,A_0)^2\right]+m_\trmi{E}^2\,{\rm Tr}\left[A_0^2\right] + i\zeta\,{\rm Tr}\left[A_0^3\right]+\lambda_\trmi{E}\,{\rm Tr}\left[A_0^4\right]+\delta{\mathcal L}_\trmi{E} \,,
\eeqa
where $D_i$ denotes the covariant derivative in the adjoint representation of the gauge group. Note that the above Lagrangian is only valid for $N_\trmi{c}\leq 3$. Indeed, for a larger number of colors, two independent operators quartic in $A_0$ have to be taken into account. We shall notice, as previously mentioned, that the parameters involved in this Lagrangian have to be determined by matching computations in full QCD. In particular, the parameter $\zeta$ is nonvanishing only in the presence of nonzero quark chemical potentials~\cite{HartLainePhilipsen}. It thus enters the quark number susceptibilities, as well as the finite density equation of state but not the zero $\mu_f$ one. Finally, let us mention that for the present purposes, EQCD is used to circumvent the infrared problems encountered in the evaluation of the pressure, and to resum an important class of higher order contributions (see below for a more careful explanation of that point). This turns out to have a remarkable effect on the convergence properties of the corresponding weak coupling expansion, as can be seen from our results in Chapter~\ref{chapter:Results}.

For now, let us explain how thermodynamic computations can be reorganized with the help of this effective field theory. Using the above framework, the pressure of QCD is recast into the simple form~\cite{AleksiFirstPaperPressure,BraatenEQCD,KajantieG6logG}
\beq\label{DRpressure}
p_\trmi{QCD}\left(T,\bm{\mu}\right) \equiv p_\trmi{HARD}\left(T,\bm{\mu}\right) + T\, p_\trmi{SOFT}\left(T,\bm{\mu}\right)\,,
\eeq
where the piece $p_\trmi{HARD}$ is obtained via a strict loop expansion in QCD, while $p_\trmi{SOFT}$ denotes the pressure obtained from the partition function of EQCD. The latter is still non-perturbative in the sense that it contains the dynamics of the length scales of order $1/(g^2 T)$, as the soft modes have not yet been integrated out from EQCD, to give the MQCD effective theory. Hence the function $p_\trmi{SOFT}$ cannot be determined via a perturbative expansion to all orders, as fundamentally non-perturbative contributions start to enter it at ${\cal O}\left(g^6\right)$. However, recalling that our present work is meant to be accurate through the order $g^6 \log g$, we shall not consider this a problem. We further note that the first term accounts for the contribution of the hard scale $T$ and takes the form of an even power series in $g$. The second one, on the other hand, contains all contributions from the soft $gT$ and ultrasoft $g^2T$ scales.

At nonzero quark chemical potentials, the contributions $p_\trmi{HARD}$ and $p_\trmi{SOFT}$ have been computed up to and partially including the order $g^6$. At present, the only missing contribution in the ${\mathcal O}(g^6)$ originates from the four-loop full theory diagrams that one encounters in $p_\trmi{HARD}$ (we refer the interested reader to~\cite{Gynther4LoopDiagram} for details about the evaluation of some of these integrals). Following the procedure of~\cite{MikkoYorkQuarkThresholds}, generalized to nonzero chemical potentials in~\cite{sylvain2}, those two functions can be written as\footnote{Strictly speaking, the two functions in~(\ref{pHARD}) and~(\ref{pSOFT}) do not correspond to the ones in~(\ref{DRpressure}), but only their sum does. Also, we recall that the present reorganization is only meant for computing quantities related to the finite density part of the equation of state.}
\beqa
\frac{p_\trmi{HARD}\left(T,\bm{\mu}\right)}{T^4} &=& \aE{1} + \hat g_3^2 \ \aE{2} + \frac{\hat g_3^4}{(4\pi)^2} \bigg(\aE{3}-\aE{2}\ \aE{7}-\frac{1}{4} d_\trmi{A} C_\trmi{A} \ \aE{5}\bigg) \, \nonumber \\
&+& \frac{\hat g_3^6}{(4\pi)^4}\left[d_\trmi{A} C_\trmi{A}\bigg(\aE{6}-\aE{4}\,\aE{7}\bigg)-d_\trmi{A} C_\trmi{A}^3\left(\frac{43}{3}-\frac{27}{32}\pi^2\right)\right]\log\frac{\lmsb}{4\pi T} +{\mathcal O}(g^6)\, , \ \ \ \ \ \ \ \label{pHARD} \\
\frac{p_\trmi{SOFT}\left(T,\bm{\mu}\right)}{T^3} &=& \frac{\hat m_\trmi{E}^3}{12\pi}\ d_\trmi{A} - \frac{\hat g_3^2 \ \hat m_\trmi{E}^2}{(4\pi)^2}\ d_\trmi{A} C_\trmi{A} \bigg( \log\frac{\lmsb}{2T \hat m_\trmi{E}} + \frac{3}{4}\bigg)-\frac{\hat g_3^4 \ \hat m_\trmi{E}}{(4\pi)^3}\,d_\trmi{A} C_\trmi{A}^2 \bigg(\frac{89}{24} + \frac{\pi^2}{6} - \frac{11}{6} \log 2\bigg) \, \nonumber \\
&+& \frac{\hat g_3^6}{(4\pi)^4}\,d_\trmi{A} \Bigg[C_\trmi{A}^3\left(\frac{43}{4}-\frac{491}{768}\pi^2\right)\log\frac{\lmsb}{2T\hat m_\trmi{E}}+C_\trmi{A}^3\left(\frac{43}{12}-\frac{157}{768}\pi^2\right)\log\frac{\lmsb}{2C_\trmi{A} T\hat g_3^2} \ \, \nonumber \\
& & \ \ \ \ \ \ \ \ \ \ \ \ - \frac{4}{3}\frac{N_\trmi{c}^2-4}{N_\trmi{c}}\Bigg(\,\sum_f\hat\mu_f\,\Bigg)^2\,\log\frac{\lmsb}{2T\hat m_\trmi{E}}\,\Bigg] +{\mathcal O}(g^6)\,, \label{pSOFT}
\eeqa
with as usual $C_\trmi{A}\equiv N_\trmi{c}$. Moreover, we note the factor $2\pi$ of difference in the rescaled mass parameters, with respect to the HTLpt notation; see appendix~\ref{appendix:Notation} for the corresponding definitions. Also, we use the result from~\cite{HartLainePhilipsen}
\beqa
\zeta&=&\frac{\hat g_3^3}{3\pi^2}\sum_f \mu_f + {\mathcal O}(\hat g_3^5)\,.
\eeqa
Note that the matching coefficients $\alpha_\trmi{E1}-\alpha_\trmi{E7}$ in $p_\trmi{HARD}$ depend on $\mu_f/T$, $\bar{\Lambda}/T$ and on various group theory invariants, but are -- by definition -- independent of the coupling $g$. The three first ones are defined via the strict weak coupling expansion of the QCD pressure, and the others stem from
\beqa
\hat m_\trmi{E}^2&=&g^2\Bigl(\aE{4}+\aE{5}\,\epsilon+{\cal O}(\epsilon^2)\Bigr)+\frac{g^4}{(4\pi)^2}\Bigl(\aE{6}+{\cal O}(\epsilon)\Bigr)+{\cal O}(g^6) , \hspace*{0.5cm} \\
\hat g_\trmi{3}^2&=&g^2+\frac{g^4}{(4\pi)^2}\Bigl(\aE{7}+{\cal O}(\epsilon)\Bigr)+{\cal O}(g^6) \,. \label{Expressiong3}
\eeqa
While these coefficients are all listed in the appendix~\ref{appendix:EQCD_Matching_Coefficients}, we refer the reader to~\cite{KajantieG6logG} for more details on their derivation, as well as to~\cite{AleksiFirstPaperPressure} regarding their evaluation at finite density.

Finally, we should note that we have intentionally written both the hard~(\ref{pHARD}) and soft~(\ref{pSOFT}) contributions in terms of the EQCD parameters $g_3$ and $m_\trmi{E}$. The reason for doing so is that we wanted to anticipate the forthcoming higher order resummation. The explicit logarithms of the renormalization scale in~(\ref{pHARD}) have namely been chosen so that the scale dependence gets canceled to order $g^6$ in $p_\trmi{SOFT}$~\cite{MikkoYorkQuarkThresholds}. If the sum of the two contributions is re-expanded in powers of $g$, discarding all the terms from order $g^6$, the correct $g^6 \log g$ weak coupling expression is clearly obtained. However, this procedure leaves us with some freedom in the treatment of higher order contributions. This can be done by considering $p_\trmi{HARD}$ and $p_\trmi{SOFT}$ functions of the three-dimensional gauge coupling and the electric screening mass\footnote{That is to say by keeping the corresponding parameters $g_3$ and $m_\trmi{E}$ unexpanded in $g$.}, indeed resumming certain higher order contributions in the result.

The ensuing resummed expression was first suggested, for the pressure at zero chemical potentials, in~\cite{ConvergenceWeakCouplingExpansion} then later successfully applied in~\cite{MikkoYorkQuarkThresholds}. In the present case, the resulting expressions relevant to the finite density aspects of thermodynamics turn to exhibit drastic improvements of the convergence properties, as can be seen from the results presented in Chapter~\ref{chapter:Results}. This is also reflected in a substantial reduction of the renormalization scale dependence, ultimately improving the convergence properties. Finally, we would like to refer the reader to~\cite{sylvain2} for more details about the implementation of this resummation at finite chemical potentials, as well as to~\cite{KajantieEQCD,BraatenEQCD} for the original references on EQCD and~\cite{AleksiFirstPaperPressure,IppAllMuPressure,AleksiThesis,HartLainePhilipsen} for its generalization to finite density.

Next, we will investigate in more detail the hard-thermal-loop perturbation theory framework.

\section{Hard-thermal-loop perturbation theory}\label{section:HTLpt}

While the weak coupling expansion amounts to expanding around an ideal gas of massless excitations, hard-thermal-loop perturbation theory shifts the ground state to an ideal gas of massive thermal quasiparticles. This conceptual difference, clearly physically motivated, provides a net improvement in the convergence properties of the weak coupling expansion, even in certain simple resummations of the sub-leading infrared divergences.

The basic idea in HTLpt is to introduce an improvement term for the Lagrangian density in the form of monomial(s) in the fields that is added and subtracted from the original Lagrangian. This term, being of course physically motivated, usually takes the form of a mass term\footnote{It should be noted that in theories with a gauge symmetry, one cannot add and subtract a local mass term for the gauge fields without explicitly breaking gauge invariance. A way out is to make use of a non-local term instead.}. In the case of HTLpt, the so-called hard thermal loop effective action~\cite{FrenkelTaylorHTL,BraatenPisarskiHTL} plays the role of the non-local, thus momentum dependent, mass improvement term. The added piece is then treated with the non-interacting part of the Lagrangian, while the subtracted one is implemented along with the interaction terms of the theory, order by order in perturbation theory. This allows for interpolating between the original theory and a theory which has dressed propagators and vertices instead of the conventional ones. Provided it is worked out consistently to a given order in a the perturbative expansion, the correct original theory is recovered at the end. However, such a variationally improved perturbative framework allows for resumming important higher order contributions, as it amounts to reorganizing the perturbative series. We refer the reader to~\cite{RefOptKneurAndOthers,RefOptPeterForSPT} and references therein, for recent works on other variationally improved perturbation theories, relevant to non-perturbative quantities in QCD. As far as perturbative QCD is concerned, the approach leading to hard-thermal-loop perturbation theory was first applied at zero quark chemical potentials in~\cite{JensMikeBraatenFirstHTLpt,JensMikeBraatenFirstHTLptbis,JensMikeFirstTwoLoop}, then later in~\cite{NanHTLpt1,NanHTLpt2}. Recent developments at finite density, can be found in~\cite{sylvain1,sylvain2} as well as in~\cite{HTLptFiniteMUThreeLoop1,HTLptFiniteMUThreeLoop2}.

Concretely, the above shift is achieved by writing the Lagrangian density of QCD in the form
\beqa\label{HTLptLagrangianDensity}
{\cal L_{\rm HTLpt}}&=&\left({\cal L}_{\rm QCD}+{\cal L}_{\rm HTL}\right)\Big|_{g\rightarrow\sqrt{\delta}g}+\Delta{\cal L}_{\rm HTL}\,,
\eeqa
${\cal L}_{\rm QCD}$ being the Lagrangian of QCD that was worked out in~(\ref{EffectiveLagrangianQCD}). ${\cal L}_{\rm HTL}$ is our gauge invariant improvement term, obtained from the hard thermal loop effective action as explained above, and $\delta$ is a formal expansion parameter that we introduced for bookkeeping purposes. The latter is set to one after having Taylor expanded the path integral up to some order, before computing the corresponding Feynman diagrams, as in this limit ${\cal L_{\rm HTLpt}}$ reduces to the Lagrangian density of QCD. Finally, $\Delta{\cal L}_{\rm HTL}$ is a piece which contains all the needed counter-terms which are not necessarily present in perturbative QCD, as they cancel the ultraviolet divergences introduced by the subsequent reorganization. Note that, strictly speaking, the general form of the counter terms is not known, and we shall assume that they can be cast into a finite sum of monomials in the fields. We refer the reader to the end of Section~\ref{section:Renormalizing_Result} for more details on the question of renormalizability in HTLpt.

Regarding the gauge invariant improvement term, which is necessary to probe QCD with dynamical massless quarks, we consider the following highly non-local term
\beqa
{\cal L}_{\rm HTL}&=&-\frac{1}{2}(1-\delta)\,m_\trmi{D}^2\,{\rm Tr}\left(G_{\mu\alpha}\bigg\langle\frac{y^{\alpha}y^{\beta}}{(y\cdot D)^2}\bigg\rangle_{\!\!\!y}G^{\mu}_{\,\ \beta}\right) + (1-\delta)\,i\,\sum_f^{N_\trmi{f}} m^2_\trmi{q$_f$}\bar{\psi}_f\gamma^{\mu}\bigg\langle\frac{y_{\mu}}{y\cdot D}\bigg\rangle_{\!\!\!y}\psi_f, \ \ \ \ \ \ \ 
\eeqa
where $D^{\mu}=\partial^{\mu}-i gA^{\mu}$ denotes covariant derivatives in the adjoint and fundamental representations. Furthermore, $y\equiv(1,\hat{\bf y})$ is a light-like four-vector, and $\langle ... \rangle_y$ represents an average over the direction of $\hat{\bf y}$. Finally, $m_\trmi{D}$ and $m_\trmi{q$_f$}$ are the Debye and quark thermal mass parameters, which we shall set to their respective weak coupling values at leading order, before numerically evaluating a quantity.

Having defined our starting point for perturbative evaluations of physical quantities, we notice that the next step is to expand the partition function in a power series in $\delta$, before truncating at some order. Then $\delta$ is set to one and the corresponding sum-integrals are evaluated. At leading order, this amounts to having dressed propagators, which incorporate physical effects such as Debye screening and Landau damping, while beyond this, dressed vertices are also generated together with new higher order interaction terms. The latter ensure that there is no overcounting of Feynman diagrams. It should be stressed, as some confusion might appear, that a given loop order in HTLpt does not have to -- and in fact does not -- correspond to the same loop order in the weak coupling expansion. Indeed, the whole point of such a scheme is to reorganize the perturbative series. However, thanks to the higher order interaction terms\footnote{Sometimes called the HTL counter-terms; not to be confused with the ones coming from renormalization.}, the weak coupling result at a given order in $g$, is obtained within the HTLpt expansion, provided that one goes to a high enough loop order. For example, the leading order HTLpt result differs from the correct weak coupling expansion at the order $g^2$, while it does reproduce the correct $g^3$ plasmon contribution. This issue is consistently taken care of already at the next order in $\delta$, and beyond.

Moreover, we notice that if the expansion is truncated at a finite order in $\delta$, the evaluation of a quantity requires each of the mass parameters\footnote{Formally, provided the functional integrations commute with the limit $\delta\rightarrow 1$ that is taken in the action, no physical quantity should depend on the mass parameters when summing to all orders. However, the question of commutation is in practice not important, and the conventional perturbative series is recovered in the weak coupling limit.} $m_\trmi{D}$ and $m_\trmi{q$_f$}$ to be given a value. While in the case of the pressure beyond leading order, this can be achieved by means of a variational principle\footnote{That is to say by extremizing the pressure as a function of $m_\trmi{D}$ and $m_\trmi{q$_f$}$.}, at one-loop level such a procedure fails due to the absence of an explicit dependence on the coupling $g$ in the result~\cite{NanHTLpt2}. That is the reason why, as previously mentioned, we will assign these parameters their leading order weak coupling values
\beq\label{HTLptmDmqparameters}
m_\trmi{D}^2\equiv\frac{g^2}{3}\left[\left(N_\trmi{c}+\frac{N_\trmi{f}}{2}\right)\, T^2 +\frac{3}{2\pi^2}\sum_g \mu_g^2\right] \, ,\quad
m_\trmi{q$_f$}^2\equiv\frac{g^2}{16}\,\,\frac{N_\trmi{c}^2-1}{N_\trmi{c}}\,\Bigg(T^2+\frac{\mu_f^2}{\pi^2}\Bigg)\,,
\eeq
where we have kept the number of colors $N_\trmi{c}$ and flavors $N_\trmi{f}$ arbitrary. We notice that $m_\trmi{q$_f$}$ carries a dependence on the index f, running over $N_\trmi{f}$ quark flavors. The above will be, from now on, our prescription for evaluating thermodynamic quantities at leading order in HTLpt\footnote{This prescription being of course medium dependent, we refer the reader to Section~\ref{section:Thermodynamic_Consistency} for a discussion about thermodynamic consistency in general, and in particular within HTLpt.}.

Following the HTLpt framework to one-loop order, the pressure can be straightforwardly obtained in the form
\beq\label{OneLoopHTLpt}
p_\trmi{HTLpt}\left(T,\bm{\mu}\right) \equiv d_\trmi{A} \Big[(2-2\epsilon)\ p_\trmi{T}\left(T,\bm{\mu}\right) +\ p_\trmi{L}\left(T,\bm{\mu}\right) \Big] + N_\trmi{c} \sum_f\ p_\trmi{q$_f$}\left(T,\bm{\mu}\right) + \Delta p\,,
\eeq
where we recall that $d_\trmi{A}\equiv N_\trmi{c}^2-1$, and the various contributions, coming respectively from the transverse and longitudinal gluons as well as the quarks, read
{\allowdisplaybreaks
\beqa
p_\trmi{T}\left(T,\bm{\mu}\right)&=&-\frac{1}{2}\,\sumint_{K}\log\Big[K^2+\Pi_\trmi{T}(i\omega_n,k)\Big] \label{FullTransverseGluon}\,, \\
p_\trmi{L}\left(T,\bm{\mu}\right)&=&-\frac{1}{2}\,\sumint_{K}\log\Big[k^2+\Pi_\trmi{L}(i\omega_n,k)\Big] \label{FullLongitudinalGluon}\,, \\
p_\trmi{q$_f$}\left(T,\bm{\mu}\right)&=&2\,\sumint_{\{K\}}\log\Big[A_\trmi{S}^2(i\widetilde{\omega}_n+\mu_f,k)-A_\trmi{0}^2(i\widetilde{\omega}_n+\mu_f,k)\Big] \label{FullQuarks}\,,
\eeqa}
\hspace{-0.12cm}after algebraic manipulations. Notice that in~(\ref{OneLoopHTLpt}), $\Delta p$ stands for the contribution coming from the necessary counter-term, which cancel the ultraviolet divergences. Note also that the transverse gluon self-energy $\Pi_\trmi{T}$, the longitudinal gluon self-energy $\Pi_\trmi{L}$ and the functions $A_\trmi{S}$ and $A_\trmi{0}$ read
{\allowdisplaybreaks
\beqa
\Pi_\trmi{T}(i\omega_n,k) &\equiv& - \frac{m_\trmi{D}^2}{2-2\epsilon} \frac{\omega_n^2}{k^2} \bigg[1-\frac{\omega^2_n + k^2}{\omega_n^2} {\cal T}_\trmi{K}(i\omega_n,k) \bigg]\,, \\
\Pi_\trmi{L}(i\omega_n,k) &\equiv& m_\trmi{D}^2 \Big[1-{\cal T}_\trmi{K}(i\omega_n,k)\Big]\,, \\
A_\trmi{0}(i\widetilde{\omega}_n+\mu_f,k) &\equiv& i\widetilde{\omega}_n+\mu_f - \frac{m_\trmi{q$_f$}^2}{i\widetilde{\omega}_n+\mu_f}\  \widetilde{{\cal T}}_\trmi{K}(i\widetilde{\omega}_n+\mu_f,k)\,, \\
A_\trmi{S}(i\widetilde{\omega}_n+\mu_f,k) &\equiv& k+\frac{m_\trmi{q$_f$}^2}{k} \Big[1-\widetilde{{\cal T}}_\trmi{K}(i\widetilde{\omega}_n+\mu_f,k)\Big]\,,
\eeqa}
\hspace{-0.12cm}while in $3-2\epsilon$ spatial dimensions, the HTL functions ${\cal T}_{\rm K}$ and $\widetilde{{\cal T}}_{\rm K}$ can be written in terms of the Gauss hypergeometric function and come with the integral representations
\beqa
{\cal T}_\trmi{K}(i\omega_n,k) &\equiv& \frac{\Gamma\left(\frac{3}{2}-\epsilon\right)}{\Gamma\left(\frac{3}{2}\right)\Gamma\left(1-\epsilon\right)}\int^{1}_{0}\kern-0.5em\mathop{{\rm d}\!}\nolimits c \left(1-c^2\right)^{-\epsilon} \frac{(i\omega_n)^2}{(i\omega_n)^2-k^2 c^2} \nonumber \\
&=& {}_2F_1\left(\frac{1}{2},1;\frac{3}{2}-\epsilon;\frac{k^2}{(i\omega_n)^2}\right) \label{HTLfunctionB}\,, \\
\widetilde{{\cal T}}_\trmi{K}(i\widetilde{\omega}_n+\mu_f,k)&\equiv& \frac{\Gamma\left(\frac{3}{2}-\epsilon\right)}{\Gamma\left(\frac{3}{2}\right)\Gamma\left(1-\epsilon\right)}\int^{1}_{0}\kern-0.5em\mathop{{\rm d}\!}\nolimits c \left(1-c^2\right)^{-\epsilon} \frac{(i\widetilde{\omega}_n+\mu_f)^2}{(i\widetilde{\omega}_n+\mu_f)^2-k^2 c^2} \nonumber \\
&=& {}_2F_1\left(\frac{1}{2},1;\frac{3}{2}-\epsilon;\frac{k^2}{(i\widetilde{\omega}_n+\mu_f)^2}\right)\,. \label{HTLfunctionF}
\eeqa
The above representations in terms of hypergeometric functions will be very useful when computing the exact pressure via branch cut methods, which we will do in the next chapter.

\chapter{Exact leading order hard-thermal-loop perturbation theory}\label{chapter:Exact_LO_HTLpt}

In this chapter, we give details on the derivation of the exact leading order pressure, at finite temperature and chemical potentials, obtained within hard-thermal-loop perturbation theory. Technical aspects of this work are based upon~\cite{sylvain1,sylvain2}, but we also refer the reader to earlier works on the matter~\cite{JensMikeBraatenFirstHTLpt,JensMikeBraatenFirstHTLptbis}, as they are relevant for this chapter\footnote{Note that~\cite{sylvain2} incorporates a numerical refinement of two of the renormalization constants that were originally computed in~\cite{JensMikeBraatenFirstHTLpt,JensMikeBraatenFirstHTLptbis}. This turns out to improve the results to quite some extent.}.

In the following, we first detail the computation of various contributions to the leading order pressure, coming from the transverse and longitudinal gluons as well as from the quarks. Then, we explain how to extract the divergences in order to renormalize the result. Finally, we give the exact one-loop expression before to establish the mass truncation method that will be analyzed in the next chapter.

\section{Transverse gluon contribution to the pressure}\label{section:Transverse_Gluon_Contribution}

\begin{figure}[!t]
\begin{center}
\includegraphics[scale=0.16]{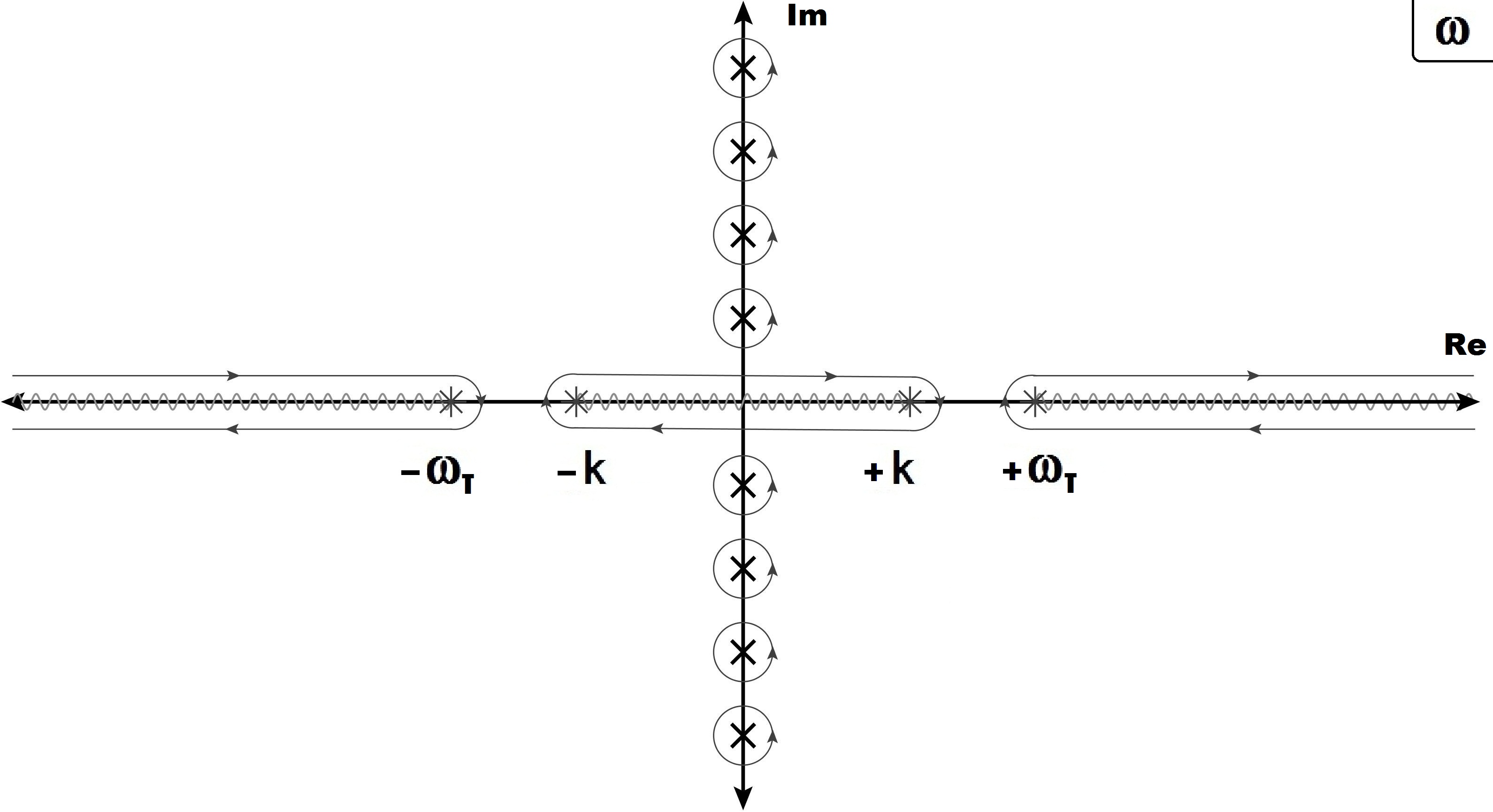}
\caption{The contours $C$ and $C_\trmi{T}$ for the evaluation of the transverse gluon contribution, together with the corresponding branch cuts. $C$ encloses all the points $\omega=i\omega_n$, except for $n=0$, while $C_\trmi{T}$ wraps around the branch cuts.}
\end{center}
\label{fig:ContourTransverseGluons}
\end{figure}

We first start from the transverse gluon part of the one-loop pressure~(\ref{FullTransverseGluon}) which reads
\beq
p_\trmi{T}\left(T,\bm{\mu}\right)=-\frac{1}{2}\,\sumint_{K}\log\Big[K^2+\Pi_\trmi{T}(i\omega_n,k)\Big]\, ,
\eeq
and rewrite it as
\beq\label{Eachterminsum}
p_\trmi{T}=-\frac{1}{2}\sumint_{K}\log(k^2)-\frac{1}{2}T\int_{\bf k}\sum_{n\neq 0}\log\left[\frac{k^2+\omega_n^2+\Pi_\trmi{T}(i\omega_n,k)}{k^2}\right]\,,
\eeq
using the fact that $\Pi_\trmi{T}(0,k)=0$ in order to drop the $n=0$ contribution to the term involving the self-energy function. The other term vanishes in dimensional regularization, since their is no scale in the integrand\footnote{See Section~\ref{section:Renormalization_Running} for more details on that point.}. We then use the well known contour trick, and represent the infinite sum over Matsubara frequencies by a series of residues. It allows us to turn the Matsubara sum into a contour integral, in the complex energy plane, by means of the residue theorem, and we get
{\allowdisplaybreaks
\beqa
p_\trmi{T}&=&-\frac{1}{2}T\,\int_{\bf k}\sum_{n\neq 0}\log\left[\frac{k^2+\omega_n^2+\Pi_\trmi{T}(i\omega_n,k)}{k^2}\right] \nonumber \\
&=&-\frac{1}{4}\,\int_{\bf k}\intec{C}{0.25em}{0em}\log\left[\frac{k^2-\omega^2+\Pi_\trmi{T}(\omega,k)}{k^2}\right]\,\coth\left(\frac{\beta\,\omega}{2}\right) \,,
\label{SumtoC}
\eeqa}
\hspace{-0.12cm}where the contour $C$ is shown in Figure~\ref{fig:ContourTransverseGluons}.

%%%%%%%%%%%%%%%%%%%%%%%%%%%%%%%%%%%%%%%%%%%%%%%%%%%%%
%%%%%%%%%%%%%%%%%%%%%%%%%%%%%%%%%%%%%%%%%%%%%%%%%%%%%

The above integrand has two types of branch cut. The first one originates from the logarithm, which leads to two cuts in the complex plane, running from $-\omega_\trmi{T}(k)$ and $+\omega_\trmi{T}(k)$ to (complex) infinity. Note that $\omega_\trmi{T}(k)$ is the quasiparticle dispersion relation in $3-2\epsilon$ spatial dimensions\footnote{It is a generalization of the transcendental equations that one encounters when working in three dimensions~\cite{KlimovDispersionRelations,WeldonDispersionRelations}.}, for a transverse gluon and therefore satisfies
\beq\label{GeneralTgluondispersionrelation}
k^2-\omega_\trmi{T}^2+\Pi_\trmi{T}(\omega_\trmi{T},k)=0 \,.
\eeq
The second one originates from the self-energy itself, via the Gauss hypergeometric function, and leads to two cuts in the complex plane, running from $-k$ and $+k$ to again (complex) infinity. For simplicity, we choose every finite branch point to be connected to a real infinity, via a straight line along the real axis. Thus, after eventual cancellations of overlapping cuts\footnote{provided that they originate from the same function, that is in the present case, the logarithm or the self-energy.}, we have the branch cuts as indicated in Figure~\ref{fig:ContourTransverseGluons}. This choice amounts to connecting $\pm\omega_\trmi{T}(k)$ to $\pm \infty$, and $\pm k$ to $+\infty$ or $-\infty$. The contour $C$ is then continuously deformed into a contour $C_\trmi{T}$ wrapping around the branch cuts, as can be seen from the Figure~\ref{fig:ContourTransverseGluons}, and we write
\beq\label{SumtoInteGluonT}
p_\trmi{T}=-\frac{1}{4}\,\int_{\bf k}\intec{C_\trmi{T}}{0.25em}{0em}\log\left[\frac{k^2-\omega^2+\Pi_\trmi{T}(\omega,k)}{k^2}\right]\,\coth\left(\frac{\beta\,\omega}{2}\right) \,.
\eeq

We therefore need to evaluate the above expression, and naturally opt for separating it into contributions coming from different type of branch cuts. The contribution coming from the branch cut running between $-k$ and $+k$ is then identified with the so-called Landau damping contribution to the pressure, as the self-energy $\Pi_\trmi{T}(\omega,k)$ gains an imaginary part along this cut. Collapsing the contour $C_\trmi{T}$ onto this branch cut, and using the notation $\disc$\!\! to denote the discontinuity across a branch cut, we write this contribution as
\beqa
p_{\trmi{T}_\trmi{Ld}}&\equiv&-\int_{\bf k} \! \inteddwbis \! \Bigg\{\!\disc\!\arctan\!\!\left[\frac{\frac{m_\trmi{D}^2}{2-2\epsilon}\frac{k^2-\omega^2}{k^2} \ \ImPart{\left\{_2F_1\left(\mbox{$\frac{1}{2},1;\frac{3}{2}-\epsilon;\frac{k^2}{\omega^2}$}\right)\right\}}}{k^2-\omega^2+\frac{m_\trmi{D}^2}{2-2\epsilon} \frac{\omega^2}{k^2} \left[ 1 + \frac{k^2-\omega^2}{\omega^2} \ \RePart{\left\{_2F_1\left(\mbox{$\frac{1}{2},1;\frac{3}{2}-\epsilon;\frac{k^2}{\omega^2}$}\right)\right\}}\right]}\right]\nonumber\\
&& \hspace*{8truecm} \times\frac{1}{2\,\pi}\,\bigg[\frac{1}{e^{\beta\omega}-1}+
\frac{1}{2}\bigg]\Bigg\} \label{GeneralTgluoncontributionaftercollapseLd}\,,
\eeqa
where we used the symmetry of the self-energy function for $\omega\rightarrow -\omega$, together with the usual relation between the Bose-Einstein distribution function and the cotangent one, as given by~(\ref{CothRelation}). The above arctangent will be referred to as the Landau damping angle of the transverse gluon contribution. Note that the discontinuity reads
{\allowdisplaybreaks
\beqa
\disc\phi_{\trmi{L},\epsilon}&\equiv&\disc\!\!\arctan\left[\frac{\frac{m_\trmi{D}^2}{2-2\epsilon}\frac{k^2-\omega^2}{k^2} \ \ImPart{\left\{_2F_1\left(\mbox{$\frac{1}{2},1;\frac{3}{2}-\epsilon;\frac{k^2}{\omega^2}$}\right)\right\}}}{k^2-\omega^2+\frac{m_\trmi{D}^2}{2-2\epsilon} \frac{\omega^2}{k^2} \left[ 1 + \frac{k^2-\omega^2}{\omega^2} \ \RePart{\left\{_2F_1\left(\mbox{$\frac{1}{2},1;\frac{3}{2}-\epsilon;\frac{k^2}{\omega^2}$}\right)\right\}}\right]}\right] \, \nonumber \\
&=&\arctan\left[\frac{\frac{m_\trmi{D}^2}{2-2\epsilon}\frac{k^2-\omega^2}{k^2} \ \ImPart{\left\{_2F_1^{\oplus}\left(\mbox{$\frac{1}{2},1;\frac{3}{2}-\epsilon;\frac{k^2}{\omega^2}$}\right)\right\}}}{k^2-\omega^2+\frac{m_\trmi{D}^2}{2-2\epsilon} \frac{\omega^2}{k^2} \left[ 1 + \frac{k^2-\omega^2}{\omega^2} \ \RePart{\left\{_2F_1^{\oplus}\left(\mbox{$\frac{1}{2},1;\frac{3}{2}-\epsilon;\frac{k^2}{\omega^2}$}\right)\right\}}\right]}\right] \, \nonumber \\
&-&\arctan\left[\frac{\frac{m_\trmi{D}^2}{2-2\epsilon}\frac{k^2-\omega^2}{k^2} \ \ImPart{\left\{_2F_1^{\ominus}\left(\mbox{$\frac{1}{2},1;\frac{3}{2}-\epsilon;\frac{k^2}{\omega^2}$}\right)\right\}}}{k^2-\omega^2+\frac{m_\trmi{D}^2}{2-2\epsilon} \frac{\omega^2}{k^2} \left[ 1 + \frac{k^2-\omega^2}{\omega^2} \ \RePart{\left\{_2F_1^{\ominus}\left(\mbox{$\frac{1}{2},1;\frac{3}{2}-\epsilon;\frac{k^2}{\omega^2}$}\right)\right\}}\right]}\right]\,,
\eeqa}
\hspace{-0.12cm}where, here and in the following, in order to shorten the expressions, we define
{\allowdisplaybreaks
\beqa
{}_2F_1^{\oplus}\left(\mbox{\small $\frac{1}{2},1;\frac{3}{2}-\epsilon;\frac{k^2}{\omega^2}$}\right)&\equiv& {}_2F_1\left(\mbox{\small $\frac{1}{2},1;\frac{3}{2}-\epsilon;\frac{k^2}{\omega^2}$}\right) \, , \label{DiscPlus} \\
{}_2F_1^{\ominus}\left(\mbox{\small $\frac{1}{2},1;\frac{3}{2}-\epsilon;\frac{k^2}{\omega^2}$}\right)&\equiv& e^{2i\pi\epsilon} \, {}_2F_1\left(\mbox{\small $\frac{1}{2},1;\frac{3}{2}-\epsilon;\frac{k^2}{\omega^2}$}\right) \nonumber \\
&+& \frac{2i\pi \, e^{i\pi\epsilon} \, \Gamma\left(\frac{3}{2}-\epsilon\right)}{\Gamma\left(1-\epsilon\right)\Gamma\left(\frac{1}{2}-\epsilon\right)\Gamma\left(1+\epsilon\right)}\,\,{}_2F_1\left(\mbox{\small $\frac{1}{2},1;1+\epsilon;\frac{\omega^2-k^2}{\omega^2}$}\right) \,. \label{DiscMinus}
\eeqa}
\hspace{-0.12cm}These are nothing but the limits of the Gauss hypergeometric function, when approaching a branch cut from above and from below.

Then, the contribution from the cuts connected to the dispersion relation is identified with the quasiparticle contribution to the pressure. Collapsing the contour $C_\trmi{T}$ onto these branch cuts, we get
\beqa
p_{\trmi{T}_\trmi{qp}}&\equiv&-\int_{\bf k}\,\,\bigg\{\frac{1}{2}\omega_\trmi{T}(k)+T\,\log\left(1-e^{-\beta\omega_\trmi{T}}\right)\bigg\} \,\label{GeneralTgluoncontributionaftercollapseqp} 
\eeqa
where we performed the integral over the frequency, taking advantage of the simple form for the discontinuity of the logarithm across a branch cut.

Using the above results, we write the total contribution from the transverse gluon part of the pressure as
\beq\label{GeneralTgluoncontributionaftercollapse}
p_\trmi{T}=p_{\trmi{T}_\trmi{qp}}+p_{\trmi{T}_\trmi{Ld}}\,.
\eeq

\section{Longitudinal gluon contribution to the pressure}\label{section:Longitudinal_Gluon_Contribution}

\begin{figure}[!t]
\begin{center}
\includegraphics[scale=0.17]{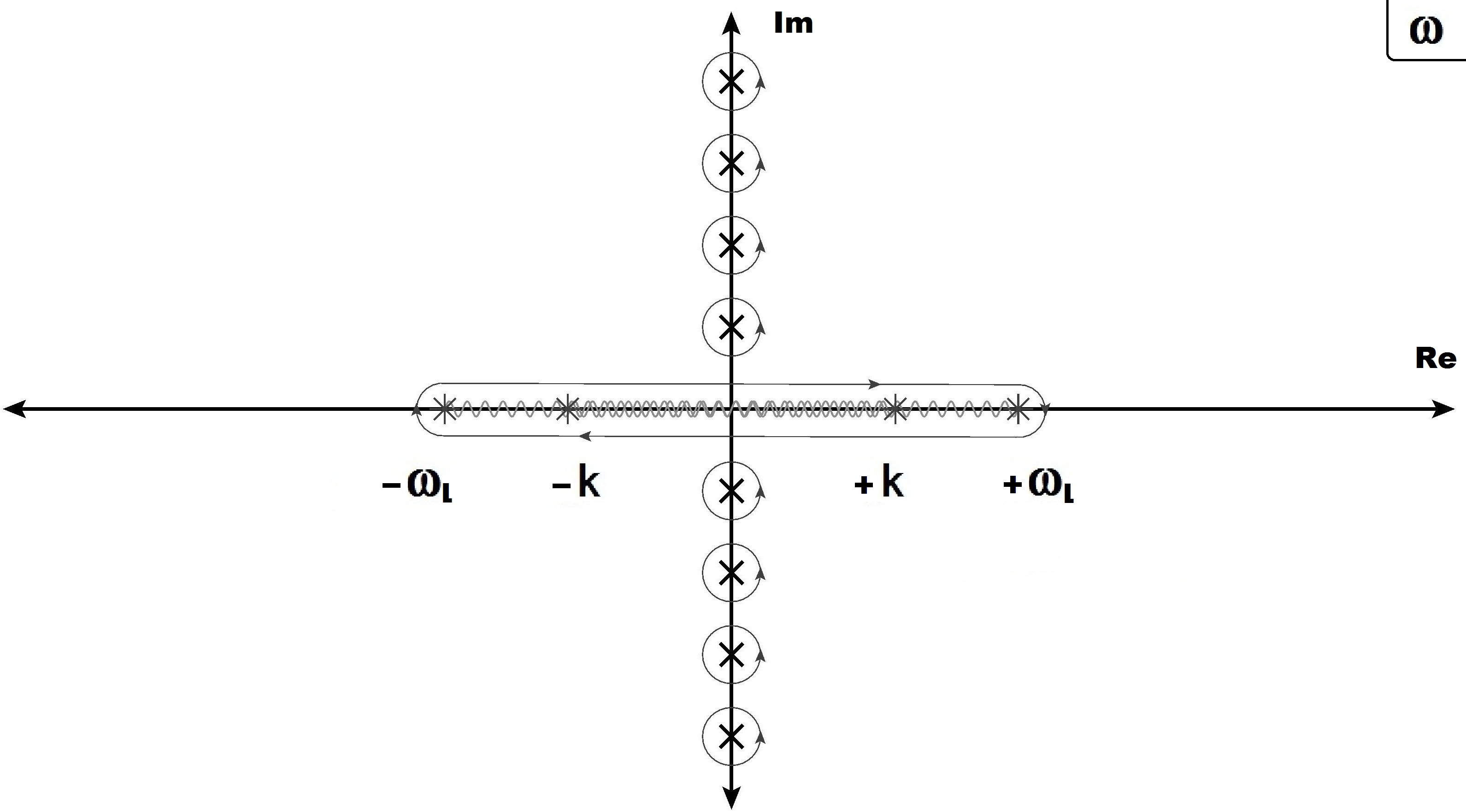}
\caption{The contours $C$ and $C_\trmi{L}$ for the evaluation of the longitudinal gluon contribution, together with the corresponding branch cuts. $C$ encloses all the points $\omega=i\omega_n$, except for $n=0$, while $C_\trmi{L}$ wraps around the branch cuts.}
\end{center}
\label{fig:ContourLongitudinalGluons}
\end{figure}

Next, we look at the longitudinal part of the one-loop pressure~(\ref{FullLongitudinalGluon})
\beq
p_\trmi{L}\left(T,\bm{\mu}\right)=-\frac{1}{2}\,\sumint_{K}\log\Big[k^2+\Pi_\trmi{L}(i\omega_n,k)\Big]\,
\eeq
that we rewrite as
\beq\label{PLongitudinal}
p_\trmi{L}=-\frac{1}{2}T\int_{\bf k}\,\sum_{n\neq 0}\log\left[k^2+\Pi_\trmi{L}(i\omega_n,k)\right] - \frac{1}{2}T\int_{\bf k}\,\log\left[k^2+m_\trmi{D}^2\right]\,,
\eeq
where we isolated the contribution from the $n=0$ mode. Then, using the same trick as for the transverse contribution, we rewrite these two terms as a single contour integral\footnote{Notice that the numerator and denominator of the argument of the logarithm originate from two separate but equal contour integrals with opposite winding numbers.} that encircles the points $\omega=i\omega_n$, except for $n=0$
\beqa\label{ContourInteGluonL}
p_\trmi{L}&=&-\frac{1}{4}\int_{\bf k}\,\intec{C}{0.25em}{0em}\log\left[\frac{k^2+\Pi_\trmi{L}(\omega,k)}{k^2+m_\trmi{D}^2}\right]\,\coth\left(\frac{\beta\,\omega}{2}\right)\,,
\eeqa
$C$ being given in Figure~\ref{fig:ContourLongitudinalGluons}.

%%%%%%%%%%%%%%%%%%%%%%%%%%%%%%%%%%%%%%%%%%%%%%%%%%%%%
%%%%%%%%%%%%%%%%%%%%%%%%%%%%%%%%%%%%%%%%%%%%%%%%%%%%%

Once again, the above integrand has branch points in $-\omega_\trmi{L}(k)$ and $+\omega_\trmi{L}(k)$ coming from the logarithm, and in $-k$ and $+k$ coming from the self-energy. Following this time a different choice\footnote{This choice is arbitrary, but in the present case makes the quasiparticle contribution finite at high momentum.} than the one we made for the transverse gluon contribution, we connect $\pm\omega_\trmi{L}(k)$ to $+\infty$ or $-\infty$, and $\pm k$ to $+\infty$ or $-\infty$, leading to the branch cuts as defined in Figure~\ref{fig:ContourLongitudinalGluons}. Note that $\omega_\trmi{L}$ is the dispersion relation for a longitudinal gluon in $3-2\epsilon$ spatial dimensions, and satisfies
\beq\label{GeneralLgluondispersionrelation}
0\equiv k^2+\Pi_\trmi{L}(\omega_\trmi{L},k)\,.
\eeq

We then continuously deform the contour $C$ into $C_\trmi{L}$, as shown in Figure~\ref{fig:ContourLongitudinalGluons}. In analogy with the transverse gluon contribution, we write two different contributions, that is to say the quasiparticle and the Landau damping parts, and obtain respectively
\beqa
p_{\trmi{L}_\trmi{qp}}&=&-\int_{\bf k}\,\,\bigg\{\frac{1}{2}\left(\omega_\trmi{L}(k)-k\right)+T\,\log\left(\frac{1-e^{-\beta\omega_\trmi{L}}}{1-e^{-\beta k}}\right)\bigg\} \,,\label{GeneralLgluoncontributionaftercollapseqp}
\eeqa
and
{\allowdisplaybreaks
\beqa
p_{\trmi{L}_\trmi{Ld}}&=&\int_{\bf k}\,\inteddwbis\left\{\disc\arctan\left[\frac{m_\trmi{D}^2 \ \ImPart{\left\{_2F_1\left(\mbox{$\frac{1}{2},1;\frac{3}{2}-\epsilon;\frac{k^2}{\omega^2}$}\right)\right\}}}{k^2 + m_\trmi{D}^2 - m_\trmi{D}^2 \ \RePart{\left\{_2F_1\left(\mbox{$\frac{1}{2},1;\frac{3}{2}-\epsilon;\frac{k^2}{\omega^2}$}\right)\right\}}}\right]\right.\nonumber\\
&& \hspace*{8truecm} \,\times\left.\frac{1}{2\,\pi}\bigg[\frac{1}{e^{\beta\omega}-1}+\frac{1}{2}\bigg]\Bigg.\Bigg.\setlength{\delimitershortfall}{-5pt}\right\} \label{GeneralLgluoncontributionaftercollapseLd}\, . \ \ \ \ \ \ \ \ \ \ 
\eeqa}
\hspace{-0.12cm}Note that here, the discontinuity of the Landau damping angle reads
{\allowdisplaybreaks
\beqa
\disc\phi_{\trmi{L},\epsilon}&\equiv&\disc\arctan\left[\frac{m_\trmi{D}^2 \ \ImPart{\left\{_2F_1\left(\mbox{$\frac{1}{2},1;\frac{3}{2}-\epsilon;\frac{k^2}{\omega^2}$}\right)\right\}}}{k^2 + m_\trmi{D}^2 - m_\trmi{D}^2 \ \RePart{\left\{_2F_1\left(\mbox{$\frac{1}{2},1;\frac{3}{2}-\epsilon;\frac{k^2}{\omega^2}$}\right)\right\}}}\right] \, \nonumber \\
&=&\arctan\left[\frac{m_\trmi{D}^2 \ \ImPart{\left\{_2F_1^{\oplus}\left(\mbox{$\frac{1}{2},1;\frac{3}{2}-\epsilon;\frac{k^2}{\omega^2}$}\right)\right\}}}{k^2 + m_\trmi{D}^2 - m_\trmi{D}^2 \ \RePart{\left\{_2F_1^{\oplus}\left(\mbox{$\frac{1}{2},1;\frac{3}{2}-\epsilon;\frac{k^2}{\omega^2}$}\right)\right\}}}\right] \, \nonumber \\
&-&\arctan\left[\frac{m_\trmi{D}^2 \ \ImPart{\left\{_2F_1^{\ominus}\left(\mbox{$\frac{1}{2},1;\frac{3}{2}-\epsilon;\frac{k^2}{\omega^2}$}\right)\right\}}}{k^2 + m_\trmi{D}^2 - m_\trmi{D}^2 \ \RePart{\left\{_2F_1^{\ominus}\left(\mbox{$\frac{1}{2},1;\frac{3}{2}-\epsilon;\frac{k^2}{\omega^2}$}\right)\right\}}}\right] \,,
\eeqa}
\hspace{-0.12cm}and the quasiparticle part has been handled in the same way as for the transverse gluon contribution.

We finally arrive to the longitudinal contribution to the pressure
\beq\label{GeneralLgluoncontributionaftercollapse}
p_\trmi{L}\equiv p_{\trmi{L}_\trmi{qp}}+p_{\trmi{L}_\trmi{Ld}} \,,
\eeq
given the aforementioned expressions.

\section{Quark contribution to the pressure}\label{section:Quark_Contribution}

Although the quark contribution differs from the gluonic ones, at least by finite chemical potentials $\mu_f$, its derivation via branch cuts does not change much in substance. Indeed, the shift brought by the frequencies, due to the presence of nonvanishing chemical potentials, does not influence the collapse of the contour onto the branch cuts which are located on the real axis. This is nothing but a direct consequence of the fact that the presence of a dense and hot medium does not affect the ultraviolet structure of the theory\footnote{At least, as far as we can say here, in the context of leading order HTLpt.}.

\begin{figure}[!t]
\begin{center}
\includegraphics[scale=0.17]{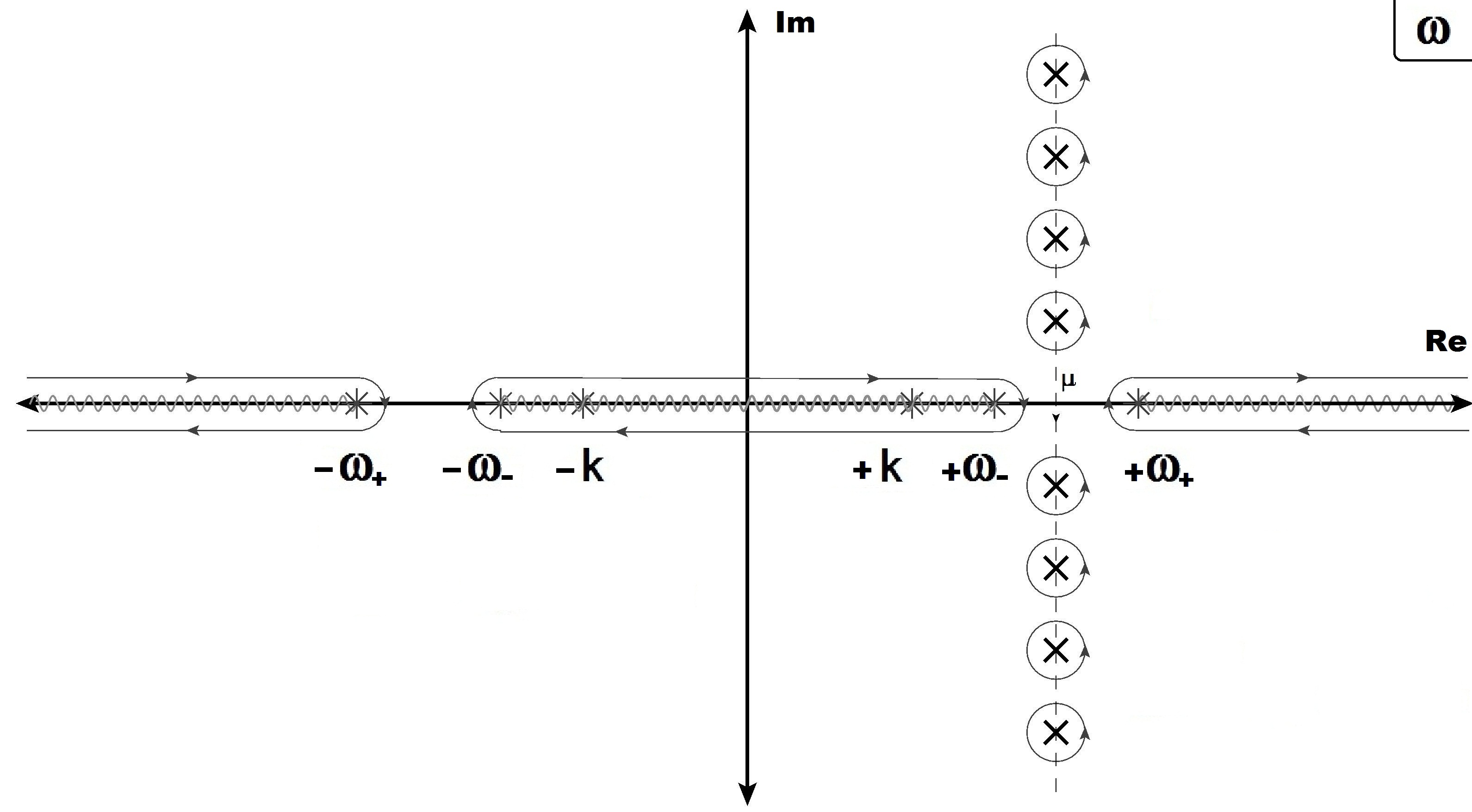}
\caption{The contours $C$ and $C_\trmi{q}$ for the evaluation of the quark contribution, together with the corresponding branch cuts. $C$ encloses all the points $\omega=i\widetilde{\omega}_n$, while $C_\trmi{q}$ wraps around the branch cuts.}
\end{center}
\label{fig:ContourQuarks}
\end{figure}

Let us finally consider the quark contribution to the one-loop pressure~(\ref{FullQuarks})
\beq
p_\trmi{q$_f$}\left(T,\bm{\mu}\right)=2\,\sumint_{\{K\}}\log\Big[A_\trmi{S}^2(i\widetilde{\omega}_n+\mu_f,k)-A_\trmi{0}^2(i\widetilde{\omega}_n+\mu_f,k)\Big]\,,
\eeq
the index $f$ running over the $N_\trmi{f}$ flavors. Using again the contour trick, we get
{\allowdisplaybreaks
\beqa
p_\trmi{q$_f$}&=&2\,\sumint_{\{K\}}\log\Big[k^2+(\widetilde{\omega}_n-i\mu_f)^2\Big] \nonumber \\
&& \ \ \ \ \ +2\int_{\bf k}\,T\,\sum_{n}\log\left[\frac{A_\trmi{S}^2(i\widetilde{\omega}_n+\mu_f,k)-A_\trmi{0}^2(i\widetilde{\omega}_n+\mu_f,k)}{k^2-(i\widetilde{\omega}_n+\mu_f)^2}\right]\; \nonumber \\
&=&\frac{\pi^2 T^4}{45}\left(\frac{7}{4}+30\hat \mu_f^2+60\hat \mu_f^4\right) \nonumber \\
&& \ \ \ \ \ +\int_{\bf k}\,\intec{C}{0.25em}{0em}\log\left[\frac{A_\trmi{S}^2(\omega,k)-A_\trmi{0}^2(\omega,k)}{k^2-\omega^2}\right]\,\tanh\left(\frac{\beta\,(\omega-\mu_f)}{2}\right) \, , \label{ContourInteQuark}
\eeqa}
\hspace{-0.12cm}only this time, we wrote explicitly the Stefan-Boltzmann quark contribution, as a matter of choice, to make the argument dimensionless. Note that the contour $C$ can be seen from Figure~\ref{fig:ContourQuarks}.

%%%%%%%%%%%%%%%%%%%%%%%%%%%%%%%%%%%%%%%%%%%%%%%%%%%%%
%%%%%%%%%%%%%%%%%%%%%%%%%%%%%%%%%%%%%%%%%%%%%%%%%%%%%

The above integrand has branch points in $\pm\omega_\trmi{\tiny $f_\pm$}(k)$ coming from the logarithm, which satisfy the quasiparticle dispersion relations for quarks and plasminos
\beq\label{GeneralQuarksdispersionrelation}
0\equiv A_\trmi{0}(\omega_\trmi{\tiny $f_\pm$},k)\mp A_\trmi{S}(\omega_\trmi{\tiny $f_\pm$},k)\,.
\eeq
The integrand also has branch points at $\pm k$, coming from the self-energy. We choose to connect $\pm\omega_\trmi{\tiny $f_+$}(k)$ to $\pm\infty$, $\pm\omega_\trmi{\tiny $f_-$}(k)$ to $+\infty$ or $-\infty$, and $\pm k$ to $+\infty$ or $-\infty$, ending up with branch cuts as in Figure~\ref{fig:ContourQuarks}. Finally, we continuously deform the contour $C$ to $C_\trmi{q}$, which is wrapping around the cuts. Collapsing the latter, we get as already seen in the previous cases, two different contributions, the quasiparticle and Landau damping ones
{\allowdisplaybreaks
\beqa
p_{\trmi{q}_{\trmi{qp}_f}}&=&2\int_{\bf k}\,\,\bigg\{T\,\log\left(1+e^{-\beta\left(\omega_{f_+}+\mu_f\right)}\right)+T\,\log\left(1+e^{-\beta\left(\omega_{f_+}-\mu_f\right)}\right) \nonumber \\
& & \ \ \ \ \ \ \ \ \ +T\,\log\left(\frac{1+e^{-\beta\left(\omega_{f_-}+\mu_f\right)}}{1+e^{-\beta (k+\mu_f)}}\right)+T\,\log\left(\frac{1+e^{-\beta\left(\omega_{f_-}-\mu_f\right)}}{1+e^{-\beta (k-\mu_f)}}\right)\bigg\} \nonumber \\
&+&2\int_{\bf k}\,\,\bigg\{\omega_{f_+}(k)+\left(\omega_{f_-}(k)-k\right)\bigg\} \,,
\label{GeneralQuarkcontributionaftercollapseqp}
\eeqa}
\hspace{-0.12cm}and
\beqa
p_{\trmi{q}_{\trmi{Ld}_f}}&=&\frac{1}{\pi}\int_{\bf k}\,\inteddwbis\left\{\disc\theta_\trmi{q$_f$,$\epsilon$}\,\bigg[\frac{1}{e^{\beta(\omega+\mu_f)}+1}+\frac{1}{e^{\beta(\omega-\mu_f)}+1}-1\bigg]\right\}\label{GeneralQuarkcontributionaftercollapseLd}\,.
\eeqa
Note that here, the Landau damping angle reads $\theta_\trmi{q$_f$,$\epsilon$}\equiv\arctan\left[\Xi_\trmi{$f$,$\epsilon$}\right]$, with the argument
\beqa\label{Xi}
{}&&\Xi_\trmi{$f$,$\epsilon$}\equiv \\
{}&&\frac{\frac{m_\trmi{q$_f$}^4}{k^2} \left[2 \ \ImPart{\left\{_2F_1\left(\mbox{$\frac{1}{2},1;\frac{3}{2}-\epsilon;\frac{k^2}{\omega^2}$}\right)\right\}}+\frac{k^2-\omega^2}{\omega^2} \ \ImPart{\left\{_2F_1\left(\mbox{$\frac{1}{2},1;\frac{3}{2}-\epsilon;\frac{k^2}{\omega^2}$}\right)^2\right\}}\right]}{k^2-\omega^2+2 m_\trmi{q$_f$}^2+\frac{m_\trmi{q$_f$}^4}{k^2}\,\left[1-2\,\RePart{\left\{_2F_1\left(\mbox{$\frac{1}{2},1;\frac{3}{2}-\epsilon;\frac{k^2}{\omega^2}$}\right)\right\}}-\frac{k^2-\omega^2}{\omega^2}\,\RePart{\left\{_2F_1\left(\mbox{$\frac{1}{2},1;\frac{3}{2}-\epsilon;\frac{k^2}{\omega^2}$}\right)^2\right\}}\right]}\nonumber\,,
\eeqa
and has the following discontinuity
\beqa
\disc\theta_\trmi{q$_f$,$\epsilon$}&\equiv&\disc\arctan\left[\Xi_\trmi{$f$,$\epsilon$}\right]=\arctan\left[\Xi^{\oplus}_\trmi{$f$,$\epsilon$}\right]-\arctan\left[\Xi^{\ominus}_\trmi{$f$,$\epsilon$}\right] \,,
\eeqa
where $\Xi^{\oplus/\ominus}_\trmi{$f$,$\epsilon$}$ is given by
{\allowdisplaybreaks
\beqa
&&\Xi^{\oplus/\ominus}_\trmi{$f$,$\epsilon$}\equiv \\
&&\frac{\frac{m_\trmi{q$_f$}^4}{k^2} \left[2 \ \ImPart{\left\{_2F_1^{\oplus/\ominus}\left(\mbox{$\frac{1}{2},1;\frac{3}{2}-\epsilon;\frac{k^2}{\omega^2}$}\right)\right\}}+\frac{k^2-\omega^2}{\omega^2} \ \ImPart{\left\{_2F_1^{\oplus/\ominus}\left(\mbox{$\frac{1}{2},1;\frac{3}{2}-\epsilon;\frac{k^2}{\omega^2}$}\right)^2\right\}}\right]}{k^2-\omega^2+2 m_\trmi{q$_f$}^2+\frac{m_\trmi{q$_f$}^4}{k^2}\,\left[1-2\,\RePart{\left\{_2F_1^{\oplus/\ominus}\left(\mbox{$\frac{1}{2},1;\frac{3}{2}-\epsilon;\frac{k^2}{\omega^2}$}\right)\right\}}-\frac{k^2-\omega^2}{\omega^2}\,\RePart{\left\{_2F_1^{\oplus/\ominus}\left(\mbox{$\frac{1}{2},1;\frac{3}{2}-\epsilon;\frac{k^2}{\omega^2}$}\right)^2\right\}}\right]}\nonumber\,.
\eeqa}

Finally, we can write the total quark contribution to the pressure
\beq\label{GeneralQuarkcontributionaftercollapse}
p_\trmi{q$_f$}\equiv 
p_{\trmi{q}_{\trmi{qp}_f}}+p_{\trmi{q}_{\trmi{Ld}_f}} \,,
\eeq
considering the above expressions.

\section{Isolating the divergences and renormalizing the result}\label{section:Renormalizing_Result}

We choose to implement the renormalization procedure within the modified minimal subtraction scheme $\MSbar$, as explained in detail in Section~\ref{section:Renormalization_Running}. This first implies isolating the potentially ultraviolet divergent terms, in order to be able to analytically compute the coefficients of the negative powers of their Laurent expansion around $\epsilon=0$. Those will be then canceled exactly, with the help of appropriate counter-terms.

We therefore take a closer look at the expressions involved in $p_\trmi{T}$, $p_\trmi{L}$, and $p_\trmi{q$_f$}$, namely~(\ref{GeneralTgluoncontributionaftercollapseLd}), (\ref{GeneralTgluoncontributionaftercollapseqp}), (\ref{GeneralLgluoncontributionaftercollapseqp}), (\ref{GeneralLgluoncontributionaftercollapseLd}), (\ref{GeneralQuarkcontributionaftercollapseqp}), and~(\ref{GeneralQuarkcontributionaftercollapseLd}). We see that most of them are ultraviolet divergent\footnote{Note that the second term in~(\ref{GeneralLgluoncontributionaftercollapseqp}) is exponentially damped at high momentum, and that the term involving the dispersion relation $\omega_\trmi{L}(k)$ comes in the combination $\omega_\trmi{L}(k)-k$. The latter being finite reflects the fact that the high momentum behavior of the longitudinal gluon is, at leading order, similar to that of a free relativistic particle~\cite{KlimovDispersionRelations,WeldonDispersionRelations}.}, i.e. carry a divergent high-$k$ behavior, when $\epsilon\rightarrow 0$. This high momentum behavior taking the form of poles in $\epsilon$, the usual trick is to subtract a simpler integral having exactly the same pole structure in order to make it finite. Of course, this has to be done while allowing for the subtracted integral to be computable analytically (at least for the poles in $\epsilon$). Then the very same integral is added back so that the total expression remains unchanged. Finally, the added part is computed analytically and the poles in $\epsilon$ removed via renormalization. However, given the complexity of the integrands involved in our expressions\footnote{Recalling that, for example, the arctangent functions which involve Gauss hypergeometric functions in their arguments, have to be summed over infinite sets of Matsubara frequencies.}, this method is likely to lead to very difficult intermediate computations, even within the first few orders of Laurent expansions. We therefore aim at using a new and slightly different method, which turns out to rely on a very simple classification of the divergent terms, making the problem much easier to solve. Indeed, the divergences can be cast in two categories, one which is explicitly dependent of the temperature and chemical potentials, and another which is explicitly independent of them\footnote{Those can be medium dependent via the choice of mass parameters $m_\trmi{D}$ and $m_\trmi{q$_f$}$. However, formally, such a medium dependence has a different origin than the explicit one, and physical results should be independent of the choice made for the parameters, when summing the perturbative series to all orders. This, in turn, just defines the ground state around which one is expanding, and we thus implement the renormalization procedure before fixing the thermal masses.}. Then, further recalling that in the zero chemical potential and temperature limit the discrete Matsubara summations turn to integrals over continuous Euclidean energies\footnote{Those benefit of symmetry properties allowing for usual rescaling, which significantly simplify the computations.}, our strategy is as follows.

We only add and subtract integrals for the potentially divergent terms which are explicitly medium dependent. We calculate the corresponding added integrals, within Laurent expansions around $\epsilon=0$, and show that accounting for the proper degrees of freedom in $3-2\epsilon$ spatial dimensions, their poles turn out to cancel each other\footnote{Had they not canceled, we would be left with medium dependent poles, to be removed via renormalization. Recalling the shift of the ground state to thermal massive quasiparticles, this would not be surprising. The important point here, is that the complicated expressions are contained in contributions which are explicitly medium independent. We then significantly simplify their computation by setting $T$ and ${\boldsymbol \mu}$ to zero.}. We then directly compute, within Laurent expansions, the contributions which are explicitly medium independent by setting both the temperature and the chemical potentials to zero. This allows for a substantial simplification of the computations, followed by a removal of the poles in $\epsilon$ by means of the renormalization procedure.

We now start by reviewing the explicitly temperature and chemical potential dependent potentially divergent expressions. Following~\cite{sylvain2}, we see that we only have two such contributions, the longitudinal and the transverse Landau damping contributions~(\ref{GeneralLgluoncontributionaftercollapseLd}) and~(\ref{GeneralTgluoncontributionaftercollapseLd}), both proportional to the Bose-Einstein distribution function from which the medium dependence originates,
{\allowdisplaybreaks
\beqa
\!\!\!\!\!\!\frac{1}{2\,\pi}&&\!\!\!\!\!\!\!\!\!\!\!\int_{\bf k}\,\!\!\inteddwbis\!\!\left\{\disc\!\arctan\!\left[\frac{m_\trmi{D}^2 \ \ImPart{\left\{_2F_1\left(\mbox{$\frac{1}{2},1;\frac{3}{2}-\epsilon;\frac{k^2}{\omega^2}$}\right)\right\}}}{k^2 + m_\trmi{D}^2 - m_\trmi{D}^2 \ \RePart{\left\{_2F_1\left(\mbox{$\frac{1}{2},1;\frac{3}{2}-\epsilon;\frac{k^2}{\omega^2}$}\right)\right\}}}\right]\,\,\!\!\!\frac{1}{e^{\beta\omega}-1}\!\Bigg.\Bigg.\setlength{\delimitershortfall}{-5pt}\right\}\!,\label{TmuDependentTermDivLongitudinal}\ \ \ \ \ \ \ \ \ \ \ \ \ \ \ \ \ \ \\
\!\!\!\!\!\!\frac{-1}{2\,\pi}&&\!\!\!\!\!\!\!\!\!\!\!\int_{\bf k}\,\!\!\inteddwbis\!\!\left\{\disc\!\arctan\!\left[\frac{\frac{m_\trmi{D}^2}{2-2\epsilon}\frac{k^2-\omega^2}{k^2} \ \ImPart{\left\{_2F_1\left(\mbox{$\frac{1}{2},1;\frac{3}{2}-\epsilon;\frac{k^2}{\omega^2}$}\right)\right\}}}{k^2-\omega^2+\frac{m_\trmi{D}^2}{2-2\epsilon} \frac{\omega^2}{k^2}\left[ 1 + \frac{k^2-\omega^2}{\omega^2} \ \RePart{\left\{_2F_1
\left(\mbox{$\frac{1}{2},1;\frac{3}{2}-\epsilon;\frac{k^2}{\omega^2}$}\right)\right\}}\right]}\right]\!\frac{1}{e^{\beta\omega}-1}\!\setlength{\delimitershortfall}{-5pt}\right\}\!.\label{TmuDependentTermDivTransverse}\ \ \ \ \ \ 
\eeqa}
\hspace{-0.12cm}In order to define the  integrals to be added and subtracted, we need to Taylor expand around $k=\infty$ and with $\omega$ fixed, the discontinuity of the imaginary part of the Gauss hypergeometric function, giving
{\allowdisplaybreaks
\beqa\label{DiscIm2F1TaylorExpand}
\disc \ImPart{\left\{_2F_1\left(\mbox{$\frac{1}{2},1;\frac{3}{2}-\epsilon;\frac{k^2}{\omega^2}$}\right)\right\}} &\underset{\underset{\mbox{\tiny $\omega$ fixed}}{k\longrightarrow \infty}}{=}& -\,4\,\, \frac{\Gamma\left(\frac{3}{2}-\epsilon\right)\,\Gamma\left(\frac{3}{2}\right)}{\Gamma\left(1-\epsilon\right)}\,\frac{\omega}{k}+{\cal O}\left(\frac{\omega^2}{k^2}\right)\,.
\eeqa}
\hspace{-0.12cm}The above formula can be easily obtained by noticing that the discontinuity and the imaginary part commute with each other. Then, using the definition of the discontinuity together with~(\ref{DiscPlus}) and~(\ref{DiscMinus}), the result is obtained by simply expanding around $k=\infty$, with $\omega$ fixed. It is then easy to obtain added integrals for~(\ref{TmuDependentTermDivLongitudinal}) and~(\ref{TmuDependentTermDivTransverse}), and we have respectively\footnote{While the negative powers of $\epsilon$, in the Laurent expansion of the integrals, must be equal to those of the potentially divergent terms, the higher orders can differ since the subtracted integrals will compensate for them.}
{\allowdisplaybreaks
\beqa
\mathcal{D}_{\trmi{L}_\trmi{Ld}}&\equiv& -\frac{2\,m_\trmi{D}^2}{\pi}\,\,\frac{\Gamma\left(\frac{3}{2}-\epsilon\right)\Gamma\left(\frac{3}{2}\right)}{\Gamma\left(1-\epsilon\right)}\int_{\bf k}\,\inteddwbis\left\{\frac{\omega}{k^3}\,\,\frac{1}{e^{\beta\omega}-1}\Bigg.\Bigg.\setlength{\delimitershortfall}{-5pt}\right\}\,,\label{InteSubtraL} \\
\mathcal{D}_{\trmi{T}_\trmi{Ld}}&\equiv& \frac{m_\trmi{D}^2}{\pi}\,\,\frac{\Gamma\left(\frac{3}{2}-\epsilon\right)\Gamma\left(\frac{3}{2}\right)}{(1-\epsilon)\,\Gamma\left(1-\epsilon\right)}\int_{\bf k}\,\inteddwbis\left\{\frac{\omega}{k^3}\,\,\frac{1}{e^{\beta\omega}-1}\Bigg.\Bigg.\setlength{\delimitershortfall}{-5pt}\right\}\,.\label{InteSubtraT}
\eeqa}
\hspace{-0.12cm}The above integrals can be easily evaluated, and the results through ${\cal O}\left(\epsilon^{0}\right)$ read
{\allowdisplaybreaks
\beqa
\mathcal{D}_{\trmi{L}_\trmi{Ld}}&=& \frac{m_\trmi{D}^2\,T^2}{48}\,\left[\frac{1}{\epsilon}+2\log\left(\frac{\Lambda}{2\pi T}\right)+2\,\,\frac{\zeta^{\prime}\left(-1\right)}{\zeta\left(-1\right)}+{\cal O}\left(\epsilon\right)\right]\,,\label{TmuDependentSubtractedInteLngitudinal}\\
\mathcal{D}_{\trmi{T}_\trmi{Ld}}&=& -\frac{m_\trmi{D}^2\,T^2}{96}\,\left[\frac{1}{\epsilon}+1+2\log\left(\frac{\Lambda}{2\pi\,T}\right)+2\,\,\frac{\zeta^{\prime}\left(-1\right)}{\zeta\left(-1\right)}+{\cal O}\left(\epsilon\right)\right]\,.\label{TmuDependentSubtractedInteTransverse}
\eeqa}
\hspace{-0.12cm}We then notice that their poles (and finite parts) cancel in the combination\footnote{Looking at~(\ref{InteSubtraL}) and~(\ref{InteSubtraT}), it is easy to see that in this combination, they even cancel at every order in $\epsilon$.}
\beq
(2-2\epsilon)\ \mathcal{D}_{\trmi{T}_\trmi{Ld}} + \mathcal{D}_{\trmi{L}_\trmi{Ld}}=0+{\cal O}\left(\epsilon\right)\,,
\eeq
which accounts for the proper gluon degrees of freedom in $3-2\epsilon$ spatial dimensions, as given by the sum of transverse and longitudinal contributions to the pressure in~(\ref{OneLoopHTLpt}). Therefore, we conclude that there is no need for making use of subtracting integrals with the medium dependent pieces in the present case. Notice that the cancellation of the above expressions at ${\cal O}\left(\epsilon^{0}\right)$ (and beyond) is merely a consequence of the choice that we made for the corresponding integrals. Another choice would have made a finite contribution survive, which would be canceled by the subtracted terms. Consequently, our choice makes the renormalization quite straightforward, and we are now left to deal only with explicitly medium independent expressions.

In order to proceed, let us gather all the contributions and write the full expression for the pressure~(\ref{OneLoopHTLpt}), accounting for the zero temperature and chemical potential limits. Thus, we have
{\allowdisplaybreaks
\beqa\label{PreResultExactHTLpt}
p_\trmi{HTLpt}\left(T,\bm{\mu}\right)&=& d_\trmi{A}\Bigg\{-(2-2\epsilon)\,T\int_{\bf k}\log\bigg(1-e^{-\beta\omega_\trmi{T}}\bigg)-\,T\int_{\bf k}\log\bigg(\frac{1-e^{-\beta\omega_\trmi{L}}}{1-e^{-\beta\,k}}\bigg) + \ p^{\star}_\trmi{L}\nonumber\\
&&\ \ \ \ +\ (2-2\epsilon)\, p^{\star}_\trmi{T} -\frac{1}{2\,\pi}\int_{\bf k}\,\inteddwbis \bigg[(2-2\epsilon)\,\disc\phi_{\trmi{T},\epsilon}-\disc\phi_{\trmi{L},\epsilon}\bigg]\,\,\frac{1}{e^{\beta\omega}-1}\Bigg\} \nonumber\\
&+&N_\trmi{c}\sum_{f,\,s=\pm 1}\Bigg\{2\,T\int_{\bf k}\log\bigg[1+e^{-\beta\left(\omega_{f_+}+s\,\mu_f\right)}\bigg]+2\,T\int_{\bf k}\log\left[\frac{1+e^{-\beta\left(\omega_{f_-}+s\,\mu_f\right)}}{1-e^{-\beta\left(k+s\,\mu_f\right)}}\right] \nonumber\\
&&\ \ \ \ \ \ \ \ \ + \ \frac{p^{\star}_\trmi{q$_f$}}{2} + \frac{1}{\pi}\,\int_{\bf k}\,\inteddwbis\,\,\disc\theta_\trmi{q$_f$,$\epsilon$}\left[\frac{1}{{e^{\beta\left(\omega+s\,\mu_f\right)}+1}}\right]\Bigg\} + \Delta p\;,
\eeqa}
\hspace{-0.12cm}where the subscript $\epsilon$ in $\phi_{\trmi{T},\epsilon}$, $\phi_{\trmi{L},\epsilon}$ and $\theta_\trmi{q$_f$,$\epsilon$}$ serves to remind that these angles have to be considered in $d=3-2\epsilon$ dimensions\footnote{Notice that the same applies to the dispersion relations.}, as we did not yet renormalize the result, hence take the limit $\epsilon\rightarrow 0$. In the above, the terms $p^{\star}_\trmi{T}$, $p^{\star}_\trmi{L}$, and $p^{\star}_\trmi{q$_f$}$ are respectively coming from $p_\trmi{T}$, $p_\trmi{L}$, and $p_\trmi{q$_f$}$, by taking the zero temperature and chemical potential limits. They contain all the potentially ultraviolet divergent pieces, and we are now going to compute them.

Switching off the explicit temperature and chemical potentials by taking the appropriate limits, the transverse gluon contribution $p_\trmi{T}$ approaches the expression $p^{\star}_\trmi{T}$, which is given by an integral over continuous Euclidean energy $\omega_\trmi{E}$
\beq
p^{\star}_\trmi{T}=-\frac{1}{4\,\pi}\inteddwters{-\infty}{+\infty}\int_{\bf k}\,\log\Big[k^2+\omega_\trmi{E}^2+\Pi_\trmi{T}(i\omega_\trmi{E},k)\Big]\,.
\eeq
By rescaling the Euclidean energy as $\omega_\trmi{E}\rightarrow k\,\omega_\trmi{E}$, we get from the above
\beq
p^{\star}_\trmi{T}=-\frac{1}{2\,\pi}\inteddwters{0}{\infty}\int_{\bf k}\,
k\log\Big[(1+\omega_\trmi{E}^2)\,k^2+\Pi_\trmi{T}(i\omega_\trmi{E},1)\Big]\,,
\eeq
which can now be easily integrated over the three-momentum. Applying the same technique to the longitudinal gluon contribution~(\ref{FullLongitudinalGluon}), as well as to the quark contribution~(\ref{FullQuarks}) while further taking the limit ${\boldsymbol \mu}\rightarrow0$, we obtain the following expressions
{\allowdisplaybreaks
\beqa
p^{\star}_\trmi{L}&=&-\frac{1}{2\,\pi}\inteddwters{0}{\infty}\int_{\bf k}\,k\,\log\Big[k^2+\Pi_\trmi{L}(i\omega_\trmi{E},1)\Big] \, , \\
p^{\star}_\trmi{q$_f$}&=&\frac{2}{\pi}\inteddwters{0}{\infty}\int_{\bf k}\,k\,\log\left[\left(1+\frac{m_\trmi{q$_f$}^2}{k^2}\,\left\{\frac{1}{1+i\,\omega_\trmi{E}}-\frac{\widetilde{{\cal T}}_\trmi{K}(i\omega_\trmi{E},1)}{i\,\omega_\trmi{E}}\right\}\right)\right. \nonumber\\
&& \ \ \ \ \ \ \ \ \ \ \ \ \ \ \ \ \ \ \ \ \ \ \ \ \ \ \ \times\left.\left(1+\frac{m_\trmi{q$_f$}^2}{k^2}\,\left\{\frac{1}{1-i\,\omega_\trmi{E}}+\frac{\widetilde{{\cal T}}_\trmi{K}(i\omega_\trmi{E},1)}{i\,\omega_\trmi{E}}\right\}\right)\right]\,. \label{QuarkZeroTandMu}
\eeqa}
\hspace{-0.12cm}Note that we added and subtracted a $\log \left(k^2+\omega_\trmi{E}^2\right)$ to~(\ref{QuarkZeroTandMu}), so that the subtracted logarithm is combined with the main expression for convenience during manipulations of its argument. The added piece vanishes in dimensional regularization, through the lack of scales in the integrand\footnote{Again, see Section~\ref{section:Renormalization_Running} for more details on that point.}. After integration over the three-momentum, we then get
{\allowdisplaybreaks
\beqa
p^{\star}_\trmi{T}&=&\frac{e^{\gamma_\trmi{\tiny E}\epsilon}\,\lmsb^{2\epsilon}}{16\,\pi^{5/2}}\,\,\frac{\Gamma\left(2-\epsilon\right)\Gamma\left(\epsilon-2\right)}{\Gamma\left(\frac{3}{2}-\epsilon\right)}\inteddwters{0}{\infty}\,\left(\frac{\Pi_\trmi{T}(i\omega_\trmi{E},1)}{1+\omega_\trmi{E}^2}\right)^{2-\epsilon}\,, \label{PTstarInte} \\
p^{\star}_\trmi{L}&=&\frac{e^{\gamma_\trmi{\tiny E}\epsilon}\,\lmsb^{2\epsilon}}{16\,\pi^{5/2}}\,\,\frac{\Gamma\left(2-\epsilon\right)\Gamma\left(\epsilon-2\right)}{\Gamma\left(\frac{3}{2}-\epsilon\right)}\inteddwters{0}{\infty}\,\bigg(\Pi_\trmi{L}(i\omega_\trmi{E},1)\bigg)^{2-\epsilon}\,,\label{PLstarInte} \\
p^{\star}_\trmi{q$_f$}&=&-\,m_\trmi{q$_f$}^{4-2\epsilon}\,\frac{e^{\gamma_\trmi{\tiny E}\epsilon}\,\lmsb^{2\epsilon}}{4\,\pi^{5/2}}\,\,\frac{\Gamma\left(2-\epsilon\right)\Gamma\left(\epsilon-2\right)}{\Gamma\left(\frac{3}{2}-\epsilon\right)} \nonumber \\
&\times&\ \inteddwters{0}{\infty}\,\Bigg[\left(\frac{1}{1+i\,\omega_\trmi{E}}-\frac{\widetilde{{\cal T}}_\trmi{K}(i\omega_\trmi{E},1)}{i\,\omega_\trmi{E}}\right)^{2-\epsilon}+\,\left(\frac{1}{1-i\,\omega_\trmi{E}}+\frac{\widetilde{{\cal T}}_\trmi{K}(i\omega_\trmi{E},1)}{i\,\omega_\trmi{E}}\right)^{2-\epsilon}\Bigg]\,.\label{PQstarInte} \ \ \ \ 
\eeqa}

Since we are interested in calculating the poles in $\epsilon$ analytically while we can compute the finite parts numerically, we notice that each of the above integrals comes with a prefactor $\Gamma(\epsilon-2)$, which has a simple pole for $\epsilon=0$. With the integrals being finite in this limit, it is sufficient to only expand the integrand through ${\cal O}\left(\epsilon\right)$
{\allowdisplaybreaks
\beqa
&&\!\!\!\!\!\!\!\!\!\!\left(\frac{\Pi_\trmi{T}(i\omega_\trmi{E},1)}{1+\omega_\trmi{E}^2}\right)^{2-\epsilon}\!\!\approx\,\left(\frac{\Pi_\trmi{T}^{{}^{\mbox{\tiny $(0)$}}}(i\omega_\trmi{E},1)}{1+\omega_\trmi{E}^2}\right)^{2}-\left(\frac{\Pi_\trmi{T}^{{}^{\mbox{\tiny $(0)$}}}(i\omega_\trmi{E},1)}{1+\omega_\trmi{E}^2}\right)^{2} \nonumber \\
&&\!\!\!\!\!\!\!\!\!\!\hspace*{6.5truecm}\times\left\{\log\bigg(\frac{\Pi_\trmi{T}^{{}^{\mbox{\tiny $(0)$}}}(i\omega_\trmi{E},1)}{1+\omega_\trmi{E}^2}\bigg)-\frac{2\,\Pi_\trmi{T}^{{}^{\mbox{\tiny $(1)$}}}(i\omega_\trmi{E},1)}{\Pi_\trmi{T}^{{}^{\mbox{\tiny $(0)$}}}(i\omega_\trmi{E},1)}\right\}\,\epsilon \,, \\
&&\!\!\!\!\!\!\!\!\!\!\bigg(\Pi_\trmi{L}(i\omega_\trmi{E},1)\bigg)^{2-\epsilon}\!\!\approx\,\bigg(\Pi_\trmi{L}^{{}^{\mbox{\tiny $(0)$}}}(i\omega_\trmi{E},1)\bigg)^{2}-\bigg(\Pi_\trmi{L}^{{}^{\mbox{\tiny $(0)$}}}(i\omega_\trmi{E},1)\bigg)^{2} \nonumber \\
&&\!\!\!\!\!\!\!\!\!\!\hspace*{6.25truecm}\times\left\{\log\bigg(\Pi_\trmi{L}^{{}^{\mbox{\tiny $(0)$}}}(i\omega_\trmi{E},1)\bigg)-\frac{2\,\Pi_\trmi{L}^{{}^{\mbox{\tiny $(1)$}}}(i\omega_\trmi{E},1)}{\Pi_\trmi{L}^{{}^{\mbox{\tiny $(0)$}}}(i\omega_\trmi{E},1)}\right\}\,\epsilon \,, \\
&&\!\!\!\!\!\!\!\!\!\!\left(\frac{1}{1\pm i\,\omega_\trmi{E}}\mp\frac{\widetilde{{\cal T}}_\trmi{K}(i\omega_\trmi{E},1)}{i\,\omega_\trmi{E}}\right)^{2-\epsilon}\!\!\approx\left(\frac{1}{1\pm i\,\omega_\trmi{E}}\pm\frac{i\,\widetilde{{\cal T}}^{{}^{\mbox{\tiny $(0)$}}}_\trmi{K}(i\omega_\trmi{E},1)}{\omega_\trmi{E}}\right)^{2}-\left(\frac{1}{1\pm i\,\omega_\trmi{E}}\pm\frac{i\,\widetilde{{\cal T}}^{{}^{\mbox{\tiny $(0)$}}}_\trmi{K}(i\omega_\trmi{E},1)}{\omega_\trmi{E}}\right)^{2} \hspace*{1.45truecm} \nonumber \\
&&\!\!\!\!\!\!\!\!\!\!\!\!\!\!\!\!\!\!\!\!\hspace*{5.25truecm}\times\left\{\log\left(\frac{1}{1\pm i\,\omega_\trmi{E}}\pm\frac{i\,\widetilde{{\cal T}}^{{}^{\mbox{\tiny $(0)$}}}_\trmi{K}(i\omega_\trmi{E},1)}{\omega_\trmi{E}}\right)\mp\frac{2\,i\,\widetilde{{\cal T}}^{{}^{\mbox{\tiny $(1)$}}}_\trmi{K}(i\omega_\trmi{E},1)}{\frac{\omega_\trmi{E}}{1\pm\,i\,\omega_\trmi{E}}\pm\,i\,\widetilde{{\cal T}}^{{}^{\mbox{\tiny $(0)$}}}_\trmi{K}(i\omega_\trmi{E},1)}\right\}\,\epsilon\,.
\eeqa}
\hspace{-0.12cm}Note that in the above, the superscripts for $\Pi_\trmi{T,L}^{{}^{\mbox{\tiny $(0)$}}},\,\Pi_\trmi{T,L}^{{}^{\mbox{\tiny $(1)$}}}$ or $\widetilde{{\cal T}}^{{}^{\mbox{\tiny $(0)$}}}_\trmi{K},\,\widetilde{{\cal T}}^{{}^{\mbox{\tiny $(1)$}}}_\trmi{K}$ denote the order of derivative with respect to $\epsilon$, before setting $\epsilon$ to zero. Also, in the last equation, notice that the sum of the first term with its complex conjugate vanishes so that there is no pole coming from the quark contribution. Computing then~(\ref{PTstarInte}),~(\ref{PLstarInte}) and~(\ref{PQstarInte}) with the help of the above expansions, we arrive at
{\allowdisplaybreaks
\beqa
(2-2\epsilon)\,p^{\star}_\trmi{T}&=&\frac{m_\trmi{D}^4}{64\,\pi^2}\,\bigg\{\frac{\log 256-5}{6}\,\left(\frac{1}{\epsilon}+\frac{7}{2}-\log 2+\log\frac{\lmsb^2}{m_\trmi{D}^2}\right)+\frac{2\,\kappa_\trmi{T}}{\pi}\bigg\}+{\cal O}\left(\epsilon\right) \label{PstarT}\,, \\
p^{\star}_\trmi{L}&=&\frac{m_\trmi{D}^4}{64\,\pi^2}\,\bigg\{\frac{8-\log 256}{6}\,\left(\frac{1}{\epsilon}+\frac{5}{2}-\log 4+\log\frac{\lmsb^2}{m_\trmi{D}^2}\right)+\frac{4\,\kappa_\trmi{L}}{\pi}\bigg\}+{\cal O}\left(\epsilon\right) \label{PstarL}\,, \\
p^{\star}_\trmi{q$_f$}&=&m_\trmi{q$_f$}^4\,\bigg\{\frac{\kappa_\trmi{q}+\kappa_\trmi{q}^{\star}}{4\,\pi^{3}}\bigg\}+{\cal O}\left(\epsilon\right) 
\label{Pstarq}\,,
\eeqa}
\hspace{-0.12cm}where $\kappa_\trmi{T}\approx 0.082875$, $\kappa_\trmi{L}\approx 0.320878$, and $\kappa_\trmi{q}+\kappa_\trmi{q}^{\star}\approx -4.53025$. Those are defined by the following one-dimensional integrals on the real axis
{\allowdisplaybreaks
\beqa
\kappa_\trmi{T}&\equiv& -\inteddwters{0}{\infty}\bigg[\omega_\trmi{E}\Big(\frac{\pi}{2}-\arctan\left(\omega_\trmi{E}\right)\Big)-\frac{\omega_\trmi{E}^2}{1+\omega_\trmi{E}^2}\bigg]^2 \\
& & \ \ \ \ \ \ \ \ \ \ \ \times\left[\frac{2\ \ {}_2F_1^{{}^{\mbox{\tiny $(0,0;1;0)$}}}\left(\frac{1}{2},1;\frac{3}{2};-\frac{1}{\omega_\trmi{E}^2}\right)}{\omega_\trmi{E}\Big(\frac{\pi}{2}-\arctan\left(\omega_\trmi{E}\right)\Big)-\frac{\omega_\trmi{E}^2}{1+\omega_\trmi{E}^2}}+\log\left[\omega_\trmi{E}\Big(\frac{\pi}{2}-\arctan\left(\omega_\trmi{E}\right)\Big)-
\frac{\omega_\trmi{E}^2}{1+\omega_\trmi{E}^2}\right]\right]\,, \nonumber\\
\kappa_\trmi{L}&\equiv& \inteddwters{0}{\infty}\bigg[1+\omega_\trmi{E}\Big(\arctan\left(\omega_\trmi{E}\right)-\frac{\pi}{2}\Big)\bigg]^2 \\
& & \ \ \ \ \ \ \ \ \, \times\left[\frac{2\ \ {}_2F_1^{{}^{\mbox{\tiny $(0,0;1;0)$}}}\left(\frac{1}{2},1;\frac{3}{2};-\frac{1}{\omega_\trmi{E}^2}\right)}{1+\omega_\trmi{E}\Big(\arctan\left(\omega_\trmi{E}\right)-\frac{\pi}{2}\Big)}-\log\Big[1+\omega_\trmi{E}\Big(\arctan\left(\omega_\trmi{E}\right)-\frac{\pi}{2}\Big)\Big]\right]\,, \nonumber \\
\kappa_\trmi{q}&\equiv& \inteddwters{0}{\infty}\left[i\,\bigg(\arctan\left(\omega_\trmi{E}\right)-\frac{\pi}{2}\bigg)-\frac{1}{1+i\,\omega_\trmi{E}}\right]^2 \\
& & \ \ \ \ \ \ \ \ \ \times\left[\log\left[i\,\bigg(\frac{\pi}{2}-\arctan\left(\omega_\trmi{E}\right)\bigg)+\frac{1}{1+i\,\omega_\trmi{E}}\right]+\frac{2\,i\,\,\,{}_2F_1^{{}^{\mbox{\tiny $(0,0;1;0)$}}}\left(\frac{1}{2},1;\frac{3}{2};-\frac{1}{\omega_\trmi{E}^2}\right)}{\frac{\omega_\trmi{E}}{1+i\,\omega_\trmi{E}}-i\,\omega_\trmi{E}\,\bigg(\arctan\left(\omega_\trmi{E}\right)-\frac{\pi}{2}\bigg)}\right]\,.\nonumber
\eeqa}
\hspace{-0.12cm}Note, for numerical convenience, the following real-valued representation
\beqa
{}_2F_1^{{}^{\mbox{\tiny $(0,0;1;0)$}}}\left(\mbox{\small $\frac{1}{2},1;\frac{3}{2};-\frac{1}{\omega_\trmi{E}^2}$}\right)&=&\omega_\trmi{E}\bigg(\frac{\pi}{2}-\arctan(\omega_\trmi{E})\bigg)\bigg(2-\log 4\bigg)-\omega_\trmi{E}\,\log 2\,\arg\left(\frac{i\omega_\trmi{E}+1}{i\omega_\trmi{E}-1}\right) \nonumber \\
&+& \frac{\omega_\trmi{E}}{2}\left[\ImPart{\Big\{\mbox{Li}_2\left(\frac{2}{1+i\omega_\trmi{E}}\right)\Big\}}-\ImPart{\Big\{\mbox{Li}_2\left(\frac{2}{1-i\omega_\trmi{E}}\right)\Big\}}\right]\,.
\eeqa

With the help of~(\ref{PstarT}),~(\ref{PstarL}), and~(\ref{Pstarq}), it is then straightforward to determine the contribution to the pressure coming from a vacuum type counter-term only, needed for canceling the ultraviolet divergences within the $\MSbar$ scheme. This one reads
\beq\label{Counterterm}
\Delta p\equiv -\,d_\trmi{A}\,\frac{m_\trmi{D}^4}{128\,\pi^2\,\epsilon}\,.
\eeq
Although no rigorous proof of the renormalizability of HTLpt has been worked out yet\footnote{That is to say, the fact that there exists a finite number of counter-terms needed to cancel the (new) ultraviolet divergences coming from the HTL improvement term, for all orders in perturbation theory.}, we note -- as a good hint -- that it is possible to render the theory finite, through three-loop order, with only a small number of counter-terms~\cite{NanHTLpt2}. In the present case, we adopt a more optimistic point of view, and shall assume that the theory is formally renormalizable, before proceeding to the analysis of our results. Notice also that, at leading order, the coupling $g$ and mass parameters $m_\trmi{D}$ and $m_\trmi{q$_f$}$ do not need to be redefined via renormalization, hence do not require additional counter-terms\footnote{Indeed, as a matter of fact, the coupling does not enter explicitly the HTLpt pressure at leading order.}. Therefore, the introduction of a vacuum counter-term, corresponding to the above cancellation, will be assumed from now on, together with the renormalization conditions for the Green's functions leading to the running of the coupling\footnote{Even though this one enters the leading order HTLpt pressure only via the mass parameters, the renormalization group equations are supposed to catch the coupling dependence whether it is explicit or not.}.

\section{Exact renormalized one-loop HTLpt pressure}\label{section:Exact_one-loop_HTLpt}

Having renormalized our theory at leading order, we are left with a finite expression for the pressure in the limit $\epsilon\rightarrow 0$. We therefore rewrite~(\ref{PreResultExactHTLpt}) using~(\ref{PstarT}),~(\ref{PstarL}),~(\ref{Pstarq}) and~(\ref{Counterterm}), while taking the appropriate limit for the dimensional regulator $\epsilon$.

To do so, we need the form of the discontinuities for various functions, in this limit. This includes, first of all, the Gauss hypergeometric function whose limits from above and below read
{\allowdisplaybreaks
\beqa
{}_2F_1^{\oplus}\left(\mbox{\small $\frac{1}{2},1;\frac{3}{2};\frac{k^2}{\omega^2}$}\right)&=& \frac{\omega}{k}\,\arctanh\left(\frac{k}{\omega}\right)=\frac{\omega}{k}\,\arctanh\left(\frac{\omega}{k}\right)-\frac{i\pi\,\omega}{2k} \, \nonumber \\
&=& \frac{\omega}{2\,k}\,\log\left(\frac{k+\omega}{k-\omega}\right)-\frac{i\pi\,\omega}{2k} \, , \label{DiscPlus3d} \\
{}_2F_1^{\ominus}\left(\mbox{\small $\frac{1}{2},1;\frac{3}{2};\frac{k^2}{\omega^2}$}\right)&=& \frac{\omega}{k}\,\arctanh\left(\frac{k}{\omega}\right) +\frac{i\pi\,\omega}{k}=\frac{\omega}{k}\,\arctanh\left(\frac{\omega}{k}\right)+\frac{i\pi\,\omega}{2k} \, \nonumber \\
&=& \frac{\omega}{2\,k}\,\log\left(\frac{k+\omega}{k-\omega}\right)+\frac{i\pi\,\omega}{2k} \,,\label{DiscMinus3d}
\eeqa}
\hspace{-0.12cm}the discontinuity being the difference of the two. Note that in the last two equalities, we made use of the fact that $k>\omega$, in order to keep the logarithms real valued. We also need the limits of the discontinuities of the Landau damping angles, in the transverse and longitudinal gluon case as well as in the quark case. Those read
{\allowdisplaybreaks
\beqa
\disc\phi_{\trmi{T},\epsilon}&{\stackrel{\epsilon\rightarrow 0}{\longrightarrow}}&-2\,\arctan\left[\frac{\frac{\pi}{4}m_\trmi{D}^2\frac{\omega}{k^3}(k^2-\omega^2)}{k^2-\omega^2+\frac{m_\trmi{D}^2}{2}\frac{\omega^2}{k^2}\Big[1+\frac{k^2-\omega^2}{2k\omega}\log\left(\frac{k+\omega}{k-\omega}\right)\Big]}\right]\,, \\
\disc\phi_{\trmi{T},\epsilon}&{\stackrel{\epsilon\rightarrow 0}{\longrightarrow}}&-2\,\arctan\left[\frac{\frac{\pi}{2}m_\trmi{D}^2\frac{\omega}{k}}{k^2+m_\trmi{D}^2\Big[1-\frac{\omega}{2k}\log\left(\frac{k+\omega}{k-\omega}\right)\Big]}\right]\,, \\
\disc\theta_\trmi{q$_f$,$\epsilon$}&{\stackrel{\epsilon\rightarrow 0}{\longrightarrow}}&-2\, \\
&\times&\arctan\left[\frac{\frac{\pi m_\trmi{q$_f$}^4}{k^2}\Big[\frac{\omega}{k}+\frac{k^2-\omega^2}{2k^2}\log\left(\frac{k+\omega}{k-\omega}\right)\Big]}{k^2-\omega^2+2m_\trmi{q$_f$}^2+\frac{m_\trmi{q$_f$}^4}{k^2}\bigg[1-\frac{\omega}{k}\log\Big(\frac{k+\omega}{k-\omega}\Big)-\frac{k^2-\omega^2}{4k^2}\bigg[\log\Big(\frac{k+\omega}{k-\omega}\Big)^2-\pi^2\bigg]\bigg]}\right]\,. \nonumber
\eeqa}

We are now left with the renormalized result for the exact one-loop HTLpt pressure~\cite{sylvain2}, using the prescription~(\ref{HTLptmDmqparameters}) for the mass parameters $m_\trmi{D}\left(T,\bm{\mu}\right)$ and $m_\trmi{q$_f$}\left(T,\bm{\mu}\right)$
{\allowdisplaybreaks
\beqa\label{FullHTLptpressure}
p_\trmi{HTLpt}\left(T,\bm{\mu}\right)&=& d_\trmi{A}\Bigg\{\frac{m_\trmi{D}^4}{64\pi^2}\left(\log\frac{\lmsb}{m_\trmi{D}}+C_\trmi{g}\right)+\frac{1}{2\pi^3}\inteddw\frac{1}{e^{\beta\omega}-1}\inteddk k^2\bigg(2\phi_\trmi{T}-\phi_\trmi{L}\bigg)\, \nonumber \\
& & \ \ \ \ -\frac{T}{2\pi^2}\intedk k^2\bigg[2\log\bigg(1-e^{-\beta\omega_\trmi{T}}\bigg)+\log\bigg(1-e^{-\beta\omega_\trmi{L}}\bigg)\bigg]-\frac{\pi^2\,T^4}{90}\Bigg\} \ \ \ \ \ \nonumber \\
&+&N_\trmi{c}\sum_{f,\,s=\pm 1}\Bigg\{\frac{C_\trmi{q}}{2}\ m_\trmi{q$_f$}^4+\frac{2\ T^4}{\pi^2}\,\,\mbox{Li}_4\bigg(-e^{s\,\beta\mu_f}\bigg)-\frac{1}{\pi^3}\intdwdk\frac{k^2\,\theta_\trmi{q$_f$}}{e^{\beta\left(\omega+s\,\mu_f\right)}+1} \, \nonumber \\
& & \ \ \ \ \ \ \ +\frac{T}{\pi^2}\intedk k^2\ \bigg[\log\bigg(1+e^{-\beta\left(\omega_{f_+}+s\,\mu_f\right)}\bigg)+\log\bigg(1+e^{-\beta\left(\omega_{f_-}+s\,\mu_f\right)}\bigg)\bigg]\Bigg\}\,, \nonumber \\
\eeqa}
\hspace{-0.12cm}for which we list now the angles $\phi_\trmi{T,L}$ and $\theta_\trmi{q$_f$}$, as well as the definitions of the dispersion relations $\omega_\trmi{T,L,{$f_\pm$}}$ and the constants $C_\trmi{g}\approx 1.17201$ and $C_\trmi{q}\approx -0.03653$. They respectively read
{\allowdisplaybreaks
\beqa
\phi_\trmi{T}&=&\arctan\left[\frac{\frac{\pi}{4}m_\trmi{D}^2\frac{\omega}{k^3}(k^2-\omega^2)}{k^2-\omega^2+\frac{m_\trmi{D}^2}{2}\frac{\omega^2}{k^2}\Big[1+\frac{k^2-\omega^2}{2k\omega}\log\left(\frac{k+\omega}{k-\omega}\right)\Big]}\right]\,, \\
\phi_\trmi{L}&=&\arctan\left[\frac{\frac{\pi}{2}m_\trmi{D}^2\frac{\omega}{k}}{k^2+m_\trmi{D}^2\Big[1-\frac{\omega}{2k}\log\left(\frac{k+\omega}{k-\omega}\right)\Big]}\right]\,, \\
\theta_\trmi{q$_f$}&=&\arctan\left[\frac{\frac{\pi m_\trmi{q$_f$}^4}{k^2}\Big[\frac{\omega}{k}+\frac{k^2-\omega^2}{2k^2}\log\left(\frac{k+\omega}{k-\omega}\right)\Big]}{k^2-\omega^2+2m_\trmi{q$_f$}^2+\frac{m_\trmi{q$_f$}^4}{k^2}\bigg[1-\frac{\omega}{k}\log\Big(\frac{k+\omega}{k-\omega}\Big)-\frac{k^2-\omega^2}{4k^2}\bigg[\log\Big(\frac{k+\omega}{k-\omega}\Big)^2-\pi^2\bigg]\bigg]}\right]\,,
\hspace*{2.75em}
\eeqa}
\hspace{-0.12cm}together with the dispersion relations in three spatial dimensions
{\allowdisplaybreaks
\beqa\label{Dispersionrelations}
\omega_\trmi{T}^2&=&k^2+\frac{1}{2}m_\trmi{D}^2\frac{\omega_\trmi{T}^2}{k^2} \left[1-\frac{\omega_\trmi{T}^2-k^2}{2\,\omega_\trmi{T} k}\log\left(\frac{\omega_\trmi{T}+k}{\omega_\trmi{T}-k}\right)\right]\,, \\
0&=&k^2+m_\trmi{D}^2\left[1-\frac{\omega_\trmi{L}}{2k}\log\left(\frac{\omega_\trmi{L}+k}{\omega_\trmi{L}-k}\right)\right]\,, \\
0&=&\omega_{f_\pm}\mp k-\frac{m_\trmi{q$_f$}^2}{2k}\left[\left(1\mp\frac{\omega_{f_\pm}}{k}\right)\log\left(\frac{\omega_{f_\pm}+k}{\omega_{f_\pm}-k}\right)\pm 2\right]\,,
\eeqa}
\hspace{-0.12cm}as well as the constants $C_\trmi{g}$ and $C_\trmi{q}$
\beqa\label{KappaT}
C_\trmi{g}&=& \frac{2\ \kappa_\trmi{T}}{\pi}+\frac{4\ \kappa_\trmi{L}}{\pi}+\frac{1}{12}\bigg[5+\Big(\log 256-3\Big)\log 4\bigg]\ \approx 1.17201\,, \\
C_\trmi{q}&=& \frac{\kappa_\trmi{q}+\kappa_\trmi{q}^{\star}}{4\pi^3}\approx -0.03653\,.
\eeqa

\section{Mass expansion and truncation of the result}\label{section:Truncated_one-loop_HTLpt}

In this section, we give details about the approximation method of mass expansions, employed in high loop order  HTLpt computations. Indeed, at leading order, we saw that the encountered sum-integrals can be computed using various combinations of analytical and numerical techniques. However, from two loops onwards the situation gets much more complicated, leading to a two-sided problem. Firstly, the established method at one loop would lead, if straightforwardly extended to two loops, to more complicated numerical integrals\footnote{Changing the most challenging pieces from two dimensional numerical integrations to five dimensional ones~\cite{JensMikeFirstTwoLoop}.}. Secondly and most importantly, our new and simplifying method for extracting the potentially divergent pieces is not guaranteed to hold beyond leading order\footnote{As can be seen from~\cite{HTLptFiniteMuTwoLoop}, the needed two-loop ultraviolet counter-terms become explicitly medium dependent.}, while the increasing complexity of the sum-integrals would surely avoid a direct calculation in $3-2\epsilon$ spatial dimensions. These arguments will then motivate the present section where we will explain in detail, at one-loop order, the systematic approximation framework that was developed along the years in order to circumvent the aforementioned technical challenges.

We are now going to compute the leading order HTLpt pressure within the mass truncated approximation scheme. This corresponds to Taylor expanding the result~(\ref{FullHTLptpressure}) in powers of $m_\trmi{D}/T$ and $m_\trmi{q$_f$}/T$ before truncating at a certain order. Given the fact that, at zero quark chemical potentials\footnote{At nonzero density, the mass parameters are of order $g$ times some functions of the temperature and chemical potentials.}, the mass parameters are of order $g T$, it is customary to truncate the result at ${\cal O}\left((m_\trmi{D}/T)^{5},\,(m_\trmi{q$_f$}/T)^{5}\right)\!\sim {\cal O}\!\left(g^5\right)$ in order to allow for a direct comparison with the known weak coupling result\footnote{Having still an unknown ${\cal O}\left(g^6\right)$ contribution, it is more suitable to choose such an order of truncation.}.

One could start directly from the exact leading order result for the pressure. However, as it involves dispersion relations, thus an implicit dependence on the mass parameters, we choose to adopt another approach, consisting of directly Taylor expanding the sum-integrals before computing them. Let us now detail the corresponding procedure, which first involves a separation of scales.

At zero quark chemical potentials, there are only two momentum scales in the sum-integrals, the hard scale $2\pi T$ and the soft scale given by the thermal masses. The hard scale region encompasses all the fermionic momenta $K=((2n+1)\pi T,{\bf k})$, together with the bosonic ones $K=(2n\pi T,{\bf k})$ provided that $n\neq0$. It also includes the $n=0$ bosonic mode, for which $k$ is of order $T$. The soft scale region, on the other hand, corresponds to the $n=0$ mode for the bosonic momenta with $k$ at most of order $g\,T$. In presence of a dense medium, i.e. at finite quark chemical potentials, this picture remains also valid, as the new hard scales $\mu_f$ only enter explicitly the fermionic momenta or implicitly the bosonic ones via the thermal mass $m_\trmi{D}$. We then use the fact that $\Pi_\trmi{T}(0,{\bf k})=0$ and $\Pi_\trmi{L}(0,{\bf k})=m_\trmi{D}^2$, in order to write the soft scale contribution to the pressure directly from~(\ref{OneLoopHTLpt}), and get
\beqa\label{SoftMassTruncp}
p_\trmi{HTLpt}^\trmi{\tiny High-T,(s)}&=&-(2-2\epsilon)\frac{d_\trmi{A}\,T}{2}\int_{\bf k}\log(k^2)-\frac{d_\trmi{A}\,T}{2}\int_{\bf k}\log\Big(k^2+m_\trmi{D}^2\Big)\,,
\eeqa
where the first integral is of the type which vanishes in dimensional regularization, as we already saw. The second integral, being dominated by momenta of order $m_\trmi{D}$, directly yields the soft contribution to the leading order pressure~\cite{BraatenPetitgirardPhiDerivable}.

Regarding the hard scale contribution, we assume the temperature to be large enough so that ratios such as $m_\trmi{D}/T$ or $m_\trmi{q$_f$}/T$ can be considered parametrically small. We then expand the sum-integrals that appear in the leading order pressure~(\ref{OneLoopHTLpt}), in powers of those ratios, and truncate at ${\cal O}\left((m_\trmi{D}/T)^{5},\,(m_\trmi{q$_f$}/T)^{5}\right)$. Recall that these ratios, thanks to the choices that we made for the mass parameters\footnote{Which are nothing but the weak coupling values for the Debye and quark thermal masses~(\ref{HTLptmDmqparameters}).}, are of order $g$. We thus expect such an expansion to be a reasonable approximation, as the temperature increases, but this is something we shall check quantitatively in Section~\ref{section:Convergence_Mass_Truncation}. This gives us the following contribution
{\allowdisplaybreaks
\beqa\label{HardMassTruncp}
p_\trmi{HTLpt}^\trmi{\tiny High-T,(h)}&=&-(1-\epsilon)\,d_\trmi{A}\,\sumint_{K}\log\left(K^2\right)+2\,N_\trmi{c}\,\sum_f\ \sumint_{\{{K\}}}\log\left(K^2\right) \nonumber \\
&-&\frac{d_\trmi{A}\,m_\trmi{D}^2}{2}\,\sumint_{K}\frac{1}{K^2}+4\,N_\trmi{c}\,\sum_f m_\trmi{q$_f$}^2\,\sumint_{\{K\}}\frac{1}{K^2} \nonumber \\
&+&\frac{d_\trmi{A}\,m_\trmi{D}^4}{8-8\epsilon}\,\sumint_{K}\left[\frac{1}{\left(K^{2}\right)^{2}}-\frac{2}{k^2\,K^2}-(6-4\epsilon)\frac{{\cal T}_\trmi{K}}{\left(k^{2}\right)^{2}}+\frac{2\,{\cal T}_\trmi{K}}{k^2\,K^2}+(3-2\epsilon)\frac{({\cal T}_\trmi{K})^2}{\left(k^{2}\right)^{2}}\right]+{\cal O}\left(m_\trmi{D}^{5}\right)\nonumber \\
&-&2\,N_\trmi{c}\,\sum_f\,m_\trmi{q$_f$}^4\,\sumint_{\{K\}}\left[\frac{2}{\left(K^{2}\right)^{2}}-\frac{1}{k^2\,K^2}+\frac{2\,\widetilde{{\cal T}}_\trmi{K}}{k^2\,K^2}-\frac{(\widetilde{{\cal T}}_\trmi{K})^2}{k^2\,\left(\widetilde{\omega}_n-i\mu_f\right)^2}\right]+{\cal O}\left(m_\trmi{q$_f$}^{5}\right). \ \ \ 
\eeqa}

Therefore, the mass expansion of the one-loop HTLpt pressure is given by the sum of~(\ref{SoftMassTruncp}) and~(\ref{HardMassTruncp}), together with the piece $\Delta p$ coming from the vacuum counter-term, computed in~(\ref{Counterterm})
\beq
p^\trmi{\tiny High-$T$}_\trmi{HTLpt} \equiv p_\trmi{HTLpt}^\trmi{\tiny High-T,(s)}\,+\,p_\trmi{HTLpt}^\trmi{\tiny High-T,(h)}\,+\,\Delta p\,,
\eeq
and the subsequent sum-integrals have to be computed. We refer to appendix~\ref{appendix:HTL_Master_Sum-Integrals} for more details on the corresponding method of computation, especially regarding the sum-integrals that involve the HTL functions ${{\cal T}_\trmi{K}}$ and ${\widetilde{{\cal T}}_\trmi{K}}$. Finally, the purely analytical mass truncated result for the finite temperature and chemical potential leading order HTLpt pressure reads~\cite{sylvain1}
{\allowdisplaybreaks
\beqa\label{HighTHTLptLeadingOrderPressure}
p_\trmi{HTLpt}^\trmi{\tiny High-T}\left(T,\bm{\mu}\right)&=&\frac{d_\trmi{A}\pi^2 T^4}{45}\Bigg\{1+\frac{N_\trmi{c}}{d_\trmi{A}}\sum_f\bigg(\frac{7}{4}+30\ \hat \mu_f^2 +60\ \hat \mu_f^4\bigg) -\frac{15}{2}\hat m_\trmi{D}^2  \, \nonumber \\
& & \ \ \ \ \ \ \ \ \ \ \ \ - \frac{30\ N_\trmi{c}}{d_\trmi{A}}\sum_f\bigg( 1+12\ \hat \mu_f^2\bigg) \hat m_\trmi{q$_f$}^2+30\ \hat m_\trmi{D}^3  \, \nonumber \\
& &  \ \ \ \ \ \ \ \ \ \ \ \ +\frac{45}{4}\bigg(\gamma_\trmi{\tiny E}-\frac{7}{2}+\frac{\pi^2}{3}+\log\frac{\lmsb}{4\,\pi\,T}\bigg)\ \hat m_\trmi{D}^4 \, \nonumber \\
& & \ \ \ \ \ \ \ \ \ \ \ \ +\frac{60\ N_\trmi{c}}{d_\trmi{A}} \bigg(6-\pi^2\bigg) \sum_f \hat m_\trmi{q$_f$}^4+{\cal O}\left(\hat m_\trmi{D}^6,\hat m_\trmi{q$_f$}^6\right)\Bigg\}\,.
\eeqa}

\chapter{Analyzing the results}\label{chapter:Results}

In this chapter, we finally proceed to the analysis of our findings for various cumulants of conserved charges in the quark-gluon plasma. We first present our results for the second-, fourth- and sixth-order diagonal quark number susceptibilities. Then, we present some ratios of those quantities, as well as observables related to the baryon number. This work is based on two peer-reviewed articles~\cite{sylvain1,sylvain2} as well as two proceedings contributions~\cite{sylvain3,sylvain4}.

In the following, we start by explaining how we fix a number of parameters appearing in our resummed perturbative calculations. Then, we proceed to compare our results with state-of-the-art lattice data, e.g.~\cite{WBLatticeData1,BielefeldLatticeDataHighT,BielefeldLatticeData1,BielefeldLatticeData2,BorsanyiDeltaP,HandLatticeDatasu6u2,GuptaLatticeData}, as well as to a recent three-loop mass expanded HTLpt calculation~\cite{HTLptFiniteMUThreeLoop1}, thereby gaining a better understanding of the relevant degrees of freedom right above the transition region. Finally, we compare our exact HTLpt results, obtained in~\cite{sylvain2,sylvain3}, with the mass expanded ones from~\cite{sylvain1,sylvain4}; this is done separately for the pressure and the second- and fourth-order diagonal quark number susceptibilities. This allows us to draw important conclusions on the convergence of this approximation, which is applied in all higher order loop calculations within HTLpt.

\section{Fixing the parameters}\label{section:Fixing_parameters}

As we saw in Section~\ref{section:Renormalization_Running}, any renormalized weak coupling expansion contains at least two parameters that need to be fixed in order to obtain a numerical prediction for the quantity under study. The first one is the renormalization scale, which is in principle somewhat arbitrary. The second one, on the other hand, is a fundamental scale of the theory and has to be determined either with experimental input or by non-perturbative computations\footnote{Of which the most prominent relies on Monte Carlo simulations. However, alternative methods have been proposed as well; see e.g.~\cite{RGImprovedOPTKneur} for a recent determination of this scale via renormalization group optimized perturbation theory.}. Besides these two parameters, there is some freedom related to the choice of the running coupling constant. In the present work, we choose to apply a two-loop running for the results of our dimensional reduction framework~\cite{MikkoYorkTwoLoopG}, as well as the standard one-loop running for the exact one-loop HTLpt computation.

In vacuum quantum field theory calculations, it is customary to fix the renormalization scale to a value of momentum typical for the relevant physical process. At finite temperature and zero density, the typical momentum of a particle in the thermal bath is of the other of $2\pi T$. Indeed, this is the thermal mass of the lowest nonzero bosonic Matsubara mode, expected to dominate at short distances. The renormalization scale is then usually chosen to be of this order\footnote{The constant of proportionality being in general close to one.}, which however violates thermodynamic consistency\footnote{As we saw in Section~\ref{section:Thermodynamic_Consistency}, such a violation is nevertheless sub-leading as far as the weak coupling series is concerned, and so is it for our dimensional reduction framework.}. We choose here to adopt a slightly different but improved choice, which is nevertheless numerically close to the usual one. It does not, however, cure the problem of thermodynamic consistency. See Section~\ref{section:Thermodynamic_Consistency} for more details on this question, especially within our setups. In order to assess the sensitivity of our results on the renormalization scale parameter, we vary this scale in all of our results by a factor of two around the central value, which is the usual procedure.

Concretely, in the dimensional reduction framework, we apply the Fastest Apparent Convergence principle to the next-to-leading order expression of the three dimensional gauge coupling of QCD~\cite{FAConNLOg3}. At zero chemical potentials, it then leads to the following values for three and two quark flavors, respectively~\cite{sylvain1}: $\bar{\Lambda}_{\rm central}\approx 1.445\times 2\pi T$ and $\bar{\Lambda}_{\rm central}\approx 1.291\times 2\pi T$. We then straightforwardly generalize these results to nonzero density, extracting a value for $\lmsb$ by demanding that the next-to-leading order term for $g_3$ in~(\ref{Expressiong3}) vanishes, i.e. by solving $\aE{7}=0$, only this time for non zero $\mu_f$. This yields~\cite{sylvain2}
\beqa
\bar{\Lambda}_{\rm central}^{\rm N_\trmi{f}=3}&\approx&\frac{0.9344\times 2\pi T}{\exp\left(\frac{1}{27}\sum_f\left[\Psi(\frac{1}{2}+i\hat\mu_f)+\Psi(\frac{1}{2}-i\hat\mu_f)\right]\right)}\,, \\
\bar{\Lambda}_{\rm central}^{\rm N_\trmi{f}=2}&\approx&\frac{0.9847\times 2\pi T}{\exp\left(\frac{1}{29}\sum_f\left[\Psi(\frac{1}{2}+i\hat\mu_f)+\Psi(\frac{1}{2}-i\hat\mu_f)\right]\right)}\,.
\eeqa
Note that the above values will be used within our HTLpt calculations as well.

Regarding the QCD scale that we choose to fix in the $\MSbar$ scheme $\Lambda_\trmi{QCD}\equiv\lQCD$, we use the recent lattice result $\alpha_\trmi{s}(1.5\;{\rm GeV})=0.326$~\cite{RunningAlpha} by requiring that both perturbative couplings agree with it for $\bar{\Lambda}_{\rm central}=1.5$ GeV. In the case of three flavors, it yields $\Lambda_{\rm\overline{MS}}=176$ MeV and 283 MeV for the one- and two-loop couplings respectively. For two flavors, it gives $\Lambda_{\rm\overline{MS}}=204$ MeV and 324 MeV. As there is no rigorous way to assess the uncertainties in fixing this scale, we account for them by varying this parameter by $30$ MeV around its central values.

Finally, we point out that the forthcoming plots incorporate all of these variations in their bands. The upper and lower edges do not necessarily correspond to the greater and smaller values of the scales. In order to obtain proper bands, it is important to scan the whole multidimensional parameter space, rather than only picking up the boundaries. Notice that in the remaining part of this chapter, we label the dimensional reduction framework results as the ``DR'' ones. We also display, as a good qualitative guidance, a thick dashed line inside every band, corresponding to the central values of both scales $\lmsb$ and $\lQCD$.

\section{The three flavor case}\label{section:Three_flavors}

\begin{figure}[!t]\centering\includegraphics[scale=0.42]{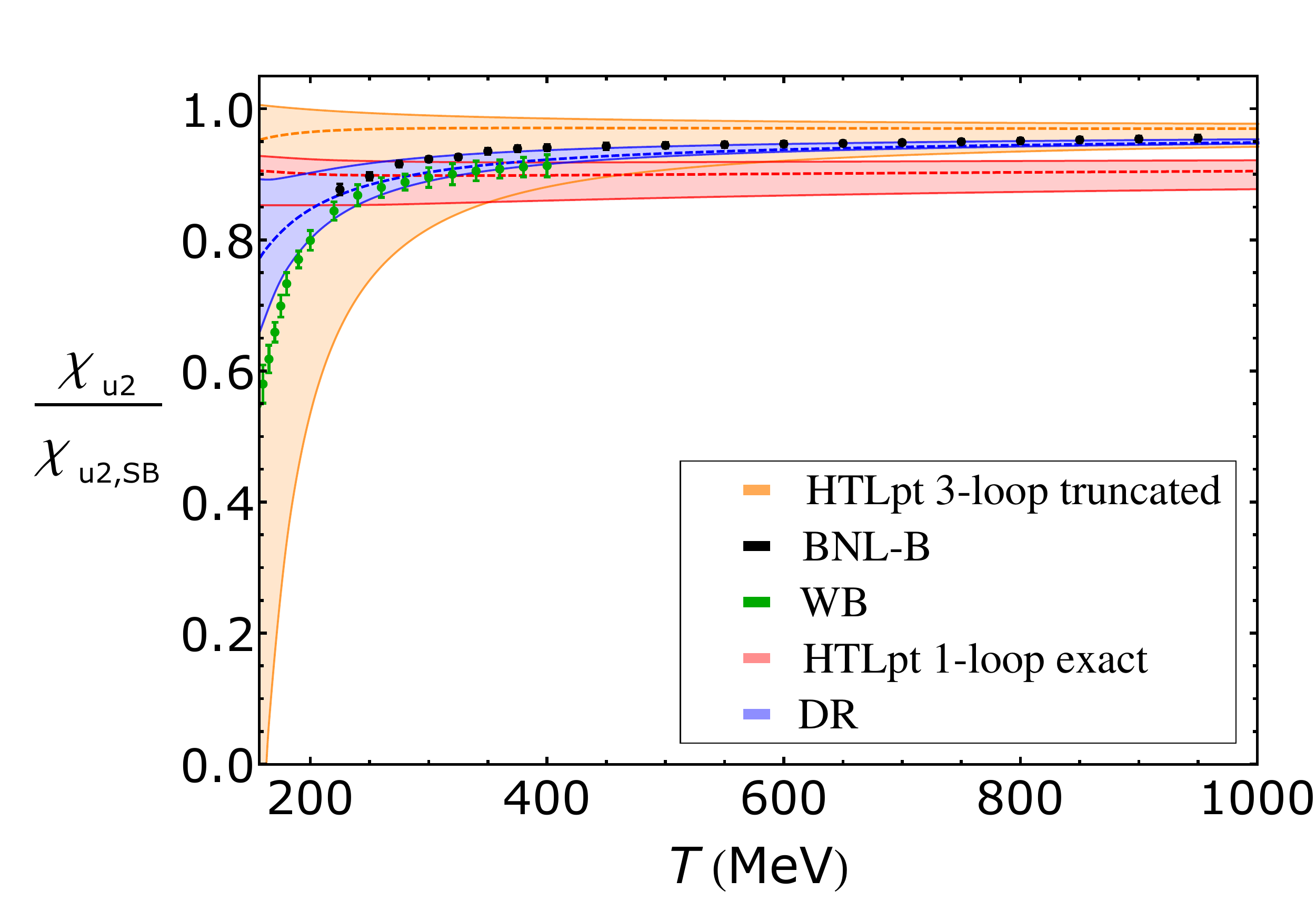}
\caption{Three flavor second-order diagonal quark number susceptibility, normalized to its non-interacting limit. The truncated three-loop HTLpt result is from~\cite{HTLptFiniteMUThreeLoop1} and the lattice data from the BNL-Bielefeld~\cite{BielefeldLatticeDataHighT} (BNL-B) as well as from the Wuppertal-Budapest~\cite{WBLatticeData1} (WB) collaborations.}
\label{fig:Chiu2Nf3}
\end{figure}

\begin{figure}[!t]\centering\includegraphics[scale=0.42]{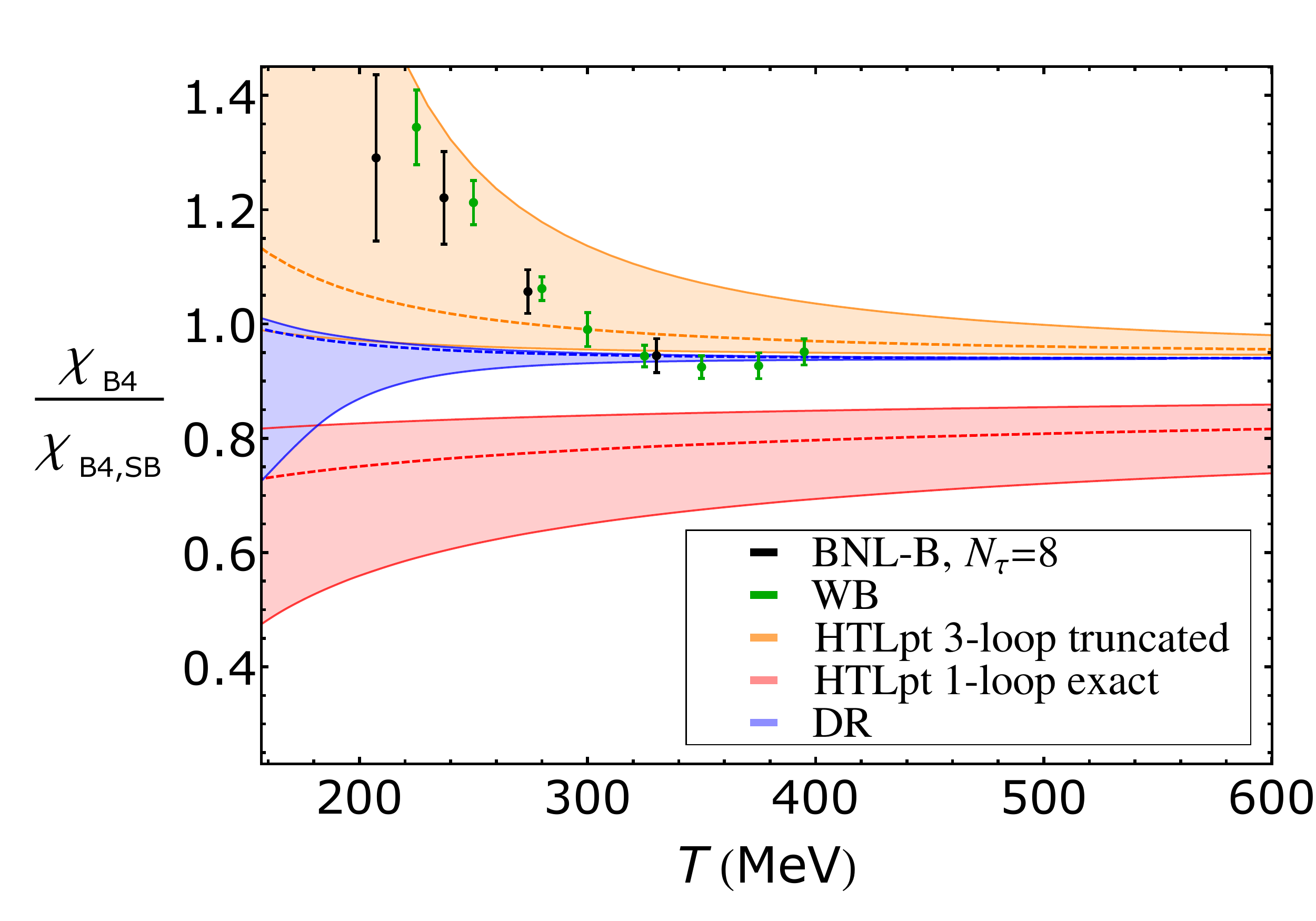}
\caption{Three flavor fourth-order diagonal baryon number susceptibility, normalized to its non-interacting limit. The truncated three-loop HTLpt result is from~\cite{HTLptFiniteMUThreeLoop1} and the lattice data from the Wuppertal-Budapest~\cite{WupBudChi4andRatio} (WB) and BNL-Bielefeld~\cite{Bazavov} (BNL-B) collaborations.}
\label{fig:ChiB4Nf3}
\end{figure}

We start our analysis from the physically most interesting case of three flavors\footnote{The relevant energy scale for modern experiments calls for considering only the up, down and strange quarks.}. We display, in Figure~\ref{fig:Chiu2Nf3}, the second-order diagonal quark number susceptibility, namely $\chi_{\rm u2}$, normalized to its non-interacting limit whose value is $\chi_{\rm u2,SB}=T^2$. Note that the blue band corresponds to the DR result while the red and orange bands are, on the other hand, the exact one-loop and truncated three-loop HTLpt results\footnote{The three-loop HTLpt band is actually obtained from the baryon number susceptibility. This is, however, not a big problem as the numerical difference between the two quantities is expected to be negligible~\cite{BielefeldLatticeDataHighT}. Also, see Figure~\ref{fig:ChiB4Nf3} and the text for more explanations on comparisons between exact and mass truncated results at different loop orders.}, respectively from~\cite{sylvain2} and~\cite{HTLptFiniteMUThreeLoop1}.
%%%%%%%%%%%%%%%%%%%%%%%%%%%%%%%%%%%%%%%%%%%%%%%%%%%%%
%%%%%%%%%%%%%%%%%%%%%%%%%%%%%%%%%%%%%%%%%%%%%%%%%%%%%
From the width of the bands, we clearly see that the DR scale dependence is extremely small for perturbatively relevant temperatures. Moreover, both HTLpt results are quite close to each other, starting from above $T\approx 500$ MeV. This can be easily noticed when comparing the relative distance between the central lines, and it is a robust indication that this quantity converges nicely at those temperatures. Notice that the corresponding two-loop HTLpt result~\cite{HTLptFiniteMuTwoLoop}, although not shown here, is quite close to the three-loop one for $T\gtrsim 500$ MeV.

We then move on to compare our results with the non-perturbative continuum extrapolated lattice data from the BNL-Bielefeld (BNL-B; black dots)~\cite{BielefeldLatticeDataHighT} and the Wuppertal-Budapest (WB; green dots)~\cite{WBLatticeData1} collaborations. We note that the DR and three-loop HTLpt results are in excellent agreement with both lattice data sets starting from $T\sim 500$ MeV. The DR central band even overlaps with the high$-T$ BNL-B results at these temperatures. Although both lattice data sets are inside the bands down to very low temperatures above the transition region, we note that the central lines of the resummed DR and three-loop truncated HTLpt result differ. Thus, the DR result seems to agree better with the lattice data\footnote{Note that one has to be careful in interpreting perturbative predictions at temperatures close to the phase transition. Indeed, none of the relevant symmetries, e.g. the $Z(N_\trmi{c})$ center symmetry even if softly broken by the quarks, or topological contributions are caught by a perturbative expansion.}.

\begin{figure}[!t]\centering\includegraphics[scale=0.4165]{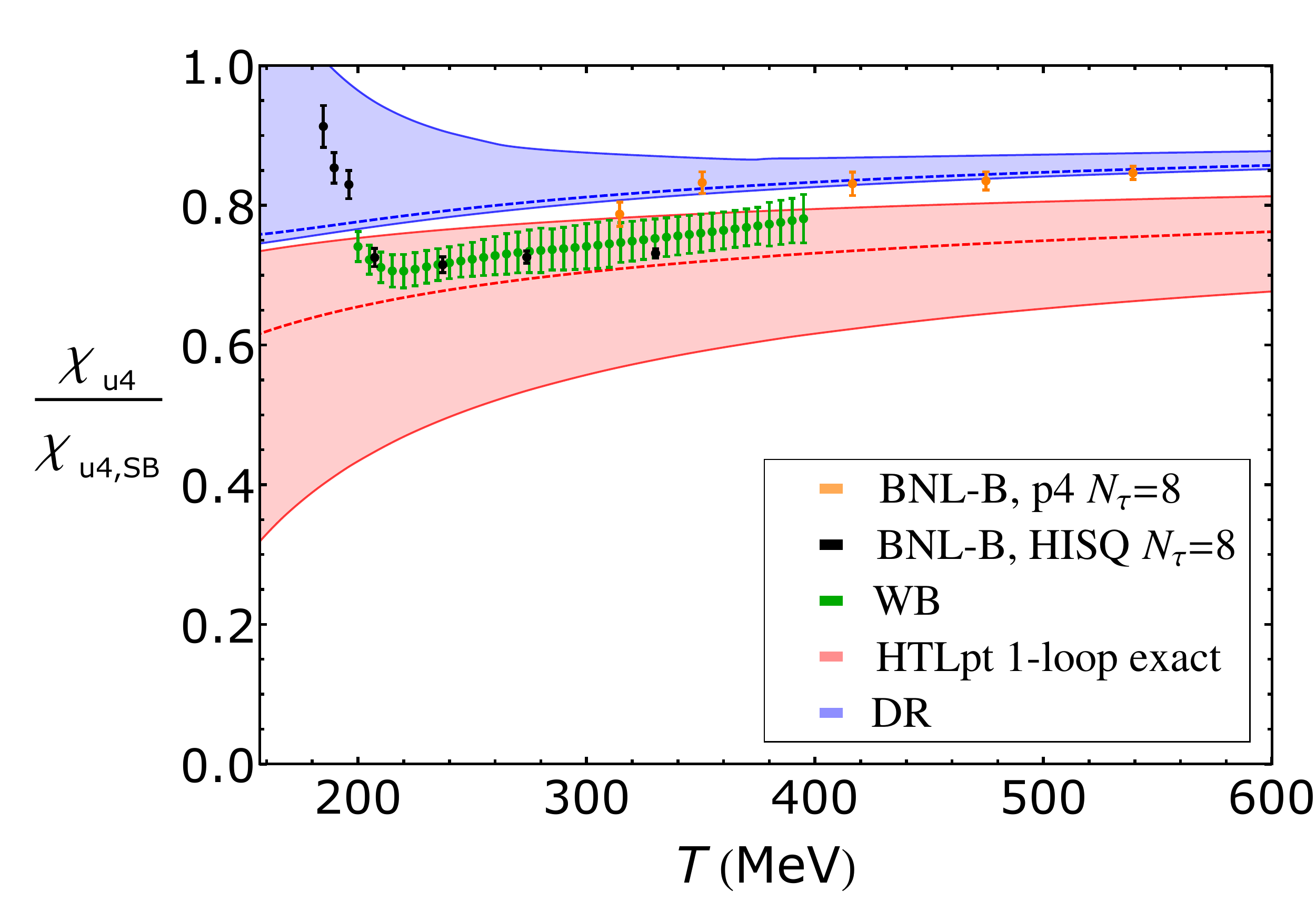}
\caption{Three flavor fourth-order diagonal quark number susceptibility, normalized to its Stefan-Boltzmann value. The lattice data are again from the BNL-Bielefeld~\cite{BielefeldLatticeData1,BielefeldLatticeData2} (BNL-B) as well as the Wuppertal-Budapest~\cite{WupBudChi4andRatio} (WB) collaborations.}
\label{fig:Chiu4Nf3}
\end{figure}

We now turn to our next result, namely the fourth-order diagonal quark number susceptibility $\chi_\trmi{u4}$, that we display in Figure~\ref{fig:Chiu4Nf3} normalized to its Stefan-Boltzmann value $\chi_{\rm u4,SB}=6/\pi^2$. We recall again that the width of the bands assesses the sensitivity of the results with respect to both the renormalization and the QCD scales. This time, the continuum extrapolated Wuppertal-Budapest (WB; green dots) lattice data  are taken from~\cite{WupBudChi4andRatio}, and the $N_\tau=8$ BNL-Bielefeld (BNL-B; black dots) results, using a highly improved staggered quark (HISQ) action, from~\cite{BielefeldLatticeData1,BielefeldLatticeData2}. Note that there are no bands for the three-loop HTLpt prediction here. The reason is that as the calculations in~\cite{HTLptFiniteMUThreeLoop1} probe the baryon numbers rather than the quark ones, and this time the difference between the two quantities is not negligible due to important and undetermined off-diagonal contributions~\cite{BielefeldLatticeDataHighT}. Therefore, for completeness, we refer the interested reader to the next figure where we display the corresponding baryon number quantity.
%%%%%%%%%%%%%%%%%%%%%%%%%%%%%%%%%%%%%%%%%%%%%%%%%%%%%
%%%%%%%%%%%%%%%%%%%%%%%%%%%%%%%%%%%%%%%%%%%%%%%%%%%%%

We see from Figure~\ref{fig:Chiu4Nf3} that both our resummed DR and exact one-loop HTLpt predictions reflect the qualitative trend of the lattice data for most temperatures. However, now the two results differ in a more noticeable way than for the second-order susceptibility. It turns out that at the lowest temperatures the lattice data seem to favor the one-loop HTLpt result. Nevertheless, the DR results and the lattice points tend to get closer to each other, as the temperature increases. Therefore, it will be very interesting to see how future lattice data at higher temperature will affect these conclusions. We recall the occurrence of an overcounting with respect to the known weak coupling result at low-loop order HTLpt, which affects the order $g^2$ coefficient for the pressure as well as the second- and fourth-order susceptibilities\footnote{See Section~\ref{section:HTLpt} for more details on that point, keeping in mind that this overcounting is automatically taken care of from the next-to-leading order onwards in the HTLpt expansion.}. Finally, notice that the lattice data which are available at higher temperatures agree precisely with the central line of the DR result. This is, however, not the case with the exact one-loop HTLpt result whose band is much wider, even at lower temperatures.

We then take a closer look at Figure~\ref{fig:ChiB4Nf3}, where we display our resummed DR and exact one-loop HTLpt results for the fourth-order baryon number susceptibility $\chi_\trmi{B4}$, normalized to the appropriate Stefan-Boltzmann value $\chi_{\rm B4,SB}=2/(9\pi^2)$. We also show lattice data from the Wuppertal-Budapest (WB; green dots)~\cite{WupBudChi4andRatio} as well as the BNL-Bielefeld (BNL-B; black dots)~\cite{Bazavov} collaborations. Inspecting this quantity allows us to further compare our exact one-loop HTLpt prediction to the truncated three-loop results of~\cite{HTLptFiniteMUThreeLoop1}. We can then continue our investigation regarding the convergence of the HTLpt expansion as started with $\chi_\trmi{u2}$, and also gain more understanding about the physics involved. We note, however, that such a comparison is somewhat distorted by the fact that the three-loop band originates from a calculation involving mass truncation, which is nothing but a further approximation in the computation of the corresponding partition function. However, we shall see in Section~\ref{section:Convergence_Mass_Truncation} that the convergence of this approximation turns to be very good\footnote{In the present case, we are investigating the leading order HTLpt. See~\cite{SPTMassTruncation} for such an investigation up to three-loop order in the case of screened perturbation theory applied to a scalar $\lambda\,\phi^4$ theory.}, for all temperatures that we consider here. Our present comparison as well as the previous one regarding $\chi_\trmi{u2}$ rely on this argument.
%%%%%%%%%%%%%%%%%%%%%%%%%%%%%%%%%%%%%%%%%%%%%%%%%%%%%
%%%%%%%%%%%%%%%%%%%%%%%%%%%%%%%%%%%%%%%%%%%%%%%%%%%%%

Inspecing this figure, we notice that the central lines corresponding to the DR and three-loop HTLpt results are in good agreement with each other over the entire range of displayed temperatures. In addition, we see that the two lattice data sets overlap with these central lines, starting from $T\sim 350$ MeV onwards. We also note that the DR result is quite a bit more predictive than the three-loop HTLpt one, as demonstrated by the width of the corresponding bands. Another interesting aspect of this quantity is that the convergence of the successive orders in HTLpt is not as good as the convergence found for the second-order susceptibility. Indeed, the leading order HTLpt result of Figure~\ref{fig:ChiB4Nf3} completely misses the lattice estimates at all temperatures shown.

There could be different explanations for the poor convergence of the HTLpt expansion, but here we focus on the physical rather than a technical issues~\cite{sylvain2}. The difference between the quark and baryon number susceptibilities resides in off-diagonal contributions, which are expected to be numerically very small for the second-order susceptibility~\cite{BielefeldLatticeDataHighT}. However, in the case of the fourth-order susceptibility, the difference should not be negligible at all. Having said that, we further recall that the one-loop HTLpt result seems to agree quite well with lattice data\footnote{At least in the low temperature regime, above the transition region.} in the case of the fourth-order diagonal quark number susceptibility. Then, a possible physical reason for the difference between the convergence of the HTLpt expansions for the quark and baryon number susceptibilities might well be in the off-diagonal contributions that are completely missed by leading order HTLpt.  Although this reason seems attractive for our present problem, we would like to stress that drawing such a conclusion in a firm manner would require further investigations, which is however not in our scope.

\begin{figure}[!t]\centering\includegraphics[scale=0.3348]{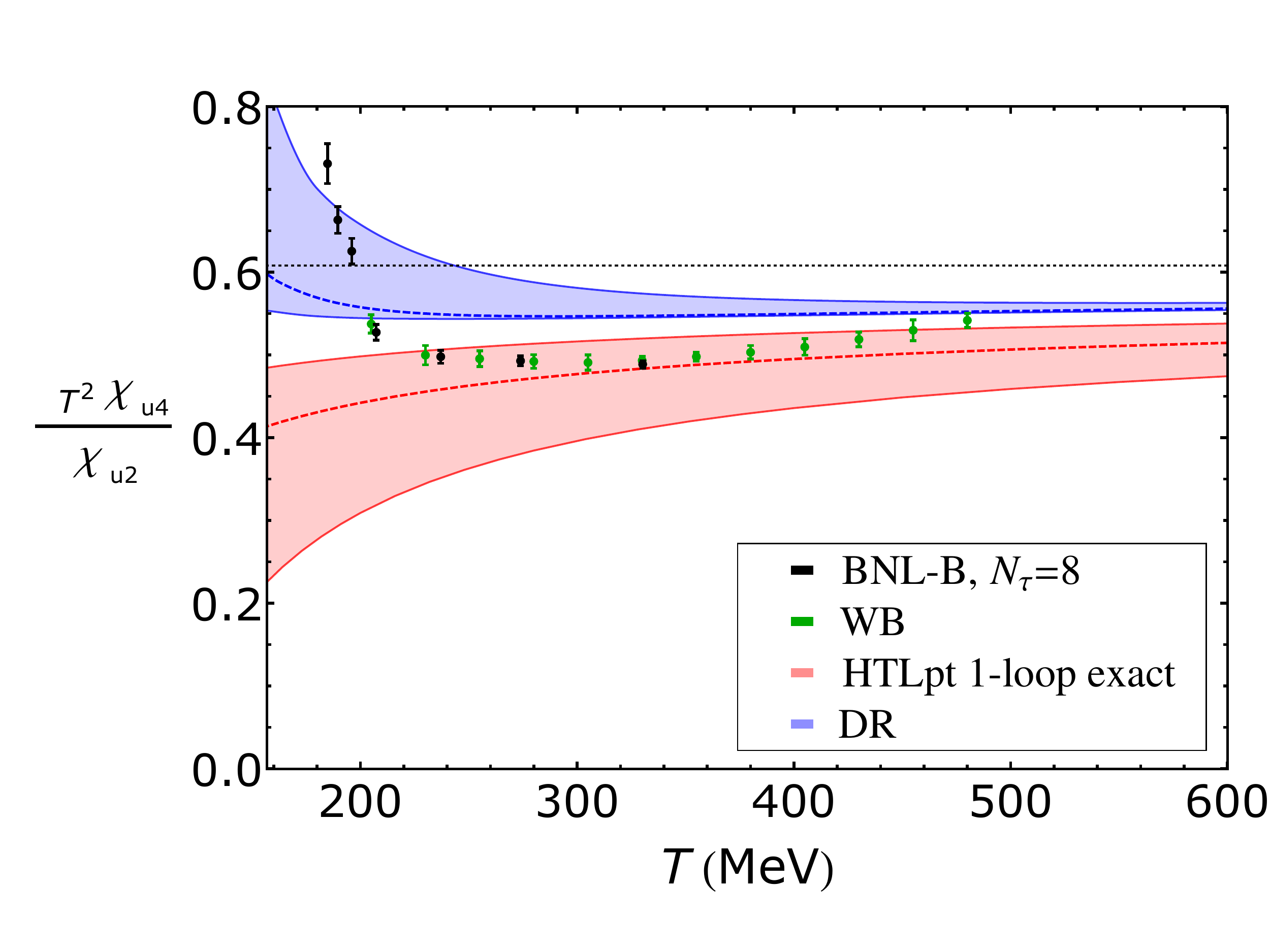}\!\!\!\!\includegraphics[scale=0.3380]{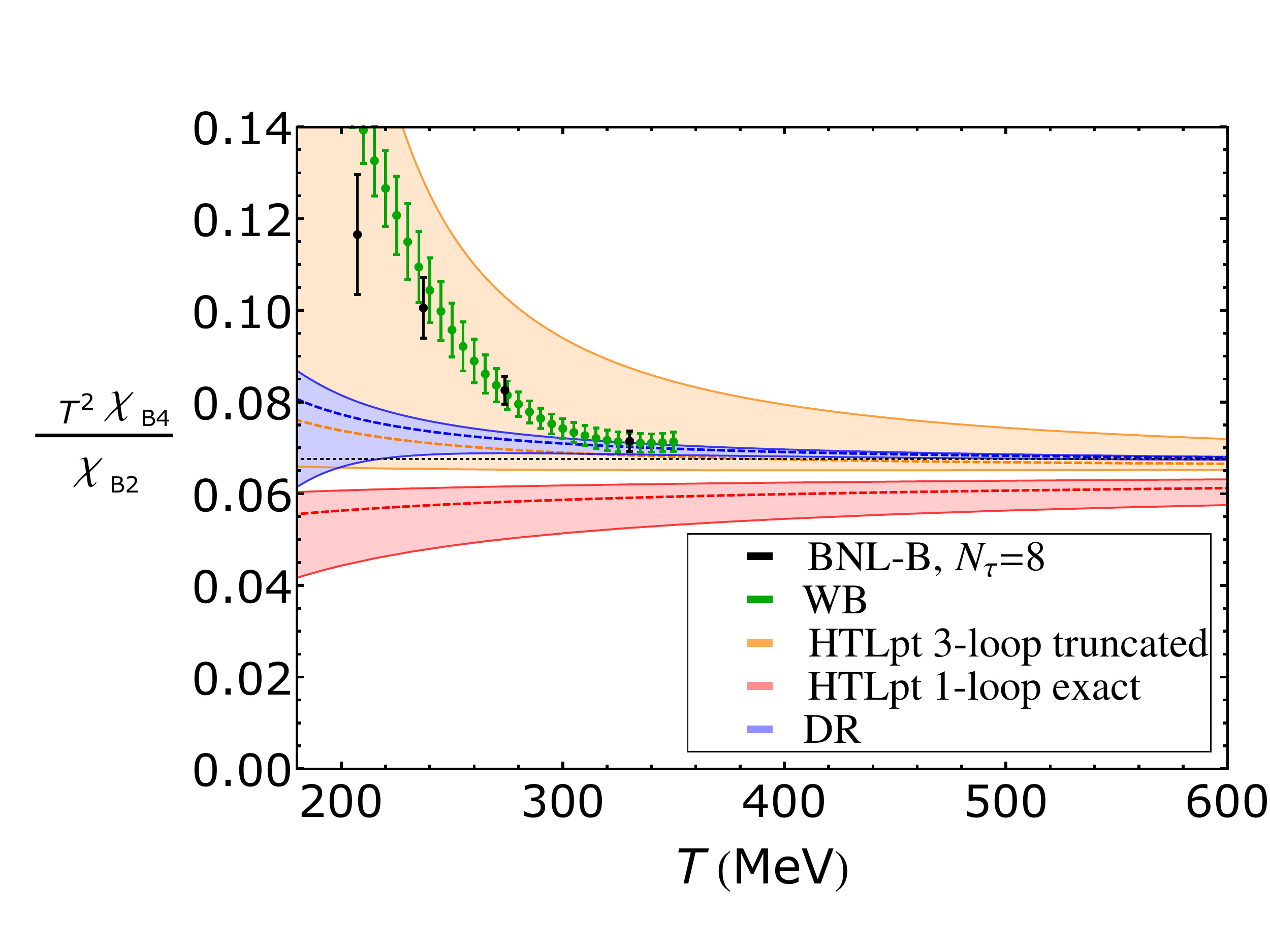}
\caption{Ratios of susceptibilities related to the quark (left) and baryon (right) numbers. The lattice data are taken from the BNL-Bielefeld~\cite{BielefeldLatticeData1,BielefeldLatticeData2} (BNL-B) and the Wuppertal-Budapest~\cite{WupBudChi4andRatio,RatioWuppBudBaryon} (WB) collaborations. Note that the three-loop HTLpt result has been obtained using the relevant susceptibilities from~\cite{HTLptFiniteMUThreeLoop1}. The black dashed straight lines denote the Stefan-Boltzmann values.}
\label{fig:Ratiosu4Overu2Andb4Overb2Nf3}
\end{figure}

Next, we proceed to the analysis of ratios of susceptibilities, related to observables such as the kurtoses. They are a measure of how strongly peaked a quantity can be, and are formally defined as the ratio of a fourth- with a second-order susceptibilities, the latter raised to the power two. However, for practical purposes when comparing our predictions with lattice data, we prefer to multiply our kurtoses by the corresponding second-order susceptibilities. This is indeed what is usually computed on the lattice, since such products are independent of the volume and thus easier to continuum extrapolate\footnote{The limit of the lattice spacing going to zero still has to be taken, as cut-off effects do not vanish for such quantities. However, the finiteness of the volume does not influence those quantities.}. Nevertheless, for thermodynamical purposes, such quantities are of primary interest as they are known to measure how fast a phase transition is\footnote{See~\cite{StephanovKurtosis} for discussion about the behavior of the kurtosis as the order parameter of the deconfinement phase transition near the critical point.}. In both Figures~\ref{fig:Ratiosu4Overu2Andb4Overb2Nf3}, we then display our results for this ratio in the case of the quark (left) and baryon (right) number susceptibilities.
%%%%%%%%%%%%%%%%%%%%%%%%%%%%%%%%%%%%%%%%%%%%%%%%%%%%%
%%%%%%%%%%%%%%%%%%%%%%%%%%%%%%%%%%%%%%%%%%%%%%%%%%%%%

More precisely, in Figure~\ref{fig:Ratiosu4Overu2Andb4Overb2Nf3} (left), we see our DR and exact one-loop HTLpt results together with lattice data originating from the BNL-Bielefeld~\cite{BielefeldLatticeData1,BielefeldLatticeData2} (BNL-B; black dots) and the Wuppertal-Budapest~\cite{WupBudChi4andRatio,RatioWuppBudBaryon} (WB; green dots) collaborations. At temperatures around $T\approx 300\,\mbox{--}\,400$ MeV, the lattice data agree better with the one-loop HTLpt band while it seems to approach the DR prediction at higher temperatures. The resummed DR result also tends to reproduce the qualitative features of the lattice data. Regarding Figure~\ref{fig:Ratiosu4Overu2Andb4Overb2Nf3} (right), both the DR and the exact one-loop HTLpt results show similar trends as in the previous figure. Only, this time, the DR prediction converges very quickly towards the non-interacting limit, and seems to meet the displayed lattice data again from the BNL-Bielefeld~\cite{BielefeldLatticeData1} (BNL-B; black dots) and the Wuppertal-Budapest~\cite{RatioWuppBudBaryon} (WB; green dots) collaborations. In the right figure, we  also show the corresponding truncated three-loop HTLpt result, as obtained from the susceptibilities of~\cite{HTLptFiniteMUThreeLoop1}. Despite the relative thickness of the band at low temperature, this result seems to reproduce the qualitative trend of the lattice data quite well. It also agrees with the DR prediction in a satisfactory way, as the respective central lines are very close to each other.

\begin{figure}[!t]\centering\includegraphics[scale=0.4264]{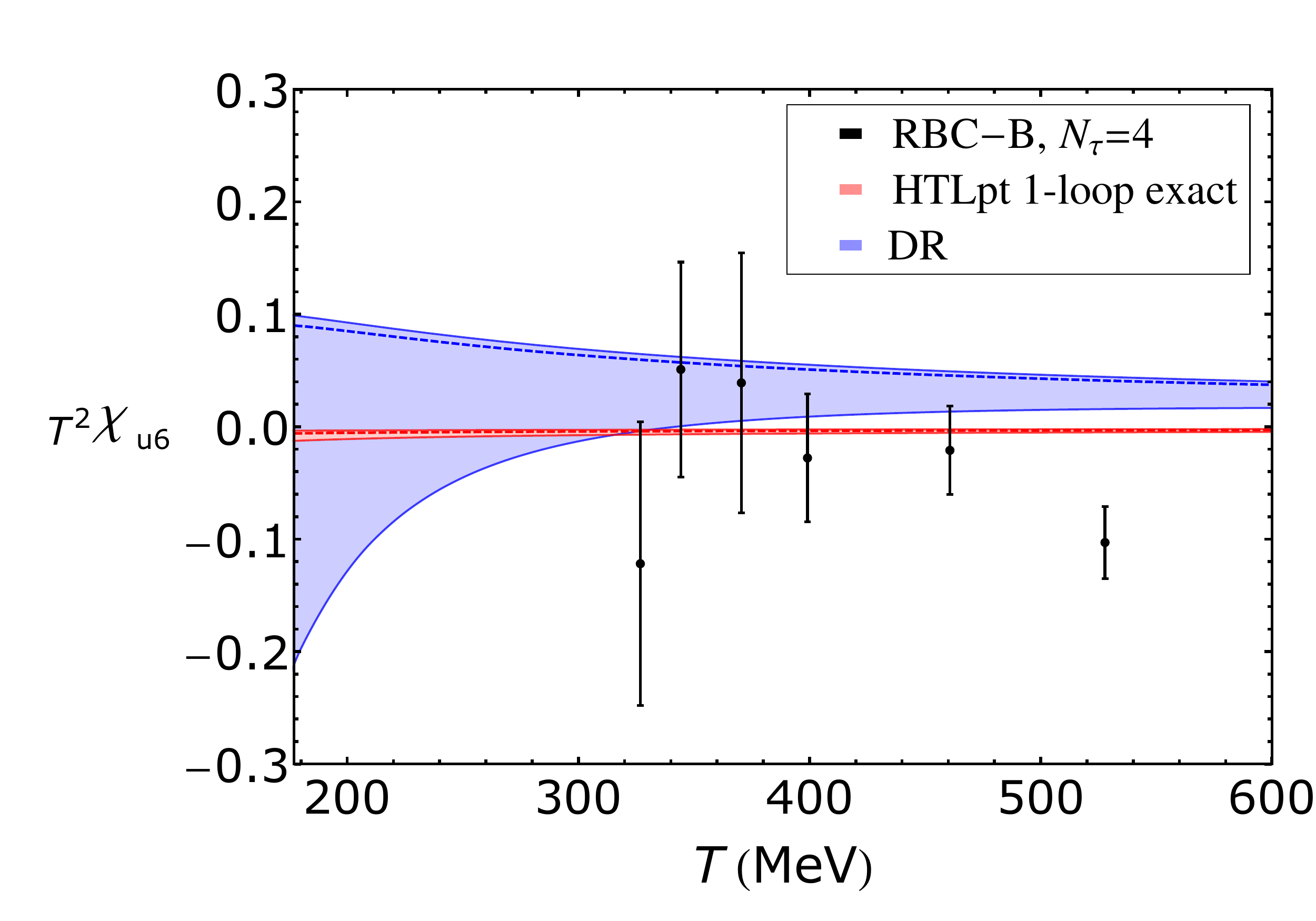}
\caption{Rescaled three flavor sixth-order diagonal quark number susceptibility. The lattice data have been taken from the RBC-Bielefeld~\cite{Peterchi6} (RBC-B) collaboration.}
\label{fig:Chiu6Nf3}
\end{figure}

Before we comment on the finite density part of the equation of state, let us have a look at some higher order cumulants. In Figure~\ref{fig:Chiu6Nf3}, we display the sixth-order diagonal quark number susceptibility. We should point out that such a high order in the susceptibilities, or equivalently in the cumulants of the partition function, is expected to be a very sensitive probe of the hadronic freeze-out~\cite{KarschFluctu}. From a purely perturbative point of view, we know that the two first orders in the weak coupling series vanish for such a quantity. Hence, the first non zero term originates from the so-called plasmon contribution at ${\cal O}(g^3)$. As a consequence, the convergence properties of the series become much poorer than in the cases of the second- and fourth-order cumulants. We nevertheless notice that our DR prediction remains positive for all temperatures\footnote{Interestingly for this quantity, the leading plasmon contribution is negative while it can be seen from a truncation of our DR results at various orders, that the sign of the result becomes positive thanks to higher order contributions.}, while the exact one-loop HTLpt result is consistently negative. The latter apparently agrees better with the non continuum extrapolated lattice data of the RBC-Bielefeld~\cite{Peterchi6} (RBC-B; black dots) collaboration. The fact that these lattice data are not continuum extrapolated renders the comparison with our predictions more difficult. It is therefore not yet possible to draw a firm conclusion about whether the sixth-order cumulant could be positive or negative in this range of temperatures. As a matter of fact, our HTLpt result being the leading order one\footnote{Note that such a quantity starts at order $g^3$, and the usual low loop order HTLpt $g^2$ overcounting is then absent.}, we further refer the reader to~\cite{HTLptFiniteMUThreeLoop2} for a recent truncated three-loop HTLpt calculation concerning this quantity. Thus, it should be noted by comparing Figure~\ref{fig:Chiu6Nf3} with the results of ~\cite{HTLptFiniteMUThreeLoop2} that our four-loop DR prediction is actually in agreement with the three-loop HTLpt one.
%%%%%%%%%%%%%%%%%%%%%%%%%%%%%%%%%%%%%%%%%%%%%%%%%%%%%
%%%%%%%%%%%%%%%%%%%%%%%%%%%%%%%%%%%%%%%%%%%%%%%%%%%%%

In general, lattice studies seem to favor negative values for the sixth-order susceptibilities~\cite{HandLatticeDatasu6u2,KarschFluctu,Peterchi6} right above the transition region. Thus, it could be that by including more perturbative orders to our state-of-the-art DR prediction, it would eventually flip its sign. This is, however, unlikely to be verified in the near future, considering the enormous technical challenge involved in a complete five-loop determination of such a quantity.

\begin{figure}[!t]\centering\includegraphics[scale=0.445]{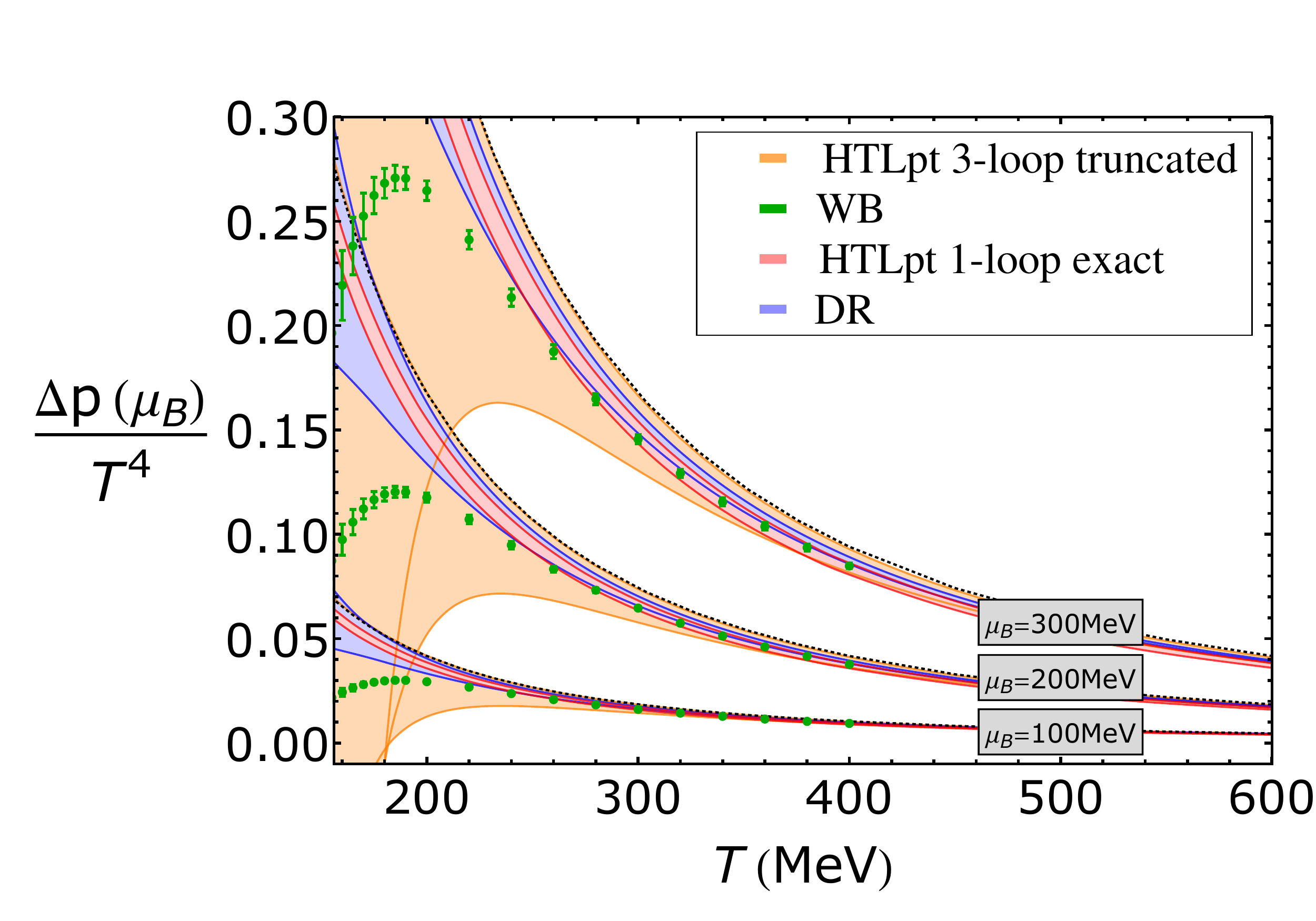}
\caption{Difference between the pressure evaluated at nonvanishing and at zero chemical potentials in the case of three flavors. Notice that the dashed lines correspond to the respective Stefan-Boltzmann values. The three-loop HTLpt results are taken from~\cite{HTLptFiniteMUThreeLoop1} and the lattice data from the Wuppertal-Budapest~\cite{BorsanyiDeltaP} (WB) collaboration.}
\label{fig:DeltaPressureNf3}
\end{figure}

Finally, we turn to inspect the finite density part of the equation of state, which is equivalent to summing the susceptibilities to all orders with the appropriate powers of chemical potentials as coefficients. We then show, in Figure~\ref{fig:DeltaPressureNf3}, the difference between the pressure at zero and nonzero density, given identical quark chemical potentials so that $\mu_\trmi{B}=100$, $200$, and $300$ MeV. The continuum ``estimated'' lattice data from the Wuppertal-Budapest (WB; green dots)~\cite{BorsanyiDeltaP} collaboration are based on an expansion of the pressure through ${\cal O}(\mu_f^2/T^2)$, while the perturbative results are accurate to all orders in $\mu_f/T$, up to possible restrictions coming from assumptions inherent to the resummation frameworks\footnote{In the DR case, the results are valid as long as $\pi T \gtrsim g\mu_f$ is satisfied~\cite{IppAllMuPressure}, which is the case in the present analysis.}.
%%%%%%%%%%%%%%%%%%%%%%%%%%%%%%%%%%%%%%%%%%%%%%%%%%%%%
%%%%%%%%%%%%%%%%%%%%%%%%%%%%%%%%%%%%%%%%%%%%%%%%%%%%%
As expected based on an earlier analysis~\cite{AleksiFirstPaperPressure}, we notice a good agreement between our results and the lattice data down to about $T\approx 250$ MeV. Note also that our DR and exact one-loop HTLpt bands are clearly distinct from the corresponding Stefan-Boltzmann values in addition to being quite narrow, which is not the case for the truncated three-loop HTLpt prediction. The latter, however, reproduces the qualitative trend of the lattice data rather well.

\section{The two flavor case}\label{section:Two_flavors}

\begin{figure}[!t]\centering\includegraphics[scale=0.417]{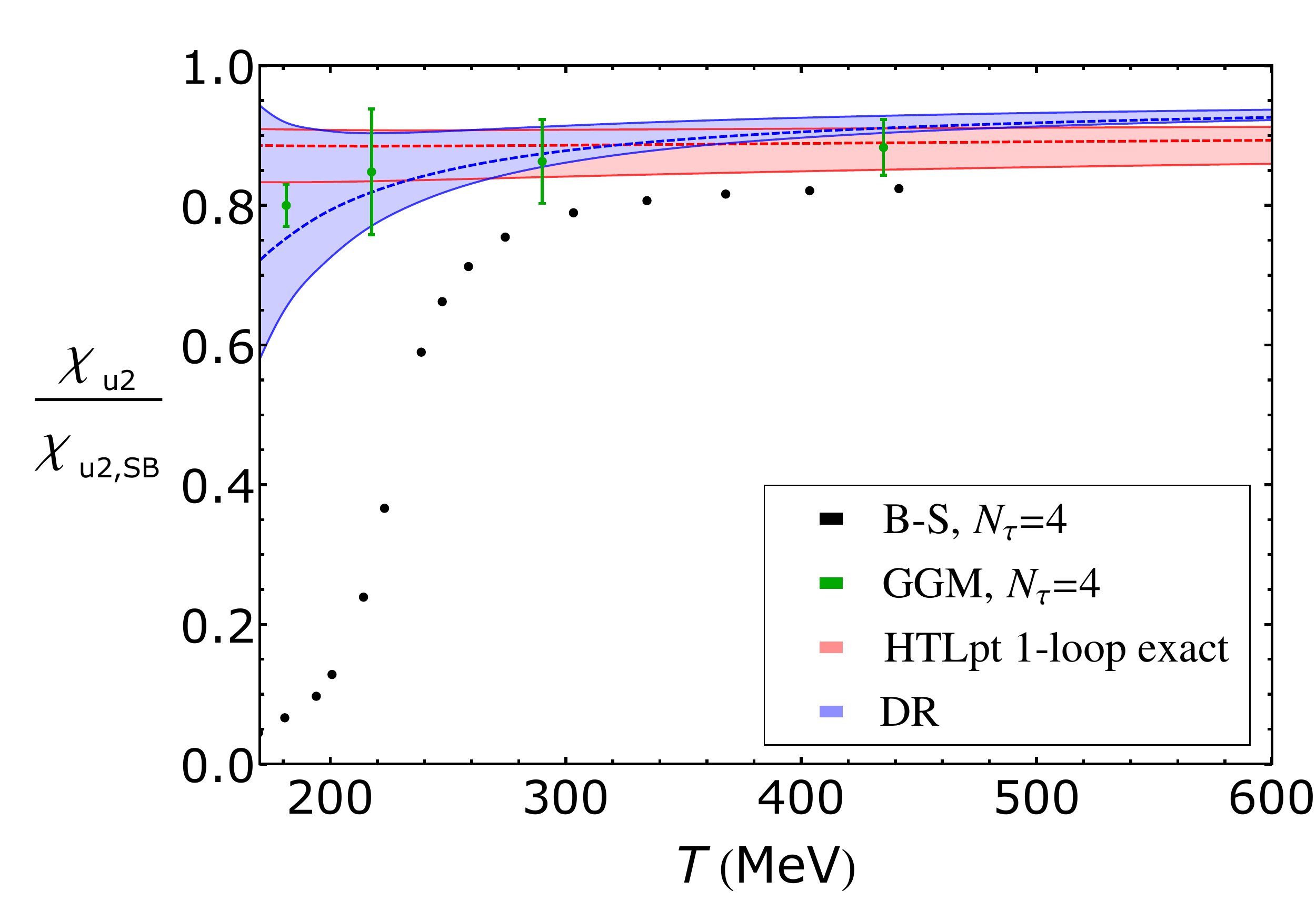}
\caption{Two flavor second-order diagonal quark number susceptibility normalized to its Stefan-Boltzmann value. The lattice data are from the Bielefeld-Swansea~\cite{HandLatticeDatasu6u2} (B-S) collaboration, and from~\cite{GuptaLatticeData} (GGM).}
\label{fig:Chiu2Nf2}
\end{figure}

We are now going to take a look at the two flavor case, motivated be the great number of related lattice studies (see e.g.~\cite{CPPACSCollaboration2FlavorQCDref1,SwanseaBielefeldCollaboration2FlavorQCDref2} for a few examples). We begin this by investigating the behavior of the second-order diagonal quark number susceptibility, normalized to its non-interacting limit and plotted in Figure~\ref{fig:Chiu2Nf2}. We notice that despite the fact that the general trend of our results is very similar to that of the three flavor case, both bands reside lower and are now slightly wider. In the same figure, we also display the lattice results from the Bielefeld-Swansea~\cite{HandLatticeDatasu6u2} (B-S; black dots) collaboration as well as results by Gavai, Gupta and Majumdar~\cite{GuptaLatticeData} (GGM; green dots). Both sets of lattice data have been obtained using $N_\tau=4$ lattices, in units of the corresponding critical temperatures. One of the difficulties in transferring such results to physical units is then the determination of critical temperatures appropriate for the used lattice settings. Regarding~\cite{GuptaLatticeData}, the problem is easily resolved by using the pseudo-critical temperature $T_\trmi{c}=145$ MeV, quoted in the very same reference. Regarding the B-S result, which was obtained using p4-improved staggered fermions, we must apply more care and therefore choose to follow~\cite{Peikert1,Peikert2}, which leads us to $T_\trmi{c}=223$ MeV. As both of these values deviate from the correct pseudo-critical temperature\footnote{Recall that in the chiral limit, the critical temperature is $T_\trmi{c}=173$ MeV~\cite{AbsoluteScale}.} in opposite directions, the lattice data points displayed in our figure have to be interpreted with some caution.
%%%%%%%%%%%%%%%%%%%%%%%%%%%%%%%%%%%%%%%%%%%%%%%%%%%%%
%%%%%%%%%%%%%%%%%%%%%%%%%%%%%%%%%%%%%%%%%%%%%%%%%%%%%
We see that both the exact one-loop HTLpt and DR bands agree very well with each other over the entire range of temperatures. However, unlike in the three flavor case, the agreement between our results and the B-S lattice data is not quite optimal. This may well have to do with the discrepancy in the values of $T_\trmi{c}$ that we use in order to convert the lattice data to physical units. The DR band in any case reproduces the qualitative trend of the B-S lattice data, and thus an alternative value of the critical temperature -- closer to the physical point -- would most likely improve the agreement between our results and the lattice data. Hence, although naively our perturbative results seem to agree better with~\cite{GuptaLatticeData}, it is difficult to draw firm conclusions without continuum extrapolated results.

\begin{figure}[!t]\centering\includegraphics[scale=0.437]{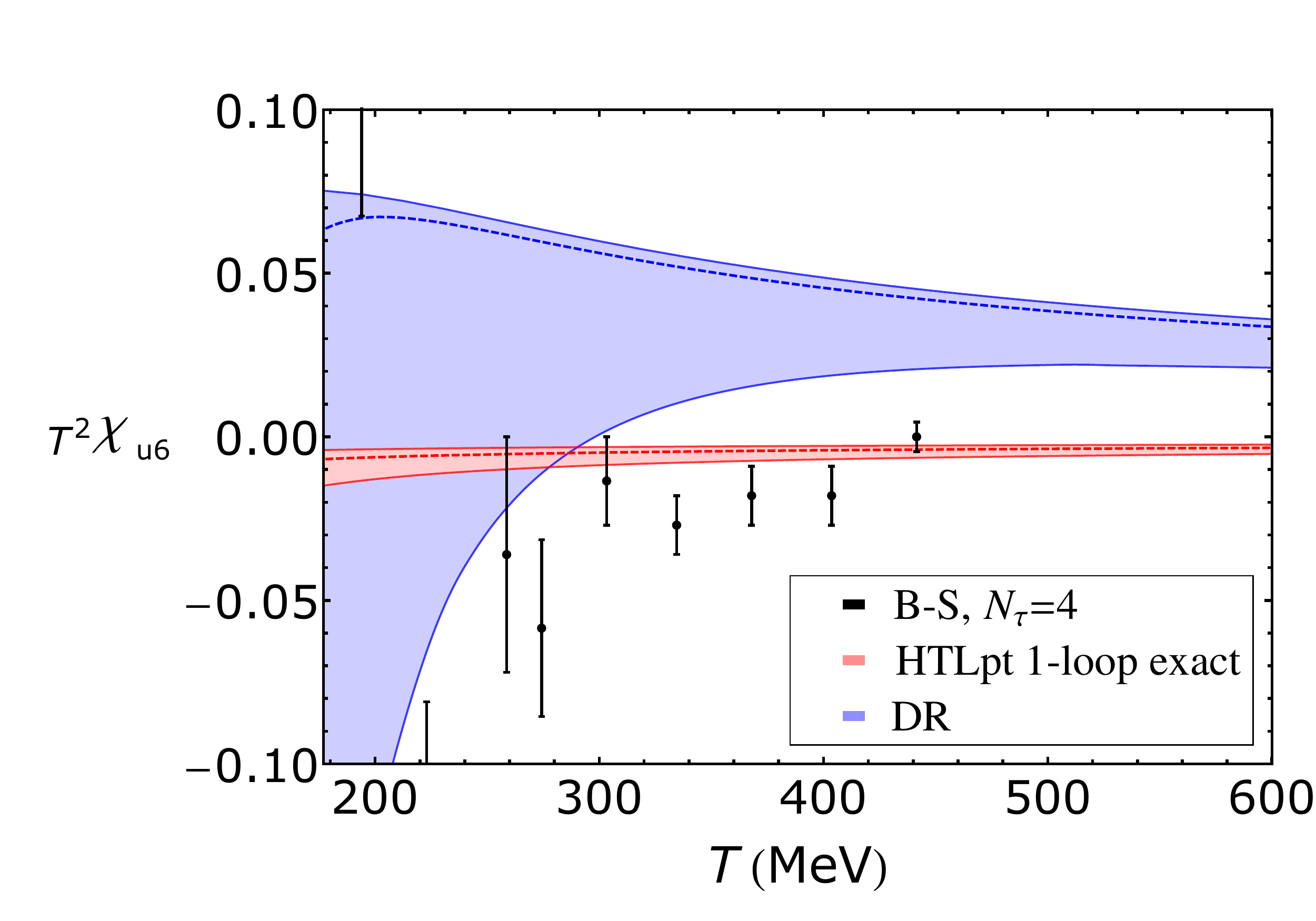}
\caption{Rescaled two flavor sixth-order diagonal quark number susceptibility. The lattice data set originates from the Bielefeld-Swansea~\cite{HandLatticeDatasu6u2} (B-S) collaboration.}
\label{fig:CHIu6Nf2}
\end{figure}

Moving on to higher orders in the susceptibilities, we show in Figure~\ref{fig:CHIu6Nf2} the sixth-order diagonal quark number susceptibility together with again lattice data from the Bielefeld-Swansea~\cite{HandLatticeDatasu6u2}(B-S; black dots) collaboration. As in the three flavor case, our exact one-loop HTLpt and DR results are seen to disagree.
%%%%%%%%%%%%%%%%%%%%%%%%%%%%%%%%%%%%%%%%%%%%%%%%%%%%%
%%%%%%%%%%%%%%%%%%%%%%%%%%%%%%%%%%%%%%%%%%%%%%%%%%%%%
Recalling the corresponding discussion in the three flavor case, this feature should not be surprising as both perturbative results start at order $g^3$, making them sensitive to higher order corrections. Although the B-S lattice results are still not continuum extrapolated, the lattice data points seem to suggest negative values for this quantity at these temperatures. This leads to a seemingly better agreement between our exact one-loop HTLpt prediction and the lattice result. The DR band, on the other hand, remains positive for most of the temperatures\footnote{Having a central line which even stands on the upper part of the band.}. Note that based on the previous encouraging comparisons between our DR predictions and the recent truncated three-loop HTLpt ones, it is likely that HTLpt would again agree with the DR results at three-loop level.

\begin{figure}[!t]\centering\includegraphics[scale=0.445]{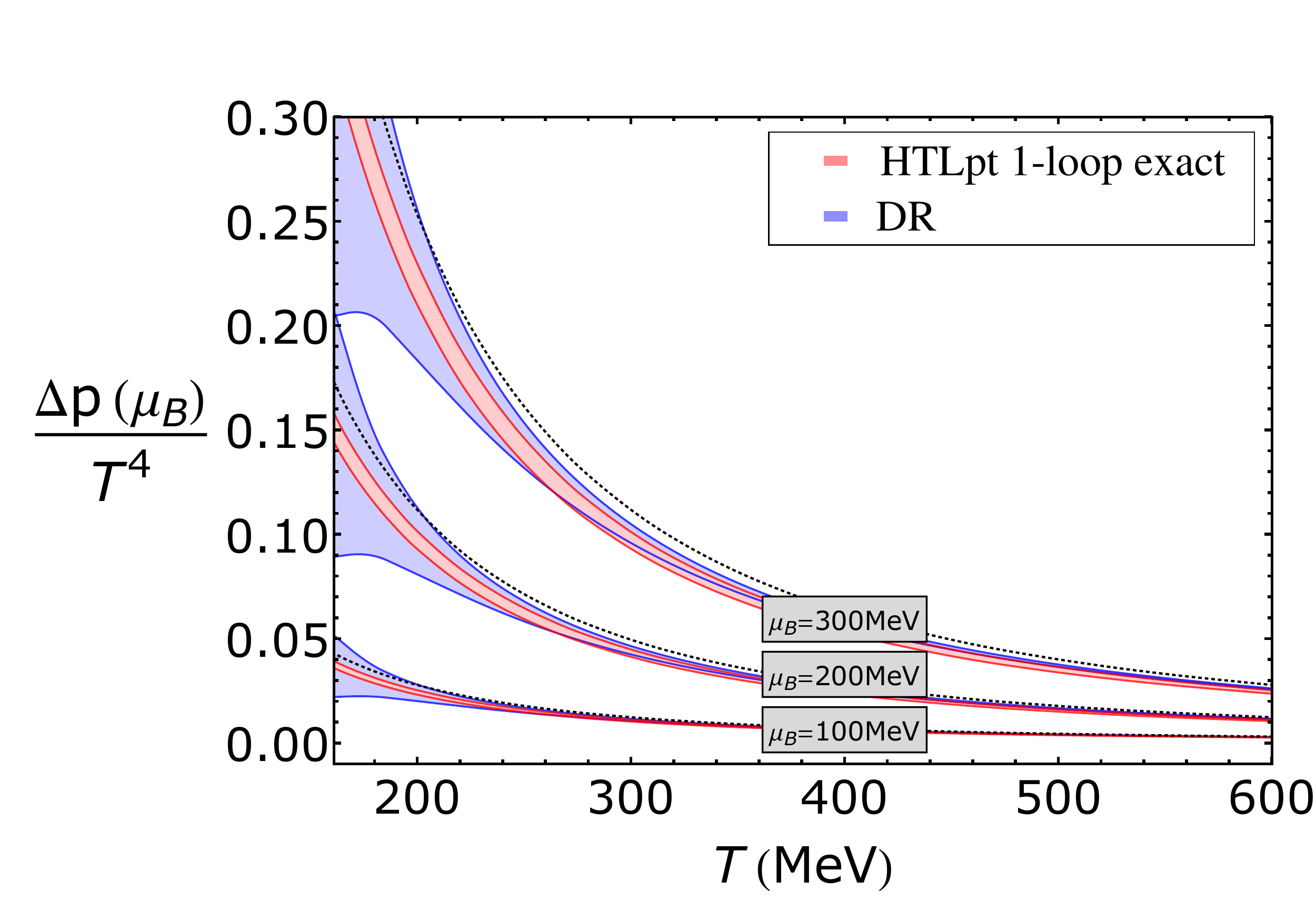}
\caption{Difference between the pressure evaluated at nonvanishing and at zero chemical potentials in the case of two flavors. The dashed lines correspond to the respective Stefan-Boltzmann values.}
\label{fig:DeltaPressureNf2}
\end{figure}

We finally investigate the full finite density behavior of the equation of state. As we saw in the three flavor case, this can be done by inspecting the difference between the pressure at nonzero density and the pressure with vanishing chemical potentials. In Figure~\ref{fig:DeltaPressureNf2}, we then display our exact one-loop HTLpt and four-loop resummed DR results. Once again, both perturbative results are in good agreement with each other. However, while the HTLpt bands are quite narrow, the DR ones are seen to widen at lower temperatures in a more pronounced way than in the three quark flavor case. We again notice that both predictions differ in a noticeable way from the non-interacting limits. Finally, based on the previous analyses, it is likely that at three-loop order the HTLpt framework would also agree with our resummed DR predictions.
%%%%%%%%%%%%%%%%%%%%%%%%%%%%%%%%%%%%%%%%%%%%%%%%%%%%%
%%%%%%%%%%%%%%%%%%%%%%%%%%%%%%%%%%%%%%%%%%%%%%%%%%%%%
We would like to refer the reader to~\cite{HandLatticeDatasu6u2,SwanseaBielefeldCollaboration2FlavorQCDref2} for extensive lattice studies of the finite density behavior of the equation of state for two flavors.

\section{Convergence of the mass expansion}\label{section:Convergence_Mass_Truncation}

Last but not least, we plan to investigate in a quantitative manner the convergence aspects of the mass expansion in our one-loop HTLpt calculation. We refer to Section~\ref{section:Truncated_one-loop_HTLpt} for more details about its implementation. In this section, we specialize to the three flavor case only, and first analyze the pressure at zero chemical potentials, after which we comment on the second- and fourth-order diagonal quark number susceptibilities. More precisely, we will plot all of these quantities as functions of the one-loop running coupling $g(\bar{\Lambda}_{\rm central})$ at the central value of the renormalization scale. All the results will then be normalized by their corresponding non interacting limits. Note that the bands are now generated by varying the renormalization scale $\lmsb$ only, by a factor of two around the central value and that we show truncations at different orders in $m/T\sim g$, as explained in the legends.

\begin{figure}[!t]\centering\includegraphics[scale=0.4045]{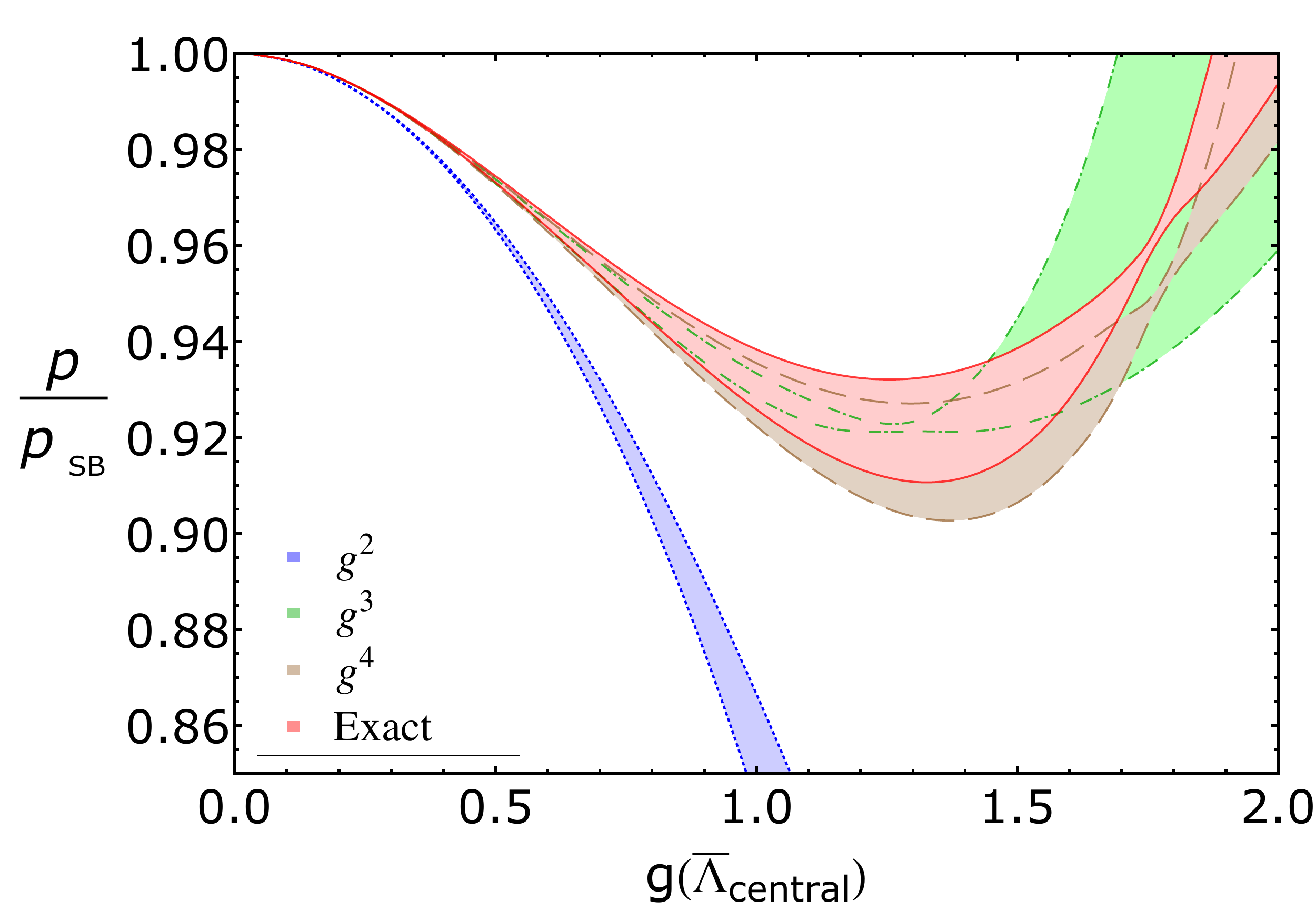}
\caption{Three flavor zero density leading order HTLpt pressure, normalized to its non-interacting limit, as a function of the coupling $g(\bar{\Lambda}_{\rm central})$. We see the exact band (red; solid), as well as various truncations at orders $g^2$ (blue; dotted), $g^3$ (green; dot-dashed), $g^4$ (brown; dashed).}
\label{PressureConv}
\end{figure}

In Figure~\ref{PressureConv}, we display the zero density pressure and see that with the exception of the most naive $g^2$ order\footnote{Which also contains the well known low loop order overcounting that appears in HTLpt.}, the truncated results are not only in good agreement with each other, but also with the exact non truncated band.
%%%%%%%%%%%%%%%%%%%%%%%%%%%%%%%%%%%%%%%%%%%%%%%%%%%%%%%%%%%%%%%%%%%%%%%%%%%%%
%%%%%%%%%%%%%%%%%%%%%%%%%%%%%%%%%%%%%%%%%%%%%%%%%%%%%%%%%%%%%%%%%%%%%%%%%%%%%
At values of $g$ relevant to our study\footnote{See the previous section, keeping in mind that we work with the one-loop perturbative running.}, the differences between all the bands are of the order of one per cent, which is indeed a positive fact. Note also that for values of $g$ below $1.5$, the widths of the bands become larger as we go to higher orders in the $m/T$ expansion. In addition, we note that the truncated series overlaps, already at order $g^4$, with the exact band to a large extent. This is a very encouraging outcome.

\begin{figure}[!t]\centering\includegraphics[scale=0.415]{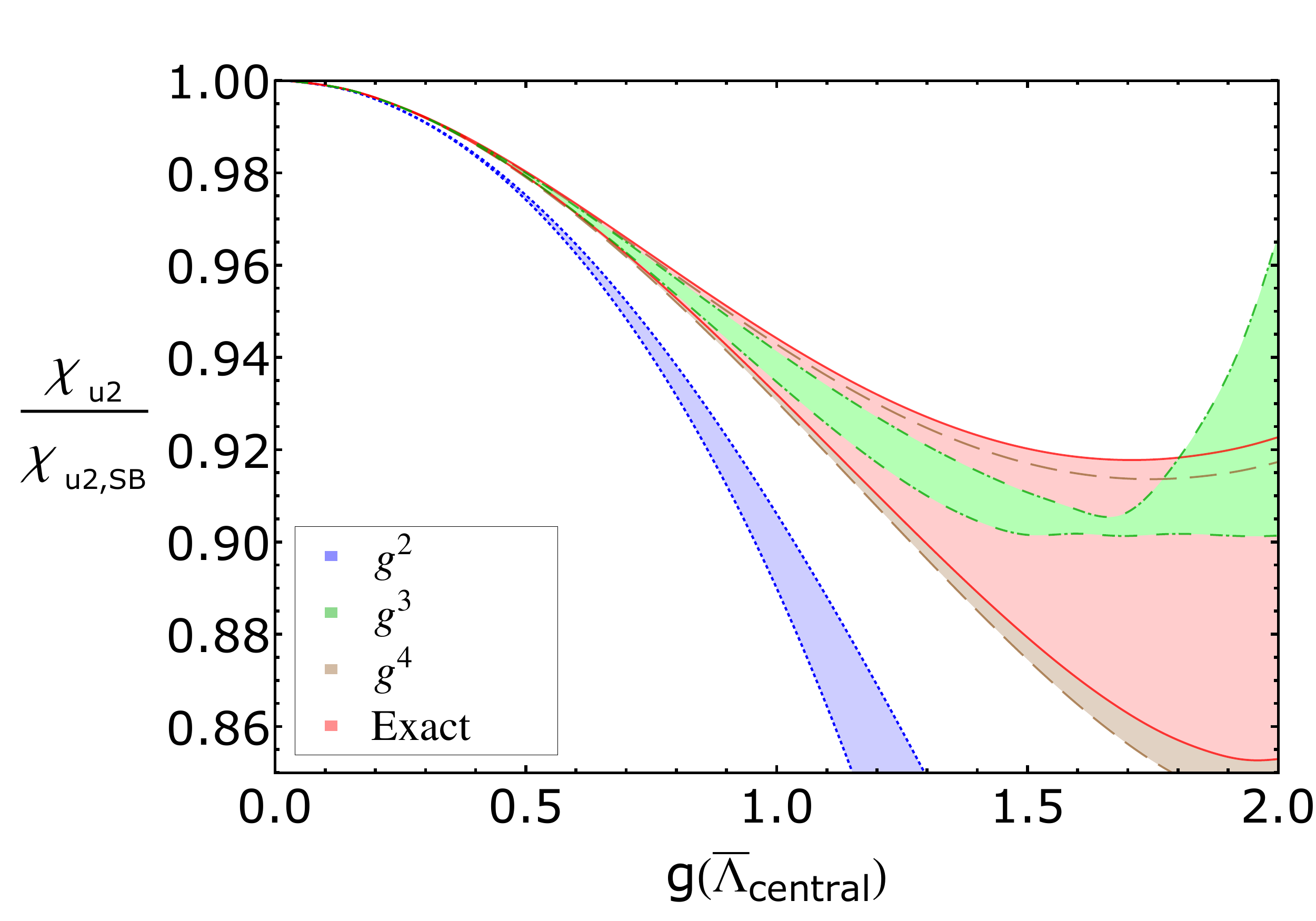}
\caption{Three flavor second-order diagonal quark number susceptibility, normalized to its Stefan-Boltzmann value, as a function of the coupling $g(\bar{\Lambda}_{\rm central})$. We show the exact band (red; solid), as well as various truncations at orders $g^2$ (blue; dotted), $g^3$ (green; dot-dashed), $g^4$ (brown; dashed).}
\label{Chiu2Conv}
\end{figure}

Let us next investigate, whether the convergence of the truncation might change at finite density by studying the second- and fourth-order diagonal quark number susceptibilities. These can be straightforwardly obtained from the leading order mass truncated expression for the pressure~(\ref{HighTHTLptLeadingOrderPressure}) by means of derivatives with respect to the chemical potentials. Note that through ${\cal O}(g^5)$, the second-order diagonal susceptibility reads
\beqa
\frac{\chi_{\rm u2}}{\chi_{\rm u2,SB}}&=&1-\frac{3\,d_\trmi{A}}{8\,N_\trmi{c}\,\pi^2}\,\,g^2+d_\trmi{A}\left(1+\frac{N_\trmi{f}}{2\,N_\trmi{c}}\right)^{1/2}\frac{\sqrt{3/N_\trmi{c}}}{8\,\pi^3}\,\,\,g^{3}+\frac{d_\trmi{A}}{32\,\pi^4}\left(1+\frac{N_\trmi{f}}{2\,N_\trmi{c}}\right) \nonumber \\
&&\ \times\Bigg[\frac{\pi^2}{3}-\frac{7}{2}+\gamma_\trmi{\tiny E}+\log\frac{\lmsb}{4\,\pi\,T}+\frac{d_\trmi{A}\,\left(6-\pi^2\right)}{4\,N_\trmi{c}\,\left(2\,N_\trmi{c}+N_\trmi{f}\right)}\,\Bigg]\,g^{4}+{\cal O}(g^6)\,.
\eeqa
This quantity is plotted to various orders in Figure~\ref{Chiu2Conv}.
%%%%%%%%%%%%%%%%%%%%%%%%%%%%%%%%%%%%%%%%%%%%%%%%%%%%%%%%%%%%%%%%%%%%%%%%%%%%%
%%%%%%%%%%%%%%%%%%%%%%%%%%%%%%%%%%%%%%%%%%%%%%%%%%%%%%%%%%%%%%%%%%%%%%%%%%%%%
We see that the bands for the orders $g^3$ and $g^4$ are in good agreement with each other, as well as with the exact result, while the order $g^2$ is again seen to be off the exact band. We in particular observe that the agreement between the order $g^4$ and the exact result is much faster than in Figure~\ref{PressureConv}.

Finally, we turn to the fourth-order diagonal quark number susceptibility to assess its convergence. The truncated expression for this quantity reads
\beqa
\frac{\chi_{\rm u4}}{\chi_{\rm u4,SB}}&=&1-\frac{3\,d_\trmi{A}}{8\,N_\trmi{c}\,\pi^2}\,\,g^2+3\,d_\trmi{A}\,\frac{N_\trmi{f}}{N_\trmi{c}}\left(1+\frac{N_\trmi{f}}{2\,N_\trmi{c}}\right)^{-1/2}\frac{\sqrt{3/\,N_\trmi{c}}}{32\,\pi^3}\,\,g^{3}+\frac{3\,d_\trmi{A}}{32\,\pi^4} \nonumber \\
&&\ \times\left(\frac{N_\trmi{f}}{2\,N_\trmi{c}}\right)\Bigg[\frac{\pi^2}{3}-\frac{7}{2}+\gamma_\trmi{\tiny E}+\log\frac{\lmsb}{4\,\pi\,T}+\frac{d_\trmi{A}\,\left(6-\pi^2\right)}{12\,N_\trmi{c}\,N_\trmi{f}}\,\Bigg]\,g^{4}+{\cal O}(g^6)\,. \ \ \ \ \ \ \ \ 
\eeqa
Naively, we expect very similar  convergence properties with the second-order susceptibility. However, this time, we observe that the $g^4$ truncation almost coincides with the exact band, the difference between the two results being not even visible.
%%%%%%%%%%%%%%%%%%%%%%%%%%%%%%%%%%%%%%%%%%%%%%%%%%%%%%%%%%%%%%%%%%%%%%%%%%%%%
\begin{figure}[!t]\centering\includegraphics[scale=0.415]{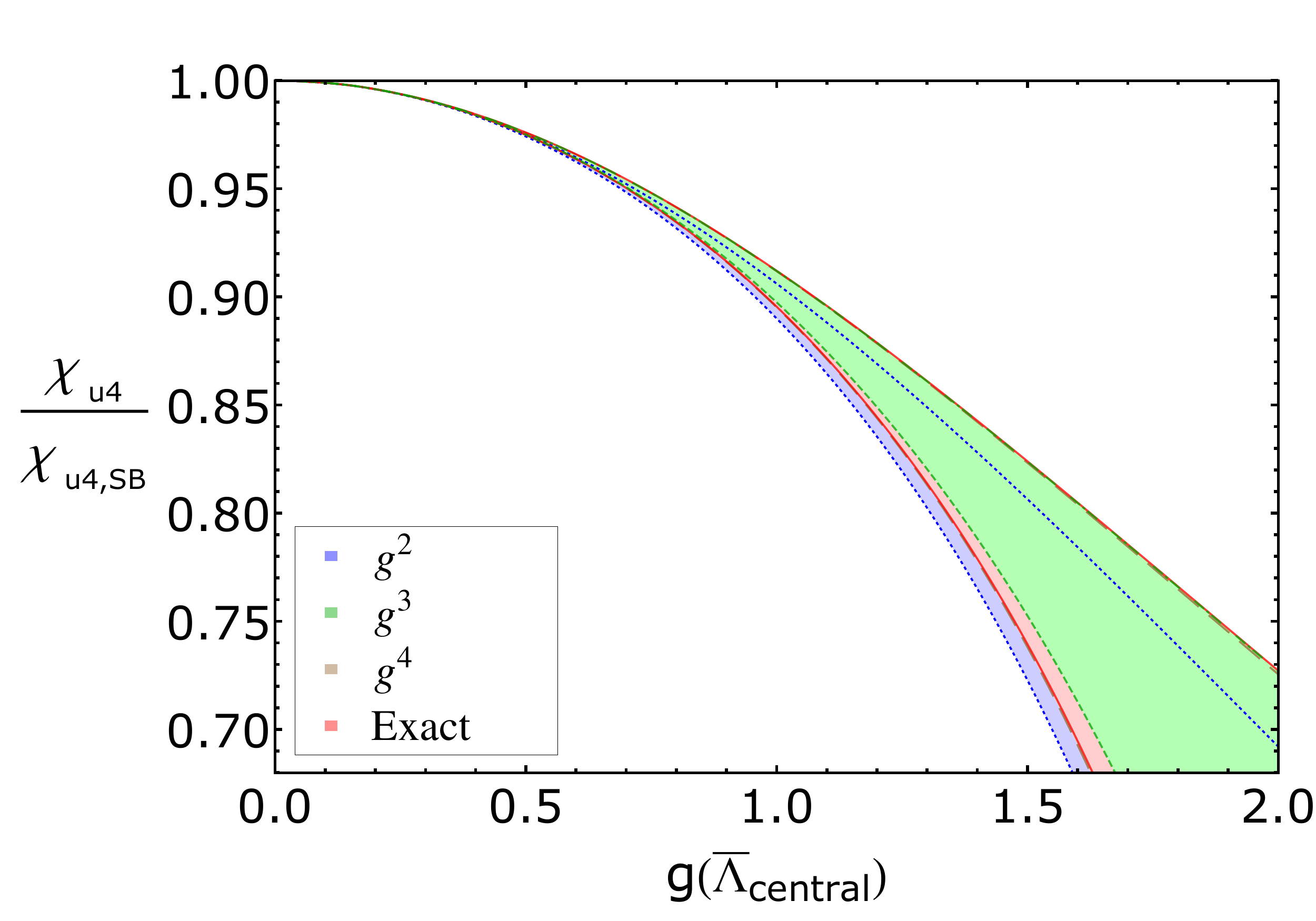}
\caption{Three flavor fourth-order diagonal quark number susceptibility, normalized to its non-interacting limit, as a function of the coupling $g(\bar{\Lambda}_{\rm central})$. Displayed are the exact band (red; solid), as well as various truncations at orders $g^2$ (blue; dotted), $g^3$ (green; dot-dashed), $g^4$ (brown; dashed).}
\label{Chiu4Conv}
\end{figure}
%%%%%%%%%%%%%%%%%%%%%%%%%%%%%%%%%%%%%%%%%%%%%%%%%%%%%%%%%%%%%%%%%%%%%%%%%%%%%
Consequently, it turns out that the finite density aspects of the truncation method shows even better convergence signs than for the zero density pressure itself.

Finally, we would like to point out that the convergence properties of the mass expansion of the one-loop HTLpt pressure is in agreement with similar observations made in the case of a scalar $\lambda\,\phi^4$ theory using screened perturbation theory~\cite{SPTMassTruncation}. The latter, however, has been carried out through three-loop order, and it was pointed out that the convergence slightly worsens as one proceeds to higher loop orders. Despite this additional observation, thanks to the promising results of Figures~\ref{PressureConv},~\ref{Chiu2Conv} and~\ref{Chiu4Conv}, it seems this method is in general a good approximation for HTLpt calculations.

\chapter{Conclusions and outlook}\label{chapter:Conclusions}

This Ph.D. thesis has dealt with the thermodynamics of the quark-gluon plasma at nonzero density,  and is based on the recently published papers~\cite{sylvain1,sylvain2} and proceedings~\cite{sylvain3,sylvain4}. We started by introducing the general framework of our work, from two different but complementary point of views: In Chapter~\ref{chapter:Aspects}, we introduced the topic from a field theoretic point of view, while in Chapter~\ref{chapter:Thermodynamics_Hot_Dense} we approached it from thermodynamics. Chapter~\ref{chapter:Resummed_Perturbative_QCD} was on the other hand devoted to introducing the computational setups that we used, and in Chapter~\ref{chapter:Exact_LO_HTLpt}, we gave some further details about the computation of the one-loop HTLpt pressure. Finally, in Chapter~\ref{chapter:Results}, we proceeded to our main goal and analyzed the thermodynamics of hot and dense quark-gluon plasma, with emphasis on quantities sensitive to finite quark number density.

The main aspect of our work was to compute the pressure of deconfined quark-gluon plasma at high temperature and nonzero density by means of two different resummed perturbative frameworks. The first one, known as hard-thermal-loop perturbation theory (HTLpt), was employed at  one-loop order, while the second one, inspired by dimensional reduction (DR), allowed us to resum the known four-loop naive weak coupling expansion of the quantity.

Having obtained the pressure at nonzero chemical potentials, we investigated the behavior of various quantities relevant for the physics of heavy ion collisions, such as the ratios of low order cumulants of the partition function for the quark and baryon conserved charges. We were able to compare our findings with the most up-to-date lattice data as well as with recent truncated three-loop HTLpt calculations. We found that for most of the quantities, e.g. the second- and fourth-order diagonal quark number susceptibilities or the chemical potential dependence of the pressure itself, the agreement between the HTLpt and DR frameworks, as well as between them and the lattice data, is very good starting from temperatures below $500$ MeV. In fact, only two types of exception were noticed. First, for the sixth-order quark number susceptibilities, studied for both three and two flavors, our one-loop HTLpt and resummed DR predictions were in clear disagreement, one being positive and the other negative for all the displayed temperatures. This behavior was largely attributed to the fact that for this quantity, weak coupling expansions only start at order $g^3$. Second, by comparing the fourth-order susceptibilities as well as their ratios with the second-order ones in the quark and baryon number channels, we noticed that convergence was somewhat slower for the baryon number related quantities. This may be due to the fact that the difference of the two quantities is proportional to off-diagonal contributions, which start to contribute to the weak coupling series only at four-loop order, but may nevertheless be numerically significant.

Finally, we studied the convergence of the mass truncation that is typically used in HTLpt calculations, which was made possible by the fact that we had computed the exact pressure to one-loop order. We saw that the truncated results converge towards the exact unexpanded ones in a very fast manner, giving us confidence on higher loop order HTLpt calculations such as~\cite{HTLptFiniteMUThreeLoop1,HTLptFiniteMUThreeLoop2,NanHTLpt1,NanHTLpt2}, for which this further approximation is unavoidable.

In summary, we have seen that the behavior of quantities such as quark number susceptibilities or the chemical potential dependent part of the pressure can often be accurately described by means of resummed perturbation theory, already at temperatures of about $T\approx 300\,\mbox{--}\,500$ MeV. Such an outcome is clearly an indication that a weakly interacting quasiparticle picture may be valid at least for some quantities, sensitive to the fermionic sector of the theory, already at temperatures relatively close to the phase transition region. We would like to point out that all of our perturbative methods are perfectly extensible for very large values of the ratios $\mu_f/T$, thereby ensuing the validity of our findings to a region of the QCD phase diagram, where there is no other first principles method available.

Finally, let us point out that we are eagerly waiting for new lattice results, in particular regarding high order cumulants which remain to be improved towards the continuum limit, but also regarding ratios of susceptibilities. Given the range of temperatures in which we have studied these quantities, it would be very interesting to investigate further the effects of nonvanishing bare quark masses, especially in the context of improvements for the HTLpt framework.

%%%%%%%%%%%%%%%%%%%%%%%%%%%%%%%%%%%%%%%%%%%%%%%%%%
%                  APPENDICES.
%%%%%%%%%%%%%%%%%%%%%%%%%%%%%%%%%%%%%%%%%%%%%%%%%%
\appendix
\chapter{Notation and useful relations}\label{appendix:Notation}

In this appendix, after presenting the notation that we are using in this manuscript, we collect some of the useful relations necessary for our calculations.

First of all, we recall that in the imaginary time (Matsubara) formalism, the Euclidean four-momentum reads $K=(K_0,{\bf k})$ and satisfies $K^2=K_0^2+{\bf k}^2$, following from the definition of the signature of the metric, whose temporal component has an opposite sign with respect to the one in the Minkowski space-time. The norm of the spatial component of $K$ is denoted by $k \equiv \left|{\bf k}\right|$, while $K_0$ becomes discrete, thanks to the compactification of the temporal direction
\beqa
K_0&=&\omega_n\;,\hspace{2cm} ({\rm bosons})\,, \\
K_0&=&\tilde{\omega}_n-i\mu_f\;,\hspace{1cm} ({\rm fermions})\,,
\eeqa
with $\omega_n=2n\pi T$, $\tilde{\omega}_n=(2n+1)\pi T$, and for which $n$ spans the whole set of integers. The quark chemical potential corresponding to the flavor $f$ is denoted by $\mu_f$. More generally, any index in {\ita{italic}} denotes a variable that eventually has to be summed over as, e.g., for $\mu_f$ where $f$ ranges over all the flavors. On the other hand, any straight index serves only the purpose of a reminder, as e.g. with $N_\trmi{f}$. In addition, the spatial component of any vector is generally denoted in {\bfseries{bold}}, while the full vector is not. When working in Euclidean space-time, raised and lowered indices are equivalently summed over.

We then define our dimensionally regularized sum-integrals as
\beqa
\sumint_K&=&\left(\frac{\lmsb^2\,e^{\gamma_\trmi{\tiny E}}}{4\pi}\right)^{\epsilon}T\sum_{\omega_n} \int\kern-0.5em\frac{\mathop{{\rm d}^{3-2\epsilon}\!}\nolimits {\bf k}}{(2\pi)^{3-2\epsilon}}\,,\\
\sumint_{\{K\}}&=&\left(\frac{\lmsb^2\,e^{\gamma_\trmi{\tiny E}}}{4\pi}\right)^{\epsilon}T\sum_{\tilde{\omega}_n} \int\kern-0.5em\frac{\mathop{{\rm d}^{3-2\epsilon}\!}\nolimits {\bf k}}{(2\pi)^{3-2\epsilon}}\,,
\eeqa
for the bosonic and the fermionic sum-integrals, respectively. For practical purposes, we shall abbreviate the three-momentum integrals by the following short-hand notation
\beqa
\int_{\bf k}&\equiv&\left(\frac{\lmsb^2\,e^{\gamma_\trmi{\tiny E}}}{4\pi}\right)^{\epsilon}\int\kern-0.5em\frac{\mathop{{\rm d}^{3-2\epsilon}\!}\nolimits {\bf k}}{(2\pi)^{3-2\epsilon}}\;.
\eeqa

Moreover, along our presentation, we rescale the electric screening mass and the three-dimensional gauge coupling of EQCD in the following manner
\beq
\hat m_\trmi{E}\equiv \frac{m_\trmi{E}}{T} \, ,\quad \hat g_3^2 \equiv \frac{g_3^2}{T}\,,
\eeq
as well as the HTLpt mass parameters and, eventually, the chemical potentials as
\beq
\hat m_\trmi{D}\equiv \frac{m_\trmi{D}}{2\pi T} \, ,\quad \hat m_\trmi{q$_f$}\equiv \frac{m_\trmi{q$_f$}}{2\pi T} \, , \quad \hat \mu_f\equiv \frac{\mu_f}{2\pi T}\,.
\eeq

Notice also that we denote by $\hat{H}$ a given quantum mechanical operator in the Heisenberg picture, while $\mathcal{\hat{H}}$ stands for its corresponding density in $d$-spatial dimensions
\beq
\hat{H}\equiv\int\!\mathop{{{\rm d}^d}\vect{x}}\ \mathcal{\hat{H}}\,.
\eeq

Finally, the Euler-Mascheroni constant intimately related to the modified minimal subtraction scheme ${\overline{\mbox{\trmi{MS}}}}$ reads $\gamma_\trmi{\tiny E}\approx 0.577216\,$. In addition, we denote the dimensionalities of the fermion and gluon representations of the SU($N_\trmi{c}$) group by $d_\trmi{F}=N_\trmi{c} N_\trmi{f}$ and $d_\trmi{A}=N_\trmi{c}^2-1$, as well as some further group theory factors by $C_\trmi{A}=N_\trmi{c}$, $C_\trmi{F}=d_\trmi{A}/(2N_\trmi{c})$ and $T_\trmi{F}=N_\trmi{f}/2$.

In the following, we collect some analytic properties of functions used for the implementation of the so-called contour trick. Indeed, we know that
\beq
\begin{array}{|c|c|}
\hline
\mbox{\scriptsize Function} & \mbox{\scriptsize Simple (isolated) pole of unit residue at each} \\
\hline
\frac{\beta}{2}\coth\left( \frac{\beta\omega}{2} \right) & \omega= i \left( 2\pi T n \right) \\ & \\
\hline
\frac{\beta}{2}\tanh\left( \frac{\beta\omega}{2} \right) & \omega= i \left( 2n+1 \right) \pi T  \\ & \\
\hline
\frac{\beta}{2}\tanh\left( \frac{\beta\left[\omega-\mu\right]}{2} \right) & \omega= i \left( 2n+1 \right) \pi T + \mu \\ & \\
\hline
\end{array}
\eeq
And we relate these hyperbolic functions to the exponential one, following
\beqa
\frac{1}{2}\coth\left( \frac{\beta\omega}{2} \right) &=& \frac{1}{2}+\frac{1}{e^{\beta\omega}-1} \label{CothRelation}\\
 &=& \frac{1}{2}\left[ \frac{1}{e^{\beta\omega}-1}-\frac{1}{e^{-\beta\omega}-1} \right] \,,
\eeqa
\beqa
\frac{1}{2}\tanh\left( \frac{\beta\omega}{2} \right) &=& \frac{1}{2}-\frac{1}{e^{\beta\omega}+1} \\
 &=& \frac{1}{2}\left[ \frac{1}{e^{-\beta\omega}+1}-\frac{1}{e^{\beta\omega}+1} \right] \,,
\eeqa
\beqa
\frac{1}{2}\tanh\left( \frac{\beta\left[\omega-\mu\right]}{2} \right) &=& \frac{1}{2}-\frac{1}{e^{\beta\left[\omega-\mu\right]}+1} \\
 &=& \frac{1}{2}\left[ \frac{1}{e^{-\beta\left[\omega-\mu\right]}+1}-\frac{1}{e^{\beta\left[\omega-\mu\right]}+1} \right] \,.
\eeqa

\chapter{EQCD matching coefficients}\label{appendix:EQCD_Matching_Coefficients}

In this appendix, we first give the matching coefficients of EQCD, followed by some more practical details. Before to do so, let us define some functions involving derivatives of the generalized Riemann Zeta function, $\zeta'(n,z)\equiv \partial_n \zeta(n,z)$, as well as the digamma function $\Psi(z)=\Gamma'(z)/\Gamma(z)$
{\allowdisplaybreaks
\beqa
\aleph(n,\mu_1,\mu_2)&\equiv&\aleph(n,\mu_1+\mu_2) \, , \label{DefAlephFunctions1} \\
\aleph(n,\mu)&\equiv&\zeta'(-n,1/2-i\ \hat \mu)+(-1)^{n+1}\zeta'(-n,1/2+i\ \hat \mu) \, , \label{DefAlephFunctions2} \\
\aleph(\mu)&\equiv&\Psi(1/2-i\ \hat \mu)+\Psi(1/2+i\ \hat \mu) \; , \label{DefAlephFunctions3}
\eeqa}
\hspace{-0.12cm}where $n$ is assumed to be a non-negative integer and the ratio $\hat\mu$ to be real. The matching coefficients, originally derived\footnote{Albeit for a typo in the $\aE{3}$ expression that is not present here. See~\cite{sylvain2} for more details.} in~\cite{AleksiFirstPaperPressure}, read
{\allowdisplaybreaks
\beqa
\aE{1}&=&\frac{\pi^2}{45\,N_\trmi{f}}\ \sum_f\bigg\{d_\trmi{A}+d_\trmi{F}\bigg(\frac{7}{4}+30\ \hat \mu_f^2 + 60\ \hat \mu_f^4\bigg)\bigg\} \, ,\\
\aE{2}&=&-\ \frac{d_\trmi{A}}{144\,N_\trmi{f}}\ \sum_f\bigg\{C_\trmi{A}+\frac{T_\trmi{F}}{2}\bigg(1+12\ \hat \mu_f^2\bigg)\bigg(5+12\ \hat \mu_f^2\bigg)\bigg\} \, ,\\
\aE{3}&=&\frac{d_\trmi{A}}{144\,N_\trmi{f}}\sum_f\Bigg\{C_\trmi{A}^2\bigg(\frac{194}{3}\log\frac{\bar{\Lambda}}{4\pi T}+\frac{116}{5}+4\gamma_\trmi{\tiny E}-\frac{38}{3}\frac{\zeta'(-3)}{\zeta(-3)}+\frac{220}{3}\frac{\zeta'(-1)}{\zeta(-1)}\bigg) \, \nonumber \\
& & \ \ \ \ \ \ \ \ \ \ \ +C_\trmi{A}\ T_\trmi{F} \bigg[\bigg(\frac{169}{3}+600\ \hat \mu_f^2-528\ \hat \mu_f^4\bigg)\log\frac{\bar{\Lambda}}{4\pi T}+\frac{1121}{60}+8\gamma_\trmi{\tiny E} \, \nonumber \\
& & \ \ \ \ \ \ \ \ \ \ \ +2\,\bigg(127+48\gamma_\trmi{\tiny E}\bigg)\ \hat \mu_f^2+ \frac{4}{3}\bigg(11+156\ \hat \mu_f^2\bigg)\frac{\zeta'(-1)}{\zeta(-1)}-644\ \hat \mu_f^4 \, \nonumber \\
& & \ \ \ \ \ \ \ \ \ \ \ +\frac{268}{15}\frac{\zeta'(-3)}{\zeta(-3)}+24\bigg(52\ \aleph(3,\mu_f)+4\ i\ \hat \mu_f\ \aleph(0,\mu_f) \, \nonumber \\
& & \ \ \ \ \ \ \ \ \ \ \ +144\ i\ \hat \mu_f\ \aleph(2,\mu_f)+\bigg(17-92\ \hat \mu_f^2\bigg)\ \aleph(1,\mu_f)\bigg)\bigg] \ \ \ \ \ \ \ \ \ \ \, \nonumber \\
& & \ \ \ \ \ \ \ \ \ \ \ +C_\trmi{F}\,T_\trmi{F} \bigg[\frac{3}{4}\bigg(1+4\ \hat \mu_f^2\bigg)\bigg(35+332\ \hat \mu_f^2\bigg)-24\bigg(1-12\ \hat \mu_f^2\bigg)\frac{\zeta'(-1)}{\zeta(-1)} \, \nonumber \\
& & \ \ \ \ \ \ \ \ \ \ \ -144\,\bigg(12\ i\ \hat \mu_f\ \aleph(2,\mu_f)-2\bigg(1+8\ \hat \mu_f^2\bigg)\ \aleph(1,\mu_f) \, \nonumber \\
& & \ \ \ \ \ \ \ \ \ \ \ -\ i\ \hat \mu_f\bigg(1+4\ \hat \mu_f^2\bigg)\ \aleph(0,\mu_f)\bigg)\bigg] \, \nonumber \\
& & \ \ \ \ \ \ \ \ \ \ \ + T_\trmi{F}^2 \bigg[\frac{4}{3}\bigg(1+12\ \hat \mu_f^2\bigg)\bigg(5+12\ \hat \mu_f^2\bigg) \log\frac{\bar{\Lambda}}{4\pi T}+\frac{1}{3}+4\,\gamma_\trmi{\tiny E} \, \nonumber \\
& & \ \ \ \ \ \ \ \ \ \ \ +8\,\bigg(7+12\gamma_\trmi{\tiny E}\bigg)\ \hat \mu_f^2+112\ \hat \mu_f^4-\frac{32}{3}\bigg(1+12\ \hat \mu_f^2\bigg)\frac{\zeta'(-1)}{\zeta(-1)} \, \nonumber \\
& & \ \ \ \ \ \ \ \ \ \ \ -\frac{64}{15}\,\frac{\zeta'(-3)}{\zeta(-3)}-96\bigg(8\ \aleph(3,\mu_f) + 12\ i\ \hat \mu_f\ \aleph(2,\mu_f) \, \nonumber \\
& & \ \ \ \ \ \ \ \ \ \ \ - 2\,(1+2\ \hat \mu_f^2)\,\aleph(1,\mu_f)-\ i\ \hat \mu_f\ \aleph(0,\mu_f)\bigg)\bigg] \, \nonumber \\
& & \ \ \ \ \ \ \ \ \ \ \ +\frac{288\,T_\trmi{F}^2}{N_\trmi{f}}\, \sum_g\bigg\{2\ \bigg(1+\gamma_\trmi{\tiny E}\bigg)\ \hat \mu_f^2\ \hat \mu_g^2-\,\aleph(3,\mu_f,\mu_g)-\,\aleph(3,\mu_f,-\mu_g) \, \nonumber \\
& & \ \ \ \ \ \ \ \ \ \ \ \ \ \ \ \ \ \ \ \ \ \ +4\ \hat \mu_g^2\ \aleph(1,\mu_f)-4\ i\ \hat \mu_f\bigg(\aleph(2,\mu_f,\mu_g) + \aleph(2,\mu_f,-\mu_g)\bigg) \, \nonumber \\
& & \ \ \ \ \ \ \ \ \ \ \ \ \ \ \ \ \ \ \ \ \ \ +\bigg(\ \hat \mu_f+\ \hat \mu_g\bigg)^2\aleph(1,\mu_f,\mu_g)+4\ i\ \hat \mu_f\ \hat \mu_g^2\ \aleph(0,\mu_f) \, \nonumber \\
& & \ \ \ \ \ \ \ \ \ \ \ \ \ \ \ \ \ \ \ \ \ \ + \bigg(\ \hat \mu_f-\ \hat \mu_g\bigg)^2\aleph(1,\mu_f,-\mu_g)\bigg\}\Bigg\} \, , \label{AE3Corrected}\\
\aE{4}&=&\frac{1}{3\,N_\trmi{f}}\ \sum_f\bigg\{C_\trmi{A}+T_\trmi{F}\bigg(1+12\ \hat \mu_f^2\bigg)\bigg\} \, ,\\
\aE{5}&=&\frac{1}{3\,N_\trmi{f}}\ \sum_f\bigg\{2\ C_\trmi{A}\bigg(\log\frac{\bar{\Lambda}}{4\pi T}+\frac{\zeta'(-1)}{\zeta(-1)}\bigg) \, \nonumber \\
& & \ \ \ \ \ \ \ \ \ +T_\trmi{F} \bigg[\bigg(1+12\ \hat \mu_f^2\bigg)\bigg(2\ \log\frac{\bar{\Lambda}}{4\pi T}+1\bigg)+ 24\ \aleph(1,\mu_f)\bigg]\bigg\} \, ,\\
\aE{6}&=&\frac{1}{9\,N_\trmi{f}}\ \sum_f\bigg\{C_\trmi{A}^2\bigg(22\ \log\frac{e^{\gamma_\trmi{\tiny E}}\bar{\Lambda}}{4\pi T}+5\bigg)-18\,C_\trmi{F}\,T_\trmi{F} \bigg(1+12\ \hat \mu_f^2\bigg) \, \nonumber \\
& & \ \ \ \ \ \ \ \ \ + C_\trmi{A}\,T_\trmi{F} \bigg[2\,\bigg(7+132\ \hat \mu_f^2\bigg)\log\frac{e^{\gamma_\trmi{\tiny E}}\bar{\Lambda}}{4\pi T}+9+132\ \hat \mu_f^2+8\gamma_\trmi{\tiny E} + 4\ \aleph(\mu_f)\bigg] \, \nonumber \\
& & \ \ \ \ \ \ \ \ \ -4\,T_\trmi{F}^2\bigg(1+12\ \hat \mu_f^2\bigg)\bigg(2 \log\frac{\bar{\Lambda}}{4\pi T}-1-\aleph(\mu_f)\bigg)\bigg\} \, ,\\
\aE{7}&=&\frac{1}{3\,N_\trmi{f}}\sum_f\bigg\{C_\trmi{A}\bigg(22 \log\frac{e^{\gamma_\trmi{\tiny E}}\bar{\Lambda}}{4\pi T}+1\bigg)-4\ T_\trmi{F}\bigg(2 \log\frac{\bar{\Lambda}}{4\pi T}-\aleph(\mu_f)\bigg)\bigg\}\;.
\eeqa}

These expressions being analytical for real values of the chemical potentials, it is actually more practical to work with $\aleph$ functions which are already expanded in powers of $\hat \mu_f$, if one is interested in calculating susceptibilities at vanishing chemical potentials. Of course, regarding the pressure itself as a function of the chemical potentials (a priori with arbitrary values), it is better to avoid such an unnecessary approximation, and one should work with the above full expressions instead. For completeness, we list the Taylor expanded expressions here
{\allowdisplaybreaks
\beqa
\aleph(\mu_i)&=&-2 \bigg(\gamma_\trmi{\tiny E}+\log 4\bigg)+14\ \zeta(3) \ \hat \mu_i^2-62\ \zeta(5) \ \hat \mu_i^4+254\ \zeta(7) \ \hat \mu_i^6 \, \nonumber \\
&-& 1022\ \zeta(9) \ \hat \mu_i^8+4094\ \zeta(11) \ \hat \mu_i^{10}+ {\cal O}\bigg(\hat \mu_i^{12}\bigg) \, ,\\
-i\ \aleph(0,\mu_i)&=&2\ \hat \mu_i \Big(\gamma_\trmi{\tiny E}+\log 4\Big)- \frac{14}{3}\ \zeta(3)\ \hat \mu_i^3+\frac{62}{5}\ \zeta(5)\ \hat \mu_i^5-\frac{254}{7}\ \zeta(7)\ \hat \mu_i^7 \, \nonumber \\
&+& \frac{1022}{9}\ \zeta(9)\ \hat \mu_i^9+ {\cal O}\bigg(\hat \mu_i^{11}\bigg) \, ,\\
\aleph(1,\mu_i)&=&-\zeta'(-1)-\frac{\log 2}{12}+\bigg(\log 4-1+\gamma_\trmi{\tiny E}\bigg)\ \hat \mu_i^2-\frac{7}{6}\ \zeta(3)\ \hat \mu_i^4 \, \nonumber \\
&+& \frac{31}{15}\ \zeta(5)\ \hat \mu_i^6-\frac{127}{28}\ \zeta(7)\ \hat \mu_i^8+\frac{511}{45}\ \zeta(9)\ \hat \mu_i^{10}+ {\cal O}\bigg(\hat \mu_i^{12}\bigg) \, ,\\
-i\ \aleph(2,\mu_i)&=&\frac{1}{12}\Big(1+\log 4+24\ \zeta'(-1)\Big)\ \hat \mu_i-\frac{1}{3}\Big(2\ \gamma_\trmi{\tiny E}-3+\log 16\Big)\ \hat \mu_i^3 \, \nonumber \\
&+& \frac{7}{15}\ \zeta(3)\ \hat \mu_i^5-\frac{62}{105}\ \zeta(5)\ \hat \mu_i^7+\frac{127}{126}\ \zeta(7)\ \hat \mu_i^9+ {\cal O}\bigg(\hat \mu_i^{11}\bigg) \, ,\\
\aleph(3,\mu_i)&=&\frac{1}{480}\Big(\log 2-840\ \zeta'(-3)\Big)+\frac{1}{24}\Big(5+\log 64+72\ \zeta'(-1)\Big)\ \hat \mu_i^2 \, \nonumber \\
&-& \frac{1}{12}\Big(6\ \gamma_\trmi{\tiny E}-11+\log 4096\Big)\ \hat \mu_i^4+\frac{7}{30}\ \zeta(3)\ \hat \mu_i^6-\frac{31}{140}\ \zeta(5)\ \hat \mu_i^8 \, \nonumber \\
&+& \frac{127}{420}\ \zeta(7)\ \hat \mu_i^{10}+{\cal O}\bigg(\hat \mu_i^{12}\bigg) \, ,\\
\aleph(1,\mu_i,\mu_j)&=&2\ \zeta'(-1)+\bigg(\gamma_\trmi{\tiny E}-1\bigg)\ \left(\hat \mu_i+\hat \mu_j\right)^2-\frac{\zeta(3)}{6}\ \left(\hat \mu_i+\hat \mu_j\right)^4 \, \nonumber \\
&+& \frac{\zeta(5)}{15}\ \left(\hat \mu_i+\hat \mu_j\right)^6-\frac{\zeta(7)}{28}\ \left(\hat \mu_i+\hat \mu_j\right)^8 \, \nonumber \\
&+& \frac{\zeta(9)}{45}\ \left(\hat \mu_i+\hat \mu_j\right)^{10}+ {\cal O}\bigg(\hat \mu_i^{12},\hat \mu_j^{12}\bigg) \, ,\\
-i\ \aleph(2,\mu_i,\mu_j)&=&-\bigg(4\ \zeta'(-1)+\frac{1}{6}\bigg)\left(\hat \mu_i+\hat \mu_j\right)+\bigg(1-\frac{2}{3}\gamma_\trmi{\tiny E}\bigg)\left(\hat \mu_i+\hat \mu_j\right)^3 \, \nonumber \\
&+& \frac{\zeta(3)}{15}\ \left(\hat \mu_i+\hat \mu_j\right)^5-\frac{2\ \zeta(5)}{105}\left(\hat \mu_i+\hat \mu_j\right)^7 \, \nonumber \\
&+& \frac{\zeta(7)}{126}\left(\hat \mu_i+\hat \mu_j\right)^9+{\cal O}\bigg(\hat \mu_i^{11},\hat \mu_j^{11}\bigg) \, ,\\
\aleph(3,\mu_i,\mu_j)&=&2\ \zeta'(-3)-\bigg(6\ \zeta'(-1)+\frac{5}{12}\bigg)\left(\hat \mu_i+\hat \mu_j\right)^2 \, \nonumber \\
&+& \bigg(\frac{11}{12}-\frac{\gamma_\trmi{\tiny E}}{2}\bigg)\left(\hat \mu_i+\hat \mu_j\right)^4+\frac{\zeta(3)}{30}\left(\hat \mu_i+\hat \mu_j\right)^6 \, \nonumber \\
&-& \frac{\zeta(5)}{140}\left(\hat \mu_i+\hat \mu_j\right)^8+\frac{\zeta(7)}{420}\left(\hat \mu_i+\hat \mu_j\right)^{10} +{\cal O}\bigg(\hat \mu_i^{12},\hat \mu_j^{12}\bigg) \, . \ \ \ \ \ \ \ 
\eeqa}

\chapter{One-loop truncated HTL master sum-integrals}\label{appendix:HTL_Master_Sum-Integrals}

In this appendix, we provide details about the computation of the one-loop HTLpt sum-integrals in the context of the mass truncated approximation. We list the needed results for evaluating the pressure up to ${\cal O}\left(g^5\right)$. First, let us define the general forms of one-loop sum-integrals that are encountered
{\allowdisplaybreaks
\beqa
\mathcal{I}^{u,m}_{w,\,l}&\equiv&\sumint_{K}\left[\frac{\left(\omega_n\right)^u\,({\cal T}_\trmi{K})^m}{\left(k^{2}\right)^{w}\,\left(K^{2}\right)^{l}}\right] \label{MasterSumIntB} \, , \\
\widetilde{\mathcal{I}}^{u,m}_{w,\,l}&\equiv&\sumint_{\{K\}}\left[\frac{\left(\widetilde{\omega}_n-i\mu_f\right)^u\,(\widetilde{{\cal T}}_\trmi{K})^m}{\left(k^{2}\right)^{w}\,\left(K^{2}\right)^{l}}\right] \label{MasterSumIntF} \, ,
\eeqa}
\hspace{-0.12cm}where ${\cal T}_\trmi{K}$ and $\widetilde{{\cal T}}_\trmi{K}$ are the so-called HTL functions~(\ref{HTLfunctionB}) and~(\ref{HTLfunctionF}), respectively.

For $m=0$, these sum-integrals reduce to the known one-loop bosonic and fermionic results, which appear within the naive weak coupling expansion. Their evaluation, using here dimensional regularization, can be done analytically. First integrating over spatial momenta, then using the (generalized) Riemann Zeta function to represent the discrete summation over Matsubara frequencies, we get
{\allowdisplaybreaks
\beqa
\mathcal{I}^{u,0}_{w,\,l}&=&e^{\gamma_\trmi{\tiny E}\epsilon}\left(\frac{\lmsb}{4\pi T}\right)^{2\epsilon}\left[\frac{\left(2\pi T\right)^{4+u-2(l+w)}}{(2\pi)^{3}}\right]\left[\frac{\Gamma\left(\frac{3}{2}-\epsilon-w\right)\Gamma\left(\epsilon-\frac{3}{2}+w+l\right)\Gamma\left(1-\epsilon\right)}{\Gamma\left(2-2\epsilon\right)\Gamma(l)} \right] \nonumber \\
& & \ \ \ \ \ \ \ \ \ \ \ \ \ \ \ \ \ \ \ \ \ \times\bigg[\big(1+(-1)^u\big)\,\zeta\Big(2\epsilon-3-u+2l+2w\Big)\bigg] \, , \\
\widetilde{\mathcal{I}}^{u,0}_{w,\,l}&=&e^{\gamma_\trmi{\tiny E}\epsilon}\left(\frac{\lmsb}{4\pi T}\right)^{2\epsilon}\left[\frac{\left(2\pi T\right)^{4+u-2(l+w)}}{(2\pi)^{3}}\right]\left[\frac{\Gamma\left(\frac{3}{2}-\epsilon-w\right)\Gamma\left(\epsilon-\frac{3}{2}+w+l\right)\Gamma\left(1-\epsilon\right)}{\Gamma\left(2-2\epsilon\right)\Gamma(l)} \right] \nonumber \\
& & \ \ \ \ \ \ \ \ \ \ \ \ \ \ \ \ \ \ \ \ \ \times\left[\zeta\Big(2\epsilon-3-u+2l+2w\,;\frac{1}{2}-\frac{i\mu_f}{2\pi T}\Big) \right. \nonumber \\
& & \ \ \ \ \ \ \ \ \ \ \ \ \ \ \ \ \ \ \ \ \ \ \ \ \ \ \ \ \ \ \left.+\,(-1)^{u}\,\,\zeta\Big(2\epsilon-3-u+2l+2w\,;\frac{1}{2}+\frac{i\mu_f}{2\pi T}\Big)\right] \,,
\eeqa}
\hspace{-0.12cm}where $T$ and $\lmsb$ have been assumed to be real and positive. For completeness, we list the $m=0$ results, expanded around $\epsilon=0$ to a sufficient order for the present purpose
{\allowdisplaybreaks
\beqa
\sumint_{K}\log(K^2)\!\!\!\!&=&\!\!\!\!-\frac{\pi^2T^4}{45}\left(\frac{\bar{\Lambda}}{4\pi T}\right)^{2\epsilon}\!\bigg[1+{\cal O}(\epsilon)\bigg]\;,\\
\sumint_{\{K\}}\log(K^2)\!\!\!\!&=&\!\!\!\!\frac{7\pi^2}{360}\left(\frac{\bar{\Lambda}}{4\pi T}\right)^{2\epsilon}\!\left[T^4+\frac{30\mu_f^2T^2}{7\pi^2}+\frac{15\mu_f^4}{7\pi^4}+{\cal O}(\epsilon)\right]\;,\\
{\cal I}^{0,0}_{0,1}\!\!\!\!&=&\!\!\!\!\frac{1}{12}\left(\frac{\bar{\Lambda}}{4\pi T}\right)^{2\epsilon}\!\bigg[T^2+{\cal O}(\epsilon)\bigg]\;,\\
{\cal \widetilde{I}}^{0,0}_{0,1}\!\!\!\!&=&\!\!\!\!-\frac{1}{24}\left(\frac{\bar{\Lambda}}{4\pi T}\right)^{2\epsilon}\!\left[T^2+\frac{3\mu_f^2}{\pi^2}+{\cal O}(\epsilon)\right]\\
{\cal I}^{0,0}_{0,2}\!\!\!\!&=&\!\!\!\!\frac{1}{(4\pi)^2}\left(\frac{\bar{\Lambda}}{4\pi T}\right)^{2\epsilon}\!\left[\frac{1}{\epsilon}+2\gamma_E+{\cal O}(\epsilon)\right]\;,\\
{\cal \widetilde{I}}^{0,0}_{0,2}\!\!\!\!&=&\!\!\!\!\frac{1}{(4\pi)^2}\left(\frac{\bar{\Lambda}}{4\pi T}\right)^{2\epsilon}\!\left[\frac{1}{\epsilon}-\Psi\left(\frac{1}{2}+\frac{i\mu_f}{2\pi T}\right)-\Psi\left(\frac{1}{2}-\frac{i\mu_f}{2\pi T}\right)+{\cal O}(\epsilon)\right]\,,\\
{\cal I}^{0,0}_{1,1}\!\!\!\!&=&\!\!\!\!\frac{2}{(4\pi)^2}\left(\frac{\bar{\Lambda}}{4\pi T}\right)^{2\epsilon}\!\left[\frac{1}{\epsilon}+2\gamma_E+2+{\cal O}(\epsilon)\right]\;,\\
{\cal \widetilde{I}}^{0,0}_{1,1}\!\!\!\!&=&\!\!\!\!\frac{2}{(4\pi)^2}\left(\frac{\bar{\Lambda}}{4\pi T}\right)^{2\epsilon}\!\left[\frac{1}{\epsilon}+2-\Psi\left(\frac{1}{2}+\frac{i\mu_f}{2\pi T}\right)-\Psi\left(\frac{1}{2}-\frac{i\mu_f}{2\pi T}\right)+{\cal O}(\epsilon)\right]\;.\ \ \ \ \ \ \ \ \ 
\eeqa}

For $m\neq0$, the sum-integrals get more complicated, the HTL functions being nothing but Gauss hypergeometric functions. A closed form can be obtained, provided that $m$ and $l$ are positive integers. For this purpose, we make use of an integral representation on the real axis for the Gauss hypergeometric function, in which terms like $1/[(\omega_n^2+k^2)(\omega_n^2+c^2k^2)]$ appear. Due to such terms, one cannot straightforwardly rescale the variable $k$ and proceed by applying the same computational techniques as in the $m=0$ case. However, one can use the following decomposition in order to separate the denominator which is $c$-independent from the one which is $c$-dependent
\beq
\frac{1}{(\omega_n^2+k^2)(\omega_n^2+c^2k^2)}=\frac{1}{k^2\,(c^2-1)}\left[\frac{1}{(\omega_n^2+k^2)}-\frac{1}{(\omega_n^2+c^2k^2)}\right]\;.
\eeq
The above relation is generalizable for arbitrary $l$ integer powers of the $c$-independent term
\beq
\frac{1}{(\omega_n^2+k^2)^l(\omega_n^2+c^2k^2)}=\frac{(1-c^2)^{-l}}{(k^2)^l(\omega_n^2+c^2k^2)}-\sum_{r=1}^l\left[\frac{(1-c^2)^{-r}}{(k^2)^r\,(\omega_n^2+k^2)^{l-r+1}}\right] \,,
\eeq
which gives a useful identity for the $m=1$ case. The subsequent generalization, for higher $m$ integer powers, related to $m$ terms being $(c_1,...,c_m)$-dependent, gives in the $m=2$ case
\beqa
\frac{1}{(\omega_n^2+k^2)^l(\omega_n^2+c_1^2k^2)(\omega_n^2+c_2^2k^2)}&=&\frac{(1-c_1^2)^{-l}}{(c_2^2-c_1^2)}\,\,\frac{1}{(k^2)^{1+l}}\left[\frac{1}{(\omega_n^2+c_1^2k^2)}-\frac{1}{(\omega_n^2+c_2^2k^2)}\right]\, \nonumber \\
&-&\sum_{r=1}^l\left[\frac{(1-c_1^2)^{-r}}{(1-c_2^2)^{1+l-r}}\right]\,\frac{1}{(k^2)^{1+l}\,(\omega_n^2+c_2^2k^2)}\, \\
&+&\sum_{r=1}^l\sum_{s=1}^{1+l-r}\left[\frac{(1-c_1^2)^{-r}}{(1-c_2^2)^{s}}\,\,\frac{1}{(k^2)^{r+s}\,(\omega_n^2+k^2)^{2+l-r-s}}\right]\,. \nonumber
\eeqa
Having done so, it is straightforward to rescale each term separately, and then perform the remaining one dimensional $c_i$-integrations. This leaves us with sums of the usual $m=0$ sum-integrals where some coefficients appear, which typically are expressible in terms of analytical functions.

For the purpose of obtaining the one-loop pressure at order $g^5$, we need sum-integrals with $m=1$ and $m=2$ only, for which we give the following decompositions in terms of basic $m=0$ sum-integrals
\beqa
\mathbfcal{I}^{u,1}_{w,\,l}&=&\mathcal{J}_{w,l}^{\mbox{\tiny$(1)$}}\,\,\mathbfcal{I}^{u+2,0}_{w+l,\,1}-\sum_{r=1}^l\bigg[\mathcal{J}_{r}^{\mbox{\tiny$(1)$}}\,\,\mathbfcal{I}^{u+2,0}_{w+r,\,l-r+1}\bigg]\,, \\
\mathbfcal{I}^{u,2}_{w,\,l}&=&\mathcal{J}_{w,l}^{\mbox{\tiny$(2)$}}\,\,\mathbfcal{I}^{u+4,0}_{w+l+1,\,1}-\sum_{r=1}^l\bigg[\mathcal{J}_{r,l,w}^{\mbox{\tiny$(2)$}}\bigg]\,\mathbfcal{I}^{u+4,0}_{w+l+1,\,1}+\sum_{r=1}^l\sum_{s=1}^{1+l-r}\bigg[\mathcal{J}_{r,s}^{\mbox{\tiny$(2)$}}\,\,\mathbfcal{I}^{u+4,0}_{w+r+s,\,2+l-r-s}\bigg]\,,
\eeqa
where $\mathbfcal{I}$ stands for $\mathcal{I}$ or $\widetilde{\mathcal{I}}$. In the above, we have defined the following notation\footnote{Notice the slight difference of notation compared to the original reference~\cite{sylvain2}. The reason is that we go through the $m=2$ sum-integrals here, and want to keep our notation self-explanatory.} for the coefficients
{\allowdisplaybreaks
\beqa
\mathcal{J}_{w,l}^{\mbox{\tiny$(1)$}}&\equiv&\frac{\Gamma\left(\frac{3}{2}-\epsilon\right)}{\Gamma\left(\frac{3}{2}\right)\Gamma\left(1-\epsilon\right)}\int^{1}_{0}\kern-0.5em\mathop{{\rm d}\!}\nolimits c \ \frac{c^{2\epsilon-3+2(l+w)}}{(1-c^2)^{l+\epsilon}}=\frac{\Gamma\left(\frac{3}{2}-\epsilon\right)\Gamma\left(1-\epsilon-l\right)\Gamma\left(\epsilon-1+l+w\right)}{\Gamma\left(\frac{1}{2}\right)\,\Gamma\left(w\right)\Gamma\left(1-\epsilon\right)}\,, \\
\mathcal{J}_{r}^{\mbox{\tiny$(1)$}}&\equiv&\frac{\Gamma\left(\frac{3}{2}-\epsilon\right)}{\Gamma\left(\frac{3}{2}\right)\Gamma\left(1-\epsilon\right)}\int^{1}_{0}\kern-0.5em\mathop{{\rm d}\!}\nolimits c \ (1-c^2)^{-(r+\epsilon)}=\frac{\Gamma\left(\frac{3}{2}-\epsilon\right)\Gamma\left(1-\epsilon-r\right)}{\Gamma\left(1-\epsilon\right)\Gamma\left(\frac{3}{2}-\epsilon-r\right)} \,, \\
\!\!\!\!\!\!\!\!\!\!\!\!\!\!\!\!\!\!\mathcal{J}_{w,l}^{\mbox{\tiny$(2)$}}\!\!\!\!&\equiv&\!\!\!\!\frac{\Gamma\left(\frac{3}{2}-\epsilon\right)^2}{\Gamma\left(\frac{3}{2}\right)^2\Gamma\left(1-\epsilon\right)^2}\int^{1}_{0}\kern-0.5em\mathop{{\rm d}\!}\nolimits{c_1}\!\mathop{{\rm d}\!}\nolimits{c_2} \ \frac{(1-c_2^2)^{-\epsilon}}{(1-c_1^2)^{l+\epsilon}(c_2^2-c_1^2)}\,\,\bigg[(c_1^2)^{w+\epsilon-\frac{1}{2}+l}-(c_2^2)^{w+\epsilon-\frac{1}{2}+l}\bigg]\, \nonumber \\
&=&\!\!\!\!\frac{2\,\Gamma\left(\frac{3}{2}-\epsilon\right)^2\Gamma\left(1-l-2\epsilon\right)\Gamma\left(l+w-\frac{1}{2}+\epsilon\right)\cot\left(\pi\left(l+w+\epsilon\right)\right)}{\Gamma\left(1-\epsilon\right)^2\Gamma\left(w+\frac{1}{2}-\epsilon\right)}\, \\
&+&\!\!\!\!\frac{\sqrt{\pi}\,\Gamma\left(\frac{3}{2}-\epsilon\right)\Gamma\left(1-l-\epsilon\right)\csc\left(\pi\left(l+w+\epsilon\right)\right)}{\Gamma\left(1-\epsilon\right)\Gamma\left(w\right)\Gamma\left(2-l-w-\epsilon\right)}\,\, {}_3F_2\left(\frac{1}{2},1,1-w;\frac{3}{2}-\epsilon,2-l-w-\epsilon;1\right) \, \nonumber\\
&+&\!\!\!\!\frac{\sqrt{\pi}\,\Gamma\left(\frac{3}{2}-\epsilon\right)^2\Gamma\left(1-l-\epsilon\right)\csc\left(\pi\left(l+w+\epsilon\right)\right)}{\Gamma\left(\frac{3}{2}-l-\epsilon\right)\Gamma\left(1-\epsilon\right)\Gamma\left(l+w\right)\Gamma\left(2-l-w-\epsilon\right)}\,\, {}_3F_2\left(\frac{1}{2},1,1-w-l;\frac{3}{2}-l-\epsilon,2-l-w-\epsilon;1\right)\!, \nonumber\\
\!\!\!\!\!\!\!\!\!\!\!\!\!\!\!\!\!\!\mathcal{J}_{r,l,w}^{\mbox{\tiny$(2)$}}\!\!\!\!&\equiv&\!\!\!\!\frac{\Gamma\left(\frac{3}{2}-\epsilon\right)^2}{\Gamma\left(\frac{3}{2}\right)^2\Gamma\left(1-\epsilon\right)^2}\int^{1}_{0}\kern-0.5em\mathop{{\rm d}\!}\nolimits{c_1}\!\mathop{{\rm d}\!}\nolimits{c_2} \ \frac{(1-c_1^2)^{-(r+\epsilon)}}{(1-c_2^2)^{1+l-r+\epsilon}}\,\,\bigg[(c_2^2)^{w+\epsilon-\frac{1}{2}+l}\bigg] \nonumber \\
&=&\frac{\Gamma\left(\frac{3}{2}-\epsilon\right)^2\Gamma\left(1-\epsilon-r\right)\Gamma\left(r-l-\epsilon\right)\Gamma\left(l+w+\epsilon\right)}{\Gamma\left(1-\epsilon\right)^2\sqrt{\pi}\,\,\Gamma\left(\frac{3}{2}-\epsilon-r\right)\Gamma\left(r+w\right)}\,, \\
\!\!\!\!\!\!\!\!\!\!\!\!\!\!\!\!\!\!\mathcal{J}_{r,s}^{\mbox{\tiny$(2)$}}\!\!\!\!&\equiv&\!\!\!\!\frac{\Gamma\left(\frac{3}{2}-\epsilon\right)^2}{\Gamma\left(\frac{3}{2}\right)^2\Gamma\left(1-\epsilon\right)^2}\int^{1}_{0}\kern-0.5em\mathop{{\rm d}\!}\nolimits{c_1}\!\mathop{{\rm d}\!}\nolimits{c_2} \ \frac{(1-c_1^2)^{-(r+\epsilon)}}{(1-c_2^2)^{s+\epsilon}}=\frac{\Gamma\left(\frac{3}{2}-\epsilon\right)^2\Gamma\left(1-\epsilon-r\right)\Gamma\left(1-\epsilon-s\right)}{\Gamma\left(1-\epsilon\right)^2\Gamma\left(\frac{3}{2}-\epsilon-r\right)\Gamma\left(\frac{3}{2}-\epsilon-s\right)}\,.
\eeqa}

Finally, we give the expanded results for the needed $m\neq 0$ sum-integrals
\beqa
{\cal I}_{2,0}^{0,1}\!\!\!\!&=&\!\!\!\!-\frac{1}{(4\pi)^2}\left(\frac{\bar{\Lambda}}{4\pi T}\right)^{2\epsilon}\!\!\left[\frac{1}{\epsilon}+2\gamma_E+\log4+{\cal O}(\epsilon)\right]\,,\\
\!\!\!\!\!\!\!\!\!\!\!\!{\cal \widetilde{I}}_{1,1}^{0,1}\!\!\!\!&=&\!\!\!\!\frac{2}{(4\pi)^2}\left(\frac{\bar{\Lambda}}{4\pi T}\right)^{2\epsilon}\!\!\left[\log2\left\{\frac{1}{\epsilon}+\log2-\Psi\!\left(\frac{1}{2}+\frac{i\mu_f}{2\pi T}\right)-\Psi\!\left(\frac{1}{2}-\frac{i\mu_f}{2\pi T}\right)\!\right\}\!+\frac{\pi^2}{6}+{\cal O}(\epsilon)\right]\!\!,\,\,\,\,\,\,\,\,\,\,\\
{\cal I}_{1,1}^{0,1}\!\!\!\!&=&\!\!\!\!\frac{2}{(4\pi)^2}\left(\frac{\bar{\Lambda}}{4\pi T}\right)^{2\epsilon}\!\!\left[\frac{\log2}{\epsilon}+\frac{\pi^2}{6}+2\gamma_E\log2+\log^22+{\cal O}(\epsilon)\right]\,,\\
{\cal I}_{2,0}^{0,2}\!\!\!\!&=&\!\!\!\!-\frac{2}{3(4\pi)^2}\left(\frac{\bar{\Lambda}}{4\pi T}\right)^{2\epsilon}\!\!\left[\frac{1+2\log2}{\epsilon}+2\gamma_E\left(1+2\log2\right)-\frac{4}{3}+\frac{22}{3}\log2\right.+2\log^22+{\cal O}(\epsilon)\bigg]\!,\\
{\cal \widetilde{I}}_{1,0}^{\,-2,2\!\!}\!\!\!\!&=&\!\!\!\!\frac{4\log2}{(4\pi)^2}\left(\frac{\bar{\Lambda}}{4\pi T}\right)^{2\epsilon}\!\!\left[\frac{1}{\epsilon}+\log2-\Psi\left(\frac{1}{2}+\frac{i\mu_f}{2\pi T}\right)-\Psi\left(\frac{1}{2}-\frac{i\mu_f}{2\pi T}\right)+{\cal O}(\epsilon)\right] \,.
\eeqa

%%%%%%%%%%%%%%%%%%%%%%%%%%%%%%%%%%%%%%%%%%%%%%%%%%
%                  BIBLIOGRAPHY.
%%%%%%%%%%%%%%%%%%%%%%%%%%%%%%%%%%%%%%%%%%%%%%%%%%

%%%%%%%%%%%%%%%%%%%%%%%%%%%%%%%%%%%%%%%%%%%%%%%%%%
%   END of the modular document, hence the file.
%%%%%%%%%%%%%%%%%%%%%%%%%%%%%%%%%%%%%%%%%%%%%%%%%%
\end{document}